\documentclass[%
 aps,
 superscriptaddress,
 prb,
 amsmath,amssymb,
 longbibliography,
 reprint,
]{revtex4-2}

\usepackage{dcolumn}
\usepackage{bm}
\usepackage[version=3]{mhchem} 
\usepackage[utf8]{inputenc}
\usepackage[T1]{fontenc}

\usepackage{amsmath}
\usepackage{amsfonts}
\usepackage{braket}
\usepackage{siunitx}
\usepackage{dcolumn}
\usepackage{placeins}
\usepackage{booktabs}
\usepackage{graphicx}
\usepackage{dcolumn}
\usepackage{tikz}
\usepackage{array}
\usepackage{pifont,paralist,multirow,amssymb}
\usepackage[inline]{enumitem}
\usepackage{algorithm}
\usepackage{algpseudocode}
\usepackage{listings}
\usepackage{xcolor}

\usepackage{hyperref,url}
\usepackage[capitalise]{cleveref}
\hypersetup{breaklinks,colorlinks=true,linkcolor=blue,citecolor=blue,urlcolor=blue,filecolor=blue}
\usetikzlibrary{decorations.pathreplacing,calligraphy}
\usetikzlibrary{decorations.markings}
\usetikzlibrary{decorations.pathmorphing}
\usetikzlibrary{shapes.misc}

\tikzset{cross/.style={cross out, draw=black, minimum size=2*(#1-\pgflinewidth), inner sep=0pt, outer sep=0pt},
cross/.default={1pt}}

\definecolor{myblue}{RGB}{0, 102, 204} 

\newcommand{\bt}{\boldsymbol{\theta}}
\newcommand{\tr}{\text{Tr}}
\newcommand{\s}{\mathbf{s}}

\newcommand{\add}[1]{{\color{black}{#1}}}

\newcommand{\Harvardc}{\affiliation{Department of Chemistry and Chemical Biology, Harvard University, Cambridge, MA, USA}}
\newcommand{\Harvardp}{\affiliation{Department of Physics, Harvard University, Cambridge, MA, USA}}
\newcommand{\Google}{\affiliation{
Google Research, Venice, CA 90291, United States}}

\begin{document}
\title{Walking through Hilbert Space with Quantum Computers}

\author{Tong Jiang}
\thanks{These authors contributed equally to this work}
\Harvardc
\author{Jinghong Zhang}
\thanks{These authors contributed equally to this work}
\Harvardc
\author{Moritz K. A. Baumgarten}
\thanks{These authors contributed equally to this work}
\Harvardc
\author{Meng-Fu Chen}
\Harvardc
\author{Hieu Q. Dinh}
\Harvardc
\author{Aadithya Ganeshram}
\Harvardc
\author{Nishad Maskara}
\Harvardp
\author{Anton Ni}
\Harvardc
\author{Joonho Lee}
\email{joonholee@g.harvard.edu}
\Harvardc
\Google

\date{\today}

\begin{abstract}
Computations of chemical systems' equilibrium properties and non-equilibrium dynamics have been suspected of being a "killer app" for quantum computers.
This review highlights the recent advancements of quantum algorithms tackling complex sampling tasks in the key areas of computational chemistry: ground state, thermal state properties, and quantum dynamics calculations.
We review a broad range of quantum algorithms, from hybrid quantum--classical to fully quantum, focusing on the traditional Monte Carlo family, including Markov chain Monte Carlo, variational Monte Carlo, projector Monte Carlo, path integral Monte Carlo, \textit{etc.} We also cover other relevant techniques involving complex sampling tasks, such as quantum-selected configuration interaction, 
minimally entangled typical thermal states, entanglement forging, and Monte Carlo-flavored Lindbladian dynamics.
We provide a comprehensive overview of these algorithms' classical and quantum counterparts, detailing their theoretical frameworks and discussing the potentials and challenges in achieving quantum computational advantages.
\end{abstract}

\maketitle

\section{Introduction}
\begin{figure*}
      \centering
    \includegraphics[width=\textwidth]{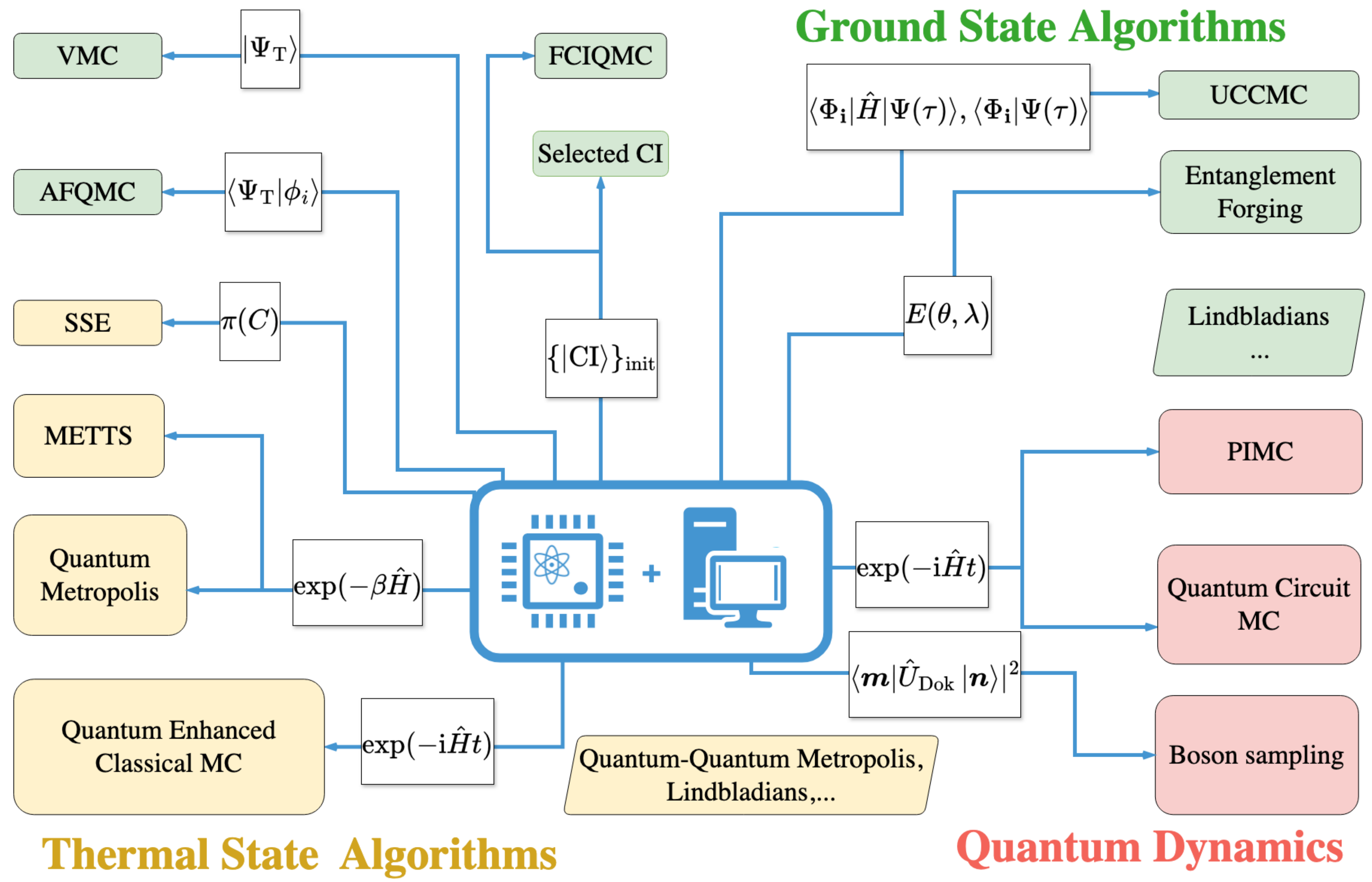}
    \caption{
    \textbf{
    Quantum Algorithms for Sampling Problems in Computational Chemistry Covered in this Review.} 
This figure shows the integration of quantum computing for  
ground state, thermal state calculations, and quantum dynamics. 
Ground state algorithms encompass hybrid quantum--classical algorithms, including auxiliary-field quantum Monte Carlo (AFQMC), full configuration interaction quantum Monte Carlo (FCIQMC), unitary coupled-cluster Monte Carlo (UCCMC),
variational Monte Carlo (VMC), selected configuration interaction (Selected CI), and entanglement forging. 
Cooling with Monte Carlo-flavored Lindbladians is covered as a fully quantum algorithm for ground state calculation. 
In the thermal state calculation category, the algorithms include quantum-enhanced classical Monte Carlo, stochastic series expansion (SSE), quantum Metropolis, and minimally entangled typical thermal states (METTS) as hybrid methods, 
complemented by quantum-quantum Metropolis and Monte Carlo-flavored Lindbladians as fully quantum algorithms.
For quantum dynamics, we covered quantum circuit Monte Carlo, hybrid path-integral Monte Carlo (PIMC), and Boson sampling.
Each hybrid method connects back to the central unit, illustrating the intermediate quantities quantum computing contributes to this methodology.
}
    \label{fig:overview}  
\end{figure*}
Quantum chemistry strives to obtain accurate, approximate solutions
to Schr{\"o}dinger's equation as efficiently as possible~\cite{helgaker2014molecular,friesner_ab_2005}.
The primary aim is to extend the capabilities of computational methods to handle calculations over larger spatial and longer temporal scales through algorithmic advances.
Despite the na{\"i}ve exponential (i.e., computationally intractable) complexity, progress in quantum chemistry has never ceased.
The range of applications enabled by quantum chemistry
has been ever growing from
drug discovery
to
materials design~\cite{Louie2021Jun,Santagati2024Apr, White2023Apr,Bauer2020Nov,Goh2017Jun}.

Owing to its inherent
relevance to practical applications,
quantum chemistry has recently received
significant interest 
from quantum computing communities, both industry and academia~\cite{Lanyon2010Feb,Cao2019Aug,Bauer2020Nov,mcclean2021foundations,McArdle2020Mar}.
A key goal is finding algorithms and applications to achieve a ``practical'' quantum advantage.
Practical quantum advantages can be assessed in two ways.
One way is to see if the end-to-end run time is shorter for a fixed error with access to the quantum computer. 
Another way is to see if we can perform more accurate quantum chemistry calculations for a fixed runtime with access to the quantum computer.
Even beyond the near-term intermediate scale (NISQ) architectures,
we do not have a definitive example
with practical quantum advantages known in the field.

One prominent example of NISQ and beyond-NISQ algorithms is variational algorithms stemming from the variational quantum eigensolver (VQE)~\cite{mcclean2016theory, Peruzzo2014Jul, Tilly2022Nov}.
The central idea of VQE is to optimize a relevant objective function for given problems 
by varying quantum circuit parameters~\cite{mcclean2016theory}.
The limitations of this approach stem from the limited expressivity of finite-depth quantum circuits
and the optimization difficulties involving many non-linear parameters accompanied by vanishing gradients exhibiting barren plateau~\cite{mcclean2018barren}.
In other words, one may fail to represent the true solution with a given circuit ansatz or to find the ground state even with an expressive circuit due to optimization difficulties~\cite{StilckFranca2021Nov}.
Nonetheless, developing different types of VQE algorithms has become one of the most active research directions in quantum chemistry over the last few years.
Interested readers are referred to review articles on this subject~\cite{Tilly2022Nov}.

Another commonly discussed quantum algorithm is called quantum phase estimation (QPE)~\cite{aspuru2005simulated}, geared towards fault-tolerant quantum computers. 
The goal of QPE is to estimate an eigenvalue (usually the lowest one) of a hermitian operator (Hamiltonian), $\hat{H}$.
In QPE, an initial state with nonzero overlap with the state of interest (typically the ground state) is prepared and then time-evolved under the Hamiltonian. The target eigenvalue is subsequently obtained through the inverse quantum Fourier transform. The probability of finding the target eigenvalue is inversely proportional to the overlap between the exact eigenstate and the initial state.
As one grows system size, this overlap decays exponentially with system size if one uses an initial state with a fixed energy per unit volume~\cite{lee2023evaluating}.
While many advancements in recent years economized the implementation of the necessary time evolution with sophisticated quantum chemistry Hamiltonians~\cite{PhysRevResearch.4.043210,Castaldo2024Jun},
the cost of state preparation associated with this overlap makes it unclear whether QPE can demonstrate quantum advantages in computational chemistry~\cite{lee2023evaluating, Ollitrault2024Apr,Fomichev2023Oct}.

This Review summarizes a relatively less discussed class of quantum algorithms developed to sample probability distributions relevant to computational chemistry. 
On a classical computer, such sampling is most commonly performed by Monte Carlo (MC) algorithms. 
Depending on the problem setup, classical MC sampling may face significant computational challenges, either due to the sample complexity (i.e., the number of samples required to obtain statistically reliable answers) or the computational cost per statistical sample.
Despite their prevalence in computational chemistry, efforts to enhance MC algorithms and develop sampling techniques using quantum computing have been relatively limited in the community.
On the one hand, our goal is to provide computational chemists with an overview of quantum algorithms that may help improve routine computational chemistry tasks.
On the other hand, we want to provide quantum information scientists with some background on the existing computational chemistry methods and connections between classical and quantum algorithms.
For the fundamental principles and basic concepts in quantum computing, we refer the readers to Nielsen and Chuang~\cite{nielsen2010quantum} and to the review papers~\cite{Bauer2020Nov,Cao2019Aug} that are more specialized for computational chemistry.
We also reference another review paper that discusses sampling problems in quantum computing~\cite{mansky2023sampling}. We hope to provide complementary viewpoints and more background for common computational chemistry sampling problems~\cite{mazzola2024quantum}.

This review is organized as follows: \cref{sec:quantum_chemistry} discusses sampling problems and approaches in computational chemistry, including ground-state, finite-temperature, and real-time dynamics problems. 
This section sets the foundation for the subsequent exploration of quantum algorithms that aim to enhance these classical approaches.
\cref{sec:quantum_for_gs} delves into quantum algorithms for ground-state calculations, exploring various quantum--classical hybrid and quantum algorithms. \cref{sec:quantum_for_thermal} introduces quantum algorithms for thermal state properties calculations. \cref{sec:quantum_for_dynamics} focuses on quantum algorithms for quantum dynamics. Finally, \cref{sec:conclusion} provides a summary and outlook.
\add{Fig.~\ref{fig:overview} provides a comprehensive overview of the algorithms discussed in this review.}

\section {Sampling problems in computational chemistry}\label{sec:quantum_chemistry}
In computational chemistry, computational kernels often provide a sample from an underlying distribution of interest.
Classically, the most common approach is Monte Carlo sampling on the classical computer to directly estimate
local observables (e.g., energy and correlation functions) from the distribution~\cite{nightingale1998quantum,austin2012quantum,becca_quantum_2017}.
Here, we provide several contexts where sampling problems arise in computational chemistry.

\subsection{Electronic ground state methods}\label{sec:2a}
A common task in computational chemistry is to determine the ground state of an electronic  Hamiltonian within the Born--Oppenheimer approximation for $N_\text{el}$ electrons and $N_\text{at}$ atoms (in atomic units),
\begin{align}
\hat{H}_\text{el}
&=\hat{H}_0
-\sum_{i}^{N_\text{el}}
\frac{\nabla_i^2}2
 -
\sum_{i}^{N_\text{el}}
\sum_{A}^{N_\text{at}}
\frac{Z_A}{|\mathbf r_i - \mathbf R_A|}
+\sum_{i>j}^{N_\text{el}}
\frac{1}{|\mathbf r_i - \mathbf r_j|},
\label{moham}
\end{align}
where 
$\{\mathbf r_i\}$ are the positions of electrons,
$\{\mathbf R_A\}$ are the positions of atoms,
$\{\mathbf Z_A\}$ are the charges of atoms, and $H_0 = \sum_{A<B} \frac{Z_A Z_B}{|\mathbf{R}_A - \mathbf{R}_B|}$ is the nuclear repulsion interaction.
The second-quantized form of the one-body and two-body interactions is
\begin{equation}
    \hat{H}_1 = \sum_{pq} h_{pq} a_p^\dagger a_q
\label{eq: one-body term}
\end{equation}
\begin{equation}
\hat{H}_2 = \frac{1}{2} \sum_{pqrs} v_{pqrs} a_p^\dagger a_q^\dagger a_s a_r,    
\label{eq: two body term}
\end{equation}
where $\{a_p^{(\dagger)}\}$ denotes the fermionic second-quantized operator, \(\hat{H}_1\) represents the one-body terms, and \(\hat{H}_2\) encapsulates the two-body terms. The coefficients \(h_{pq}\) and \(v_{pqrs}\) are the one- and two-electron integrals, respectively, defined as:
\begin{equation}
h_{pq} = \int \mathrm{d} \mathbf{r} \: \phi_p^*(\mathbf{r}) \left( -\frac{1}{2} \nabla^2 - \sum_A \frac{Z_A}{|\mathbf{r} - \mathbf{R}_A|} \right) \phi_q(\mathbf{r})
\end{equation}
\begin{equation}    
\begin{aligned}
v_{pqrs} =\int \int \mathrm{d}\mathbf{r}_1 \mathrm{d}\mathbf{r}_2 \: \phi_p^*(\mathbf{r}_1) \phi_q^*(\mathbf{r}_2) \frac{1}{|\mathbf{r}_1 - \mathbf{r}_2|} \phi_r(\mathbf{r}_1) \phi_s(\mathbf{r}_2),
\end{aligned}
\end{equation}
where $v_{pqrs}=(pr|qs)$ follows chemist's notation (or the Mulliken notation), and
\( \{ \phi_p \} \) is a single-particle basis set chosen a priori. 
With this Hamiltonian in mind, one aims to find the ground state corresponding to its lowest eigenstate (i.e., the ground state energy).
We discuss several classical heuristics that approximately compute the ground state energy and other properties.

\subsubsection{Variational Monte Carlo}
\noindent
Variational Monte Carlo (VMC) is a quantum Monte Carlo method commonly used in both \textit{ab initio} and model settings~\cite{PhysRevB.71.241103,Yokoyama1987Apr,PhysRevResearch.2.033429,Toulouse2016Jan,Choo2020May}.
VMC optimizes a set of variational parameters, $\boldsymbol{\theta}$, to minimize the variational energy expression
\begin{equation}
E(\bt)
=
\min_{\bt}
\frac{\langle \Psi({\bt}) | \hat{H} | \Psi({\bt}) \rangle}{ \langle \Psi({\bt}) |  \Psi({\bt}) \rangle}.
\end{equation}
Certain wavefunction ansätze, $|\Psi(\boldsymbol{\theta})\rangle$, cannot be efficiently stored or directly applied with $\hat{H}$ in the computational basis. Therefore, MC sampling is employed as a practical alternative to deterministic evaluations of the expectation value of the Hamiltonian.
A stochastic sampling of the variational energy is achieved by inserting a resolution-of-the-identity in a chosen orthonormal computational basis, $\{|n\rangle\}$,
\begin{equation}
\begin{aligned}
    E(\bt) = \sum_{n} E_{L}( \bt,n) P(\bt,n),
\end{aligned}
\label{eq:vmceloc}
\end{equation}
with the local energy
\begin{equation}
\begin{aligned}
E_{L}( \bt,n)
&=
\frac{\langle \Psi({\bt}) | \hat{H} |n \rangle}{ \langle \Psi({\bt}) | n \rangle}=
\frac{\sum_{m}\langle \Psi({\bt}) |m\rangle\langle m |\hat{H} | n \rangle}{ \langle \Psi({\bt}) |n \rangle}
\end{aligned}
\end{equation}
and the probability distribution
\begin{equation}
P(\bt, n)=
\frac{|\langle \Psi({\bt}) |n \rangle|^2 }{ \langle \Psi({\bt}) |  \Psi({\bt}) \rangle}.
\label{eq:pn}
\end{equation}
By drawing samples from $P(\bt, n)$, the variational energy can be approximated as 
\begin{equation}
    E(\bt)
    =
    \frac{1}{N_\text{samp}}
    \sum_i^{N_\text{samp}} E_{L}( \bt,n_i).
\end{equation}
Similarly, one can obtain the gradient of the energy with respect to the parameters,
\begin{equation}
    \frac{\partial E(\bt)}
    {\partial \bt}=
    \frac{1}{N_\text{samp}}
    \sum_i^{N_\text{samp}} 
    \frac{\partial E_{L}( \bt,n_i)}{\partial\bt}.
\end{equation}
Markov chain Monte Carlo (MCMC) samples the normalized distribution in \cref{eq:pn}, most commonly via the Metropolis--Hastings algorithm~\cite{levin2017markov}.
We explain details of the Markov chain Monte Carlo algorithm in the context of finite-temperature methods in \cref{sec:classical_mc}.

Regarding the cost of VMC, there are two primary considerations for its efficiency.
First, one should consider the cost of sampling $|n\rangle$ from the distribution, $|\langle\Psi|n\rangle|^2$. 
Within the Metropolis--Hastings algorithm, this sampling cost arises from two factors.
The first is the cost of evaluating the overlap, $\langle\Psi| n\rangle$, which determines the transition probability to a new state. Another is the mixing time, 
which quantifies how quickly the Markov chain converges to the stationary distribution.
\add{
    MCMC inherently faces challenges with long mixing and autocorrelation times,
    particularly when the initial distribution is far from the target distribution~\cite{montanaro2023accelerating}
    or when encountering critical slowing down~\cite{sorella2013variational}.
    While various techniques have been developed to improve sampling efficiency,
    such as cluster updates~\cite{Wang1990Sep,PhysRevLett.62.361},
    parallel tempering~\cite{PhysRevLett.57.2607},
    worm algorithm~\cite{Prokofev1998Feb}, and
    event-chain Monte Carlo~\cite{PhysRevE.80.056704},
    these accelerated MCMC approaches are typically system-specific
    and cannot be applied universally.
    The mixing time challenges become especially pronounced when sampling
    complex wavefunctions, such as neural network ansätze~\cite{choo2020fermionic,PhysRevResearch.2.033429,PhysRevResearch.3.L042024},
    or when studying frustrated quantum systems~\cite{Houdayer2001Aug,PhysRevE.98.053308}.
    The severity of these challenges depends on both the choice of wavefunction
    and the physical system under investigation.}  
The second consideration of VMC is the cost of the local energy evaluation in \cref{eq:vmceloc}.
Based on the overlap between $|\Psi\rangle$ and $|n\rangle$ and the Hamiltonian matrix element in the computational basis, $\langle m|\hat H |n\rangle$, one can evaluate the local energy. One should account for the sparsity of $\hat{H}$ in the cost model given that the Hamiltonian $\hat{H}$ in Eq.~\eqref{moham} is sparse.

For a VMC algorithm to be efficient, with costs scaling polynomially with system size $N$, three key elements must be computationally efficient: the overlap evaluation, the number of non-zero elements in any given column of $\hat{H}$, and the mixing time of the algorithm. Each of these components should scale as $\mathcal{O}(\textrm{poly}(N))$~\cite{Sorella2013}. 
\add{For example, consider the \textit{ab initio} electronic Hamiltonian expressed in the orthogonal 
Slater determinant basis. This Hamiltonian contains one-body terms (\cref{eq: one-body term}) 
and two-body terms (\cref{eq: two body term}). Due to double excitations, 
each column has $\mathcal O(N^4)$
non-zero elements, where $N$ is system size.}
A more detailed cost analysis can be provided once specific variational wavefunctions and Hamiltonians are determined.
The goal of quantum algorithms in VMC's context is to accelerate these computational aspects.

\subsubsection{Projector Monte Carlo}
Projector Monte Carlo (PMC) 
methods~\cite{becca_quantum_2017,zhang_constrained_1995,Zhang2003Apr,PhysRevD.24.2278,PhysRevD.27.1304,booth2009fermion,RevModPhys.73.33,PhysRevB.41.4552,austin2012quantum} stochastically perform imaginary-time evolution (ITE) to sample the ground state of many-body systems. 
In the infinite imaginary time limit, one can obtain the ground state from any initial state $\ket{\Psi_I}$ with non-zero overlap with the ground state $\ket{\Psi_0}$,
\begin{equation}
    \left|\Psi_0\right\rangle \propto \lim _{\tau \rightarrow \infty} \exp (-\tau \hat{H})\left|\Psi_I\right\rangle.
 \label{eq:pmc}
\end{equation}
which will be projected to the ground state in the long time limit, as graphically shown in Fig.~\ref{fig:qcafqmc}(a).
\begin{figure}
    \centering
    \includegraphics[width=0.4\textwidth]{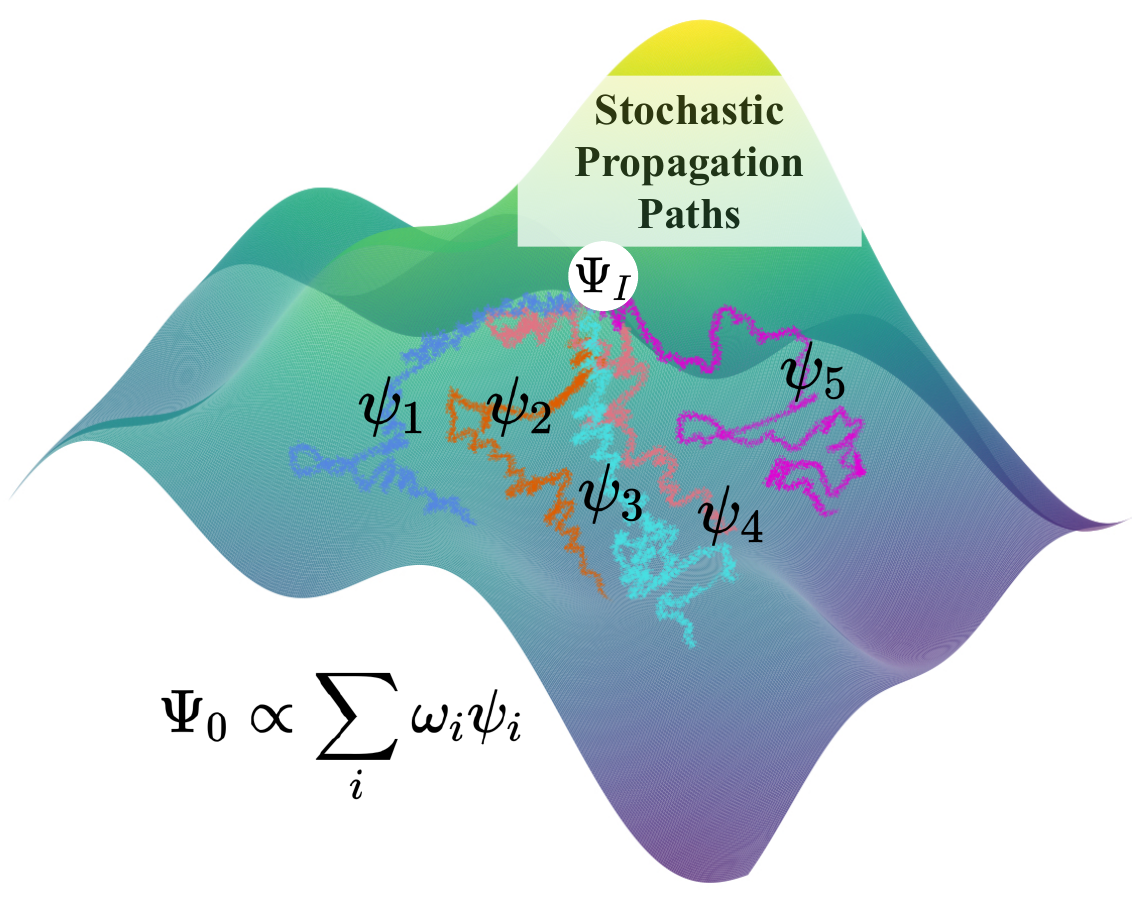}
    \caption{The stochastic propagation of walkers $\psi_i$ of projector quantum Monte Carlo methods. The ground state wavefunction $\Psi_0$ is approximated by sampling weight-averaged walkers in the long imaginary time limit.}
    \label{fig:pmc_traj}
\end{figure}
A deterministic implementation of imaginary time evolution is generally exponentially expensive. PMC performs this ground state projection stochastically to remove this exponential scaling bottleneck. 

PMC samples the action of the propagator using an ensemble of statistical samples with weights, $\{w_i\}$, and their states, $\{|\psi_i\rangle\}$. This is graphically shown in Fig.~\ref{fig:pmc_traj}.
The global wavefunction at a given imaginary time $\tau$ is written as
\begin{equation}\label{eq:global_wfn}
|\Psi(\tau)\rangle =
\sum_i
w_i(\tau)
|\psi_i(\tau)\rangle.
\end{equation}
\noindent\add{In contrast to VMC's parametrized wavefunction approach, 
PMC expands the ground state wavefunction as a linear combination in a 
chosen basis $\{|\psi_i(\tau)\rangle\}$. The basis states can be orthogonal, 
as used in FCIQMC~\cite{booth2009fermion,cleland2010communications,petruzielo2012semistochastic}, 
or non-orthogonal, as employed in 
approaches like AFQMC~\cite{zhang_constrained_1995,Zhang2003Apr} 
and path integral renormalization group~\cite{Kashima2001Aug}. 
These resemble orthogonal and non-orthogonal configuration interaction methods and will 
appear again when we discuss AFQMC and selected configuration interaction.}
The local energy estimate of Eq.~\eqref{eq:global_wfn} is written as
\begin{equation}
    E(\tau)= 
    \frac{\langle \Psi_\mathrm{T}|\hat{H}|\Psi(\tau)\rangle}
    {\langle \Psi_\mathrm{T}|\Psi(\tau)\rangle},
\end{equation}
where $|\Psi_\mathrm{T}\rangle$ is a trial wavefunction used to evaluate the energy estimate and often to importance-sample (see below).

Different PMC approaches vary in the details of the implementation of a projection operator in \cref{eq:pmc}.
It is most common to consider a discrete propagator,
\begin{equation}
    \ket{\Psi_0} = \lim_{n \to \infty} \hat{P}^n(\Delta\tau)\ket{\Psi_I},
\end{equation}
where $\Delta\tau$ is the time step to discretize the imaginary time.
Depending on the details of the discrete-time propagator and the computational basis state, one may use 
diffusion Monte Carlo (DMC)~\cite{grimm1971monte,anderson1975random,anderson1976quantum,reynolds1982fixed,toulouse2016introduction}, 
Green's function Monte Carlo~\cite{moskowitz1982new,PhysRevB.41.4552,buonaura1998numerical}, 
full configuration interaction quantum Monte Carlo~\cite{booth2009fermion,cleland2010communications,petruzielo2012semistochastic}, 
and auxiliary field quantum Monte Carlo~\cite{zhang_constrained_1995,Zhang2003Apr,motta_ab_2018}.
\add{We also note that there are other types of projectors for quantum cooling, 
such as Gaussian projector proposed in Refs.~\citenum{zeng2022universalquantumalgorithmiccooling,PRXQuantum.2.020321}. 
}

\begin{table*}
\centering
\caption{Classifications of the commonly used PMC approaches~\cite{mahajan2021taming,umrigar2015observations}}
\begin{ruledtabular}
\begin{tabular}{lccc}
\textbf{Method} & \textbf{Projector} & \textbf{Computational basis} & \textbf{Approximation} \\
\midrule
DMC & $\mathrm{e}^{\Delta\tau (E_\mathrm{T} - \hat{H})}$ & first quantization & Fixed-node \\ 
GFMC & $1-\Delta\tau(\hat{H}-E_\mathrm{T})$ & first/second quantization & Fixed-node \\ 
FCIQMC & $1-\Delta \tau(\hat{H}-E_\mathrm{T})$ & second quantization & None \\ 
AFQMC & $\mathrm{e}^{\Delta\tau (E_\mathrm{T} - \hat{H})}$ & second quantization & Phaseless \\
\end{tabular}\label{tab:PMC}
\end{ruledtabular}
\end{table*}

\begin{figure}
    \centering
    \begin{tikzpicture}
    \draw [thick, ->] (-4.5,-2.25) -- (-3.5,-2.25) node[right, black]{};
    \draw [thick, ->] (-4.5,-2.25) -- (-4.5,-1.25) node[above, black]{};
    \node[] at (-4,-2.){$\mathcal{S}(D)$};
    
    \draw[red,fill=gray!20,thick] (0,0) ellipse (4cm and 1.5cm);
    \node[] at (3cm, 1.7cm){\textcolor{red}{\footnotesize{$\langle \Psi_{\mathrm{T}}|\phi_{\mathrm{w}}\rangle$ = 0}}};

    \node[circle,fill=black,inner sep=0pt,minimum size=3pt,label=below:{\footnotesize{$\phi_1(\tau)$}}] (1a) at (-2.5,-0.5) {};
    \node[circle,fill=black,inner sep=0pt,minimum size=3pt,label=below:{\footnotesize{$\phi_1(\tau+\Delta\tau)$}}] (1b) at (-1,0.) {};
    
    \draw [->,shorten >=0.05cm, shorten <=0.05cm] (1a) to [out=90,in=135] (1b);

    \node[circle,fill=black,inner sep=0pt,minimum size=3pt,label=below:{\footnotesize{$\phi_2(\tau)$}}] (2a) at (1.5,1.) {};
    \node[circle,fill=black,inner sep=0pt,minimum size=3pt,label=below:{\footnotesize{$\phi_2(\tau+\Delta\tau)$}}] (2b) at (2.5,0.) {};
    \draw [->,shorten >=0.05cm, shorten <=0.05cm] (2a) to [out=0,in=90] (2b);

    \node[circle,fill=black,inner sep=0pt,minimum size=3pt,label=below:{\footnotesize{$\phi_3(\tau)$}}] (3a) at (-1.5,1.15) {};
    \node[cross,thick,red,inner sep=0pt,minimum size=3pt, label=above:{\footnotesize{$\phi_3(\tau+\Delta\tau)$}}] (3b) at (0.,1.9) {};
    \draw [->,shorten >=0.05cm, shorten <=0.05cm,dashed,thick] (3a) to [out=70,in=200] (3b);
    \end{tikzpicture}
    \caption{Propagation of walkers $\phi_{\mathrm{w}}$  in the manifold of Slater determinants $\mathcal{S}(D)$. Walkers $\phi_{1}$ and $\phi_{2}$ remain in the allowed hyperspace of similar signed overlap with the trial $\Psi_\mathrm{T}$, whereas walker $\phi_{3}$ will be removed from the simulation after crossing the overlap node.}
    \label{fig:sign_control}
\end{figure}
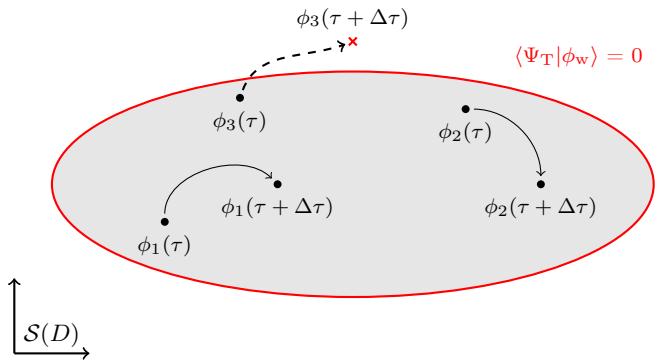

In nearly all fermionic simulations, except a few special cases~\cite{PhysRevLett.91.186402,PhysRevB.89.134422,PhysRevLett.117.267002,PhysRevB.91.241117,li2019sign,PhysRevB.71.155115}, PMC methods suffer from the fermionic sign problem. This can be accredited to the non-stoquasticity of the general fermionic Hamiltonians on a local basis~\cite{PhysRevLett.94.170201}, which generally poses no constraint on the sign of the walker weights. 
\add{Here, a Hamiltonian in a local basis is called stoquastic if all the off-diagonal elements are real and non-positive.}
Thus, for non-stoquastic cases, the expectation values have a large variance due to many samples canceling each other to produce a final signal.
These cancellations will lead to a vanishing signal-to-noise ratio of the observable estimates. Consequently, the computational scaling (i.e., the sample complexity) becomes exponentially large when one seeks to compute observables up to a certain additive error.
It was proven that solving the sign problem is \textbf{NP}-hard using classical algorithms~\cite{PhysRevLett.94.170201}. 
Approximations to PMC methods aim to control the sign problem and produce an overall polynomial-scaling algorithm. However, they inevitably introduce biases in the final statistical estimates. 
Controlling the sign problem is most commonly done by removing walkers that cross the node of the trial wavefunction as shown in \cref{fig:sign_control}, resulting in consistent signs in the weights.
Practical PMC methods are classified based on whether they control the sign problem, the computational basis (first versus second quantizations), and the form of the projector, as shown in \cref{tab:PMC}. We provide more details of PMC methods based on second quantization as they have been used in quantum algorithms in recent years.

\paragraph{Full Configuration Interaction Quantum Monte Carlo} \label{para:fciqmc}
\noindent Specifically for solving fermionic problems, the full configuration interaction quantum Monte Carlo (FCIQMC) method \add{can offer a memory-efficient framework because the imaginary time propagator is not stored explicitly, but rather is being sampled on-the-fly}~\cite{booth2009fermion,cleland2010communications,petruzielo2012semistochastic}. 
The imaginary time propagation of an initial state 
\begin{equation}    
\ket{\Psi_I} = \sum_\mathbf{i} C_\mathbf{i} |\Phi_\mathbf{i} \rangle
\end{equation}
which can be written in terms of the evolution of its coefficients,
\begin{align}
    C_\mathbf{i}(\tau) &= \sum_\mathbf{j} (\mathrm{e}^{-\tau H})_{\mathbf{ij}} C_\mathbf{j}(0) = \sum_\mathbf{j} \left[ \left(\mathrm{e}^{-\Delta\tau H_{\mathbf{ij}}}\right)^n\right]_{\mathbf{ij}} C_\mathbf{j}(0)\nonumber\\
    &\approx \sum_{\{\mathbf{j_1, ... ,j_{n-1},j}\}} (1 - \Delta\tau H_{\mathbf{ij_1}})... (1 - \Delta\tau H_{\mathbf{j_{n-1}j}}) \: C_\mathbf{j}(0).  
\end{align}
It suffices to take the first-order approximation to the imaginary time propagator, as the eigenspaces, as well as the order of eigenvalues, are preserved independently of the truncation order.
FCIQMC works in a configuration interaction (CI) basis $\{|\Phi_{\mathbf{i}}\rangle\}$, including all determinants generated by successive excitations from a reference determinant $|\Phi_0\rangle$. Indices $\mathbf{i}$ enumerate excitation tuples, e.g. a single excitation $(i,a)$, which corresponds to an excitation from orbital $i$ to orbital $a$. This reference is generally obtained from a Hartree--Fock calculation.
The matrix elements $H_{\mathbf{ij}}$ are defined in the same CI basis, $\langle \Phi_\mathbf{i} | \hat{H} | \Phi_\mathbf{j}\rangle$. 
During the propagation of walkers across the CI space via a stochastic application of the imaginary time propagator, one needs to evaluate the transition matrix elements, $H_{\mathbf{ij}}$ for each walker. These elements need not be stored and can be evaluated on the fly.
 
FCIQMC discretizes the coefficients $C_\mathbf{i}(\tau)$ by representing them as a finite number of walkers $N_\mathbf{i}(\tau)$ on determinants $|\Phi_\mathbf{i}\rangle$. 
The non-unitary nature of FCIQMC may lead to an exponential growth in the number of walkers ($N_\mathrm{w} = \sum_\mathbf{i} |N_\mathbf{i}|$) as commonly seen in other PMC methods.
To manage this, a dynamical shift $S(\tau)$ of the spectrum of the Hamiltonian is introduced. This shift is adjusted such that the total number of walkers remains approximately constant throughout the simulation, i.e. approximately setting the highest eigenvalue of the propagator to 1. The walkers are updated according to 
\begin{align} \label{eq:fciqmc_evolution}
    C_\mathbf{i}(\tau+\Delta\tau) &= C_\mathbf{i}(\tau) \underbrace{- \Delta\tau(H_{\mathbf{ii}} - S(\tau))C_\mathbf{i}(\tau)}_{\text{Kill/Clone}} \nonumber \\ &\underbrace{- \Delta\tau \sum_{\mathbf{j} \neq \mathbf{i}}H_{\mathbf{ij}} C_\mathbf{j}(\tau)}_{\text{Spawn}}.
\end{align}
A detailed description of the resulting population dynamics and implementation can be found in Ref.~\citenum{booth2009fermion}.
Each time step one can evaluate the projected energy,
\begin{equation}
    E(\tau) = \frac{\bra{\Psi_\mathrm{T}} H | \Psi(\tau) \rangle }{ \bra{\Psi_\mathrm{T}} \Psi(\tau) \rangle} = \frac{\sum_\mathbf{i} \bra{\Psi_\mathrm{T}} H | \Phi_\mathbf{i}\rangle N_\mathbf{i}(\tau) }{N_{\mathrm{T}}(\tau)},
\end{equation}
where $N_{\mathrm{T}}(\tau)$ is the total number of walkers located on determinants contributing to $\bra{\Psi_\mathrm{T}}$. One can show that for $\tau \to \infty$, the projected energy converges to the exact ground state if $\ket{\Psi(\tau)}$ faithfully represents the true ground state. This is the case in the limit of an infinite number of walkers present in the simulation~\cite{booth2009fermion,cleland2010communications}.
 
 While being exact, FCIQMC does not control the sign problem and its application is limited to relatively small systems due to the exponential-scaling sample complexity.
 Some efforts have been put forward to reduce the sign problem via the initiator approximation, but they have not changed the asymptotic scaling.
 One could hope to use quantum computers as a part of FCIQMC to reduce the sign problem, an approach we will briefly review later in \cref{subsubsec:fciqmc}.

\paragraph{Green's Function Monte Carlo}
\add{There are two different methods named Green's function Monte Carlo (GFMC) 
dealing with real space {\it ab initio} and lattice Hamiltonians, respectively. The real space method was first proposed by Kalos in 1962~\cite{Kalos1962Nov} to solve the few-body nuclear problem. In this review, we focus on the second kind of GFMC that mainly deals with lattice problems.} GFMC was widely used in solving the ground state of various systems, including Heisenberg Model~\cite{Trivedi1989Aug}, lattice fermions~\cite{vanBemmel1994Apr}, molecular systems~\cite{Ceperley1984Dec}, and bosonic systems such as Helium~\cite{Kalos1974May, Whitlock1979Jun}. 
In GFMC, the ground state is also reached by performing imaginary time evolution successively, similar to all other PMC methods. Suppose $\ket{\Psi_I}$ is the initial wavefunction that has a non-zero overlap with the exact ground state. We define
\begin{equation}
     \ket{\Psi^{(n)}} := (1 - \Delta \tau (\hat{H} - E_\mathrm{T}))^n\ket{\Psi_I},
\end{equation}
and then we expand the projection operator in a complete basis set $\{\ket{x}\}$, which yields
\begin{equation}
    \Psi^{(n + 1)}(x') = \sum_x \mathcal{G}_{x', x}\Psi^{(n)}(x)
    \label{eq: GFMC}
\end{equation}
where $\Psi(x) = \langle x | \Psi\rangle$, and $\mathcal{G}_{x', x}$ is the Green's function defined as
\begin{equation}
    \mathcal{G}_{x', x} = \langle x' | (1 - \Delta \tau (\hat{H} - E_\mathrm{T}) )| x\rangle.
\end{equation}
It is then natural to view \cref{eq: GFMC} as a master equation with $\Psi^{(n)}(x)$ being the probability distribution and $\mathcal{G}_{x', x}$ being the transition probability if all of the entries of the Green's function and the wavefunction are non-negative. However, this is not always the case. If a system under some local basis cannot guarantee the positivity of all the matrix elements in $\mathcal{G}_{x', x}$, then the sign problem will arise. For simplicity, we will only consider systems that do not exhibit sign problems in this section. We further decompose Green's function into the product of the normalization factor and the stochastic matrix describing the transition probability:
\begin{equation}
    \mathcal{G}_{x', x} = b_x p_{x', x}
\end{equation}
where $b_x$ is the normalization factor given by
\begin{equation}
    b_x = \sum_{x'} \mathcal{G}_{x', x}.
\end{equation}
For each configuration $x^{(n)}$ at time step $n$, we assign a weight $w^{(n)}$ to take account of the normalization factor, and hence the update rule of the weight is given by 
\begin{equation}
    w^{(n+1)} = b_{x^{(n)}}w^{(n)}.
\end{equation}
Then, after a sufficiently long time, the equilibrium probability distribution over $\{\ket{x}\}$ is the ground state expanded on this local basis. \add{It is noteworthy that GFMC and FCIQMC are similar except that FCIQMC carries out the annihilation step when two walkers are on the same determinant.}

\add{There is also importance sampling to reduce the variance of the Monte-Carlo estimate. One carries out a similarity transformation by redefining the probability distribution to minimize the variance. In GFMC, this is achieved by introducting the guiding function $\Psi_{\mathrm{G}}(x)$. We redefine the Green's function with importance sampling as:
\begin{equation}
    \tilde{\mathcal{G}}_{x', x} = \mathcal{G}_{x', x}\frac{\Psi_{\mathrm{G}}(x')}{\Psi_{\mathrm{G}}(x)}
\end{equation}
and the wavefunction
\begin{equation}
    \tilde{\Psi}^{(n)}(x) = \Psi_{\mathrm{G}}(x)\Psi^{(n)}(x)
\end{equation}
so that the new Green's function and wavefunction satisfies the same propagation rule Eq.~\eqref{eq: GFMC}. The guiding function also serves as a remedy for the sign problem by carefully choosing the form of $\Psi_{\mathrm{G}}(x)$ to ensure the positivity of the off-diagonal elements of the new Green's function.}

For most of the fermionic Hamiltonians, there is no guarantee that the entries of Green's function are all positive. Hence, we need to control the sign problem by using fixed-node approximation~\cite{vanBemmel1994Apr}. The fixed node approximation removes the sign problem in sampling but introduces a bias. Recently, progress has been made in enhancing GFMC with quantum computers, thereby reducing the bias introduced by the fixed-node approximation~\cite{Yang2024Jan}. Interested readers can refer to Ref.~\citenum{becca_quantum_2017} for further details on the theory of GFMC.

\paragraph{Auxiliary Field Quantum Monte Carlo}\label{sec:classic_afqmc}
In the past decades, AFQMC~\cite{zhang_constrained_1995,Zhang2003Apr} has established itself as an exceptionally accurate and efficient method for tackling many-body systems. In recent developments, AFQMC has proven to be a powerful tool for solving {\it ab initio} problems, with extensive use across a broad spectrum of quantum chemical problems~\cite{motta_ab_2018,lee_twenty_2022,Lee2020Jul,lee2021phaseless,lee2021constrained,mahajan_selected_2022,mahajan2023response,Malone2023Jan,jiang2024unbiasing,Jiang2024Jun}. 
AFQMC performs an open-ended random walk to propagate an initial representation of the wave function $\ket{\Psi_I}$ in imaginary time, 
\begin{equation}
    \ket{\Psi_0} = \lim_{N \to \infty} \left(\mathrm{e}^{-\Delta \tau(\hat{H} - E_\mathrm{T})}\right)^N\ket{\Psi_I}.
\end{equation}
Applying the Trotter decomposition~\cite{trotter1959product} and the subsequent Hubbard--Stratonovich transformation~\cite{hubbard1959calculation,stratonovich1957method}, we obtain a formula of the short-time propagator expressed as an integral over auxiliary fields
\begin{equation}
    \mathrm{e}^{-\Delta \tau \hat{H}} = \int \mathrm{d}^{N_{\text{aux}}} \mathbf{x}\ p(\mathbf{x}) \hat{B}(\mathbf{x}) + \mathcal{O}(\Delta\tau^2),
    \label{eq: HS}
\end{equation}
where $\mathbf{x}$ is a vector of $N_{\text{aux}}$ auxiliary fields subjected to Gaussian distribution, $p(\mathbf{x})$ is the standard normal distribution and $\hat{B}(\mathbf{x})$ is a one-body propagator. The integral is carried out via Monte Carlo by writing the global wavefunction at each time step $n$ as a weighted sum of $N_{\text{w}}$ so-called \textit{random walkers},
\add{(with importance sampling),}
\begin{equation}
    \ket{\Psi^{(n)}} = \sum_{i = 1}^{N_{\text{w}}}w_i^{(n)}\frac{\ket{\phi^{(n)}_i}}{\langle\Psi_\mathrm{T}|\phi^{(n)}_i\rangle},
\end{equation}
where $\ket{\phi_i^{(n)}}$ are single determinant states. \add{Here, the trial wavefunction, $|\Psi_\mathrm{T}\rangle$, defines the importance sampling distribution.} The weights of the walkers are updated according to the overlap ratio
\begin{equation}
    w_i^{(n+1)} = w_i^{(n)} \times \frac{\langle\Psi_\mathrm{T}|\hat{B}(\mathbf{x}_i^{(n)})|\phi^{(n)}_i\rangle}{\langle\Psi_\mathrm{T}|\phi^{(n)}_i\rangle}.
\end{equation}
\add{This defines the free-projection AFQMC (fp-AFQMC) with importance sampling. It may be worth noting the similarities and differences between fp-AFQMC and FCIQMC algorithms. Both fp-AFQMC and FCIQMC are formally exact but suffer from the sign problem and are hence limited to small systems. fp-AFQMC can be viewed as an algorithm in which the wavefunction is represented by a non-orthogonal determinant expansion, with both coefficients and determinants being sampled from imaginary-time propagation.} 

\add{In fp-AFQMC}, the weights are complex numbers in general, and the average of any observable will suffer from an exponentially large variance, known as the phase problem~\cite{motta_ab_2018}. One could control the phase problem by imposing constraints on the weights of the walkers. 
The phaseless approximation~\cite{Zhang2003Apr} is employed in phaseless AFQMC (ph-AFQMC) calculations to achieve this. \add{It is often used together with a complex shift (which is known as the optimal force bias) in the Hubbard--Stratonovich transformation Eq.~\eqref{eq: HS}. With the constraint, ph-AFQMC} ensures the positivity of the weights during propagation (also depicted in~\cref{fig:sign_control}) by applying
\begin{equation}
    w_i^{(n+1)} = w_i^{(n)} \times \left|\frac{\langle\Psi_\mathrm{T}|\hat{B}(\mathbf{x}_i^{(n)})|\phi^{(n)}_i\rangle}{\langle\Psi_\mathrm{T}|\phi^{(n)}_i\rangle}\right|\times \max\{0, \cos\theta_i^{(n)}\},
\end{equation}
where $\theta_i^{(n)}$ is the argument of the complex overlap ratio. This approximation introduces biases to the calculation, however, one can mitigate the bias by employing more accurate trial wavefunctions~\cite{Qin2016Aug,vitali_calculating_2019,mahajan_selected_2022,mahajan2021taming,jiang2024unbiasing}. In the limit of the trial wavefunction being the exact ground state of the Hamiltonian, the bias can be completely removed. 
The estimate of the ground state energy is given by the mixed estimator 
\begin{equation}
    E(n\Delta\tau) =  \frac{\sum_i w_i^{(n)} E_{L,i}^{(n)}}{\sum_i w_i^{(n)}}
\end{equation}
where $E_{L,i}^{(n)}$ denotes the local energy of the $i$-th walker at time step $n$,
\begin{equation}
    E_{L,i}^{(n)} = \frac{\langle \Psi_\mathrm{T}|\hat{H}|\phi_i^{(n)}\rangle}{\langle\Psi_\mathrm{T}|\phi_i^{(n)}\rangle}.
\end{equation}
Similarly to GFMC, one could hope to use a more sophisticated trial wavefunction implemented through quantum computers to reduce the phaseless bias.
This is the central intuition behind the hybrid quantum--classical AFQMC that we will discuss later in \cref{sec:qc-afqmc}
We only briefly reviewed the AFQMC method as needed to understand its quantum counterpart.
For a detailed review of AFQMC, we refer the readers to Ref~\citenum{motta_ab_2018}.

\subsubsection{Coupled Cluster Monte Carlo}
The coupled cluster (CC) wavefunction is parametrized by a cluster operator $\hat{T} = \sum_{\mathbf{i}\in\mathcal S} t_\mathbf{i} \hat{E}_{\mathbf{i}}$, giving
\begin{equation}
|\Psi_{\mathrm{CC}}\rangle = \mathrm{exp}(\hat{T})|\Phi_{\mathrm{ref}}\rangle,
\end{equation}
where $\mathcal S$ defines the set of excitation operators, $\hat{E}_\mathbf{i}$ is the $i$-th excitation operator that creates an excited determinant via $\hat{E}_\mathbf{i} |\Phi_\text{ref}\rangle = \ket{\Phi_\mathbf{i}}$, and $t_\mathbf{i}$ is the cluster amplitude for the $i$-th excitation. Here, $|\Phi_\text{ref}\rangle$ is some reference determinant, usually chosen from a previous Hartree--Fock calculation. This exponential parametrization can express the exact solution if all possible excitations are included in the definition of $\hat{T}$.
In practice, however, the cluster operator is truncated after a certain excitation order to keep computations tractable.

Similarly to FCIQMC, coupled cluster Monte Carlo (CCMC)\cite{Thom:2010aa,Spencer2016stochastic_cc} utilizes a CI expansion to stochastically solve the CC projection,
\begin{align} \label{eq:cc_projection}
    \langle \Phi_{\mathbf{i}} | \hat{H} - E | \Psi_{\mathrm{CC}}\rangle = 0,
\end{align}
for all excitations $\mathbf{i}$.  As a result of the projection equations in Eq.~\eqref{eq:cc_projection} one can write a first order imaginary-time propagator,
\begin{equation} \label{eq:ccmc_propagtion}
    \langle \Phi_{\mathbf{i}} | 1 - \tau( \hat{H} - E) | \Psi_{\mathrm{CC}}\rangle = \langle \Phi_{\mathbf{i}} | \Psi_{\mathrm{CC}}\rangle.
\end{equation}
Unlike FCIQMC, the propagation prescribed by Eq.~\eqref{eq:ccmc_propagtion} is not straightforward due to non-linearity, as generally 
\begin{equation}\label{eq:ccmc_approx}
    \langle \Phi_{\mathbf{i}} | \Psi_{\mathrm{CC}}\rangle = t_\mathbf{i} + \mathcal{O}(t^2).
\end{equation}
This can be seen when considering a double excitation $\mathbf{i} = (i,j,a,b)$,
\begin{equation}
    \langle \Phi_{\mathbf{i}} | \Psi_{\mathrm{CC}}\rangle = t^{a b}_{i j} + t^{a}_{i} t^{b}_{j} + t^{a}_{j} t^{b}_{i}.
\end{equation}
The projection of the CC wave function onto a doubly excited determinant has contributions from products of cluster amplitudes. These products correspond to composite excitors $\hat{E}_{\mathbf{i}}\hat{E}_{\mathbf{j}}$ in the cluster operator. From Eq.~\eqref{eq:ccmc_propagtion}, it is not clear how to formulate population dynamics, including composite clusters.
The CCMC evolution circumvents this problem by requiring to spawn on non-composite clusters. Hence, one only needs to store walkers on these clusters instead of storing walkers on all possible excitors and their products. This, however, also limits CCMC to systems where composite cluster amplitudes are expected to be sufficiently small such that Eq.~\eqref{eq:ccmc_approx} is a valid approximation.
The CCMC iterative procedure is then captured by 
\begin{align}
    t_{\mathbf{i}}(\tau + \Delta\tau) &= t_{\mathbf{i}}(\tau)
     - \Delta\tau \sum_{\mathbf{j}}H_{\mathbf{i}\mathbf{j}} \langle \Phi_\mathbf{j}|\Psi(\tau)\rangle.
\end{align}

Unfortunately, CCMC suffers from the sign problem~\cite{Spencer2016stochastic_cc}, which is closely related to that in FCIQMC. Due to the exponential-scaling sample complexity, it is not always clear whether there is a significant computational benefit to use CCMC over standard deterministic CC algorithms.
Extending this method, one could hope to sample a different variant of CC wavefunctions (called unitary CC) using quantum computers or reduce the sign problem with quantum computers. We will review an attempt along this line in~\cref{sec:qc-ccmc}.

\subsubsection{Selected Configuration Interaction}\label{sec:sci}
Configuration interaction (CI) is one of the conceptually simplest post-HF methods for calculating correlation energy. The basic idea is to diagonalize the Hamiltonian in a selected basis of $N_\textrm{el}$-electron Slater determinants, or in other words, construct a ground state wavefunction as a linear combination of Slater determinants with the coefficients determined variationally~\cite{szabo1996modern}.
Similarly to CC, 
\begin{equation}
|\Psi_\text{CI}\rangle
=
\hat{C}|\Phi_\text{ref}\rangle,
\end{equation}
where $\hat{C} = \sum_{\mathbf i\in\mathcal S} c_{\mathbf i} \hat{E}_{\mathbf i}$.
If one defines the set $\mathcal S$ to include all possible excitations, one recovers the exact FCI approach with an exponential scaling cost. Therefore, the most commonly seen polynomial scaling approximations truncate the space of determinants based on excitation level. These approximations include well-known methods such as CI-singles (CIS), CI-singles and doubles (CISD), etc. It is worth noting that truncated CI methods are well known to lack size consistency, limiting their usefulness~\cite{helgaker2014molecular}, unlike the CC wavefunction with the same truncation. 

Going beyond the excitation-based truncation in CI, selected CI (SCI) has seen revived interests in quantum chemitry~\cite{Buenker1975Oct,Buenker1974Mar,Cimiraglia1987Jan,Harrison1991Apr,Greer_EstimatingFullConfiguration1995,Greer_MonteCarloConfiguration1998}.
In essence, SCI methods define the excitation subspace $\mathcal S$ via a sampling protocol (not necessarily based on Monte Carlo) intending to find important determinants for electron correlation.
One of the earliest SCI methods was developed roughly 50 years ago, named configuration interaction perturbatively selected iteratively (CIPSI), where an iterative procedure based on a perturbative estimate is used to improve the choice of subspace~\cite{Huron_IterativePerturbationCalculations1973}. CIPSI has been shown to yield near-exact energies for systems much beyond the reach of the limit of FCI~\cite{Garniron2018Aug,Loos2020Nov,Jiang2024Jun}.
Furthermore, new selection schemes exploring the tradeoff between the selection efficiency and accuracy have been proposed, example being adaptive sampling configuration interaction (ASCI) \cite{Tubman_DeterministicAlternativeFull2016, Tubman_ModernApproachesExact2020}, adaptive configuration interaction (ACI)\cite{Schriber_CommunicationAdaptiveConfiguration2016,Evangelista2014Mar}, heat-bath configuration interaction (HCI)~\cite{Holmes2016Apr,Holmes2016Aug,Sharma2017Apr}, and iterative CI (iCI)~\cite{Liu2014May,Liu2016Mar,Lei2017Nov,zhang2020iterative}.  %

A Monte Carlo alternative to these SCI methods is the Monte Carlo configuration interaction (MCCI) \cite{Greer_EstimatingFullConfiguration1995, Greer_MonteCarloConfiguration1998}.
In the MCCI approach, the FCI space is divided into three parts: the trial CI vector, CI vectors that interact with the trial (only allow single or double excitations from the trial CI), and those that do not interact.
Each cycle commences with branching processes where random single and double excitations are uniformly sampled to generate new vectors, thus forming a defined subspace. Within this subspace, the Hamiltonian matrix is diagonalized to obtain the lowest eigenstate, which then serves as the trial CI vector for the next cycle. 
Following diagonalization, a selection mechanism evaluates the components of the vector to decide which should be retained or discarded based on their contributions to the system's energy and wavefunction.
The importance of a state is determined by the magnitude of the energy reduction estimated through second-order perturbation theory.
This procedure is iteratively repeated until the vector's expansion stabilizes in length or a predefined energy criterion is met.
Typically, the Hartree--Fock wavefunction is employed as the initial trial CI vector. 
The quantum version of MCCI, further discussed in \cref{sec:qsci}, extends these principles to quantum computational frameworks.

Despite significant progress in SCI-related methods and impressive results, these methods all scale exponentially with system size due to the (exponentially) growing number of important determinants. 
This ultimately limits the scope of these methods to small systems.
Nevertheless, one could hope to sample relevant determinants more efficiently using quantum computers, which is an idea we will review later in \cref{sec:qsci}.  

\subsection{Finite-temperature methods}
In addition to the ground state calculations in the previous section, 
understanding finite-temperature properties is also common in computational chemistry and materials science, as they provide insights into the equilibrium properties and responses to external perturbations of systems at finite temperatures.
The classical Monte Carlo framework gives rise to a multitude of finite temperature methods such as finite temperature AFQMC~\cite{rubenstein2012ftafqmc,zhang1999ftafqmc,He2019Jan,lee2021phaseless}, density matrix QMC~\cite{blunt2014dmqmc}, path-integral MC~\cite{Ceperley1995Apr,Selke1997May,PhysRevLett.115.130402,PhysRevLett.110.146405}, stochastic series expansion~\cite{sandvik_1991, Sandvik1992sse,Handscomb_1962,PhysRevB.59.R14157}, thermo-field Monte Carlo~\cite{Suzuki1986Jun}, \textit{etc.}
Not all of these methods have a quantum algorithm counterpart, so we will only review those with the quantum counterpart.

\subsubsection{Classical Monte Carlo}\label{sec:classical_mc}
Statistical physics provides the theoretical framework to study the thermal properties of macroscopic systems composed of many microscopic local degrees of freedom.
The Ising model serves as a fundamental model for exploring phase transitions. The demonstration of the order-to-disorder phase transition within the two-dimensional and three-dimensional Ising model with Monte Carlo methods is considered one of the significant achievements of 20th-century statistical physics~\cite{PhysRevLett.61.2635,Ferrenberg1991Sep}. The studies profoundly influenced understanding critical phenomena and the nature of phase changes in various physical systems. Most importantly, such studies are commonly performed by Monte Carlo sampling.

We briefly recapitulate the Markov chain Monte Carlo with the classical Ising model as an example.
The classical Ising model consists of discrete magnetic spins on a lattice that can be in one of two states (up or down) and interact with their nearest neighbors. The Hamiltonian gives the energy of a particular state of the system:
\begin{equation}
    E(\textbf{s}) = -\sum^n_{j<k} J_{jk}s_js_k - \sum_j^n h_js_j\label{eq:IsingModel}
\end{equation}
where each spin configuration $\textbf{s}\in\{-1,+1\}^n$.
The central object of interest that allows us to study the properties at varying temperatures is the canonical partition function,
\begin{equation}
    Z = \sum_\mathbf{s} \mathrm{e}^{-\beta E(\mathbf{s})}
\end{equation}
at temperature $1/\beta$.
Systems that obey Boltzmann statistics are ubiquitous in physics, and sampling these distributions efficiently is an important subroutine in several algorithms. 

\add{To sample from a Boltzmann distribution, one can construct a Markov chain that 
converges to the Boltzmann distribution as its stationary distribution. 
Such a Markov chain is characterized by a transition matrix $\textbf{P}$, 
where each element $\textbf{P}_{\textbf{s}'\textbf{s}} = P(\textbf{s}'|\textbf{s})$ 
represents the probability of transitioning from state $\textbf{s}$ to state $\textbf{s}'$. 
The evolution of a probability distribution $p$ under one step of the Markov chain is given 
by $p' = \textbf{P}p$.} To achieve the desired stationary condition, 
the Markov chain must be irreducible and aperiodic. 
\add{A Markov chain is \textit{irreducible} if there exists a 
sequence of transitions with non-zero probability between any pair of states. 
An irreducible Markov chain is \textit{aperiodic} if the greatest 
common divisor of all possible cycle lengths in the state transition graph is 1.
}
The detailed balance condition (a form of Bayes' theorem) of the Markov chain needs to be satisfied to converge to the desired stationary distribution $\pi$: 
\begin{equation}
P(\textbf{s}'|\textbf{s})\pi(\textbf{s}) = P(\textbf{s}|\textbf{s}')\pi(\textbf{s}')\label{eq:mc_detail_balance}
\end{equation}
\begin{equation}
    \sum_s P(\textbf{s}'|\textbf{s})\pi(\textbf{s})=\pi(\textbf{s}')
\end{equation}
with the target probability being the Boltzmann distribution,
\begin{equation}
    \pi(\mathbf{s}) = \frac{\mathrm{e}^{-\beta E(\mathbf{s})}}{Z}.
\end{equation}
If we know a function $f(\textbf{s}) \propto \pi(\textbf{s})$, the most common way to construct a Markov chain that converges to $\pi$ is through an acceptance-rejection based algorithm. In the proposal step, from the current state \textbf{s}, the next state \textbf{s}' is proposed with probability $Q(\textbf{s}'|\textbf{s})$. In the accept/reject step, the proposed step is accepted with probability $A(\textbf{s}'|\textbf{s})$. The samples stay in the same configuration if the proposal is rejected. There are multiple ways of choosing $A$: the most popular are the Metropolis--Hastings algorithm, similar to that of VMC, and the Gibbs sampling algorithm.
The Metropolis--Hastings choice for $A$ is given by \begin{equation}
    A(\textbf{s}'|\textbf{s}) = \min\left( 1,\frac{\pi(\textbf{s}')}{\pi(\textbf{s})}\frac{Q(\textbf{s}|\textbf{s}')}{Q(\textbf{s}'|\textbf{s})} \right).
\end{equation}
The mixing time, defined as the expected number of iterations required for the Markov chain's distribution to get within $\epsilon$ of the true distribution in total variation distance is given by \begin{equation}
    t_\epsilon \leq \delta^{-1}\ln\left(\frac{1}{\epsilon\min \pi(\textbf{s})} \right),
\end{equation} where $\delta = 1 - \max_{|\lambda|<1}|\lambda|$ and $\lambda$ is an eigenvalue of the transition matrix $P$~\cite{levin2017markov}.

\add{A key challenge in MCMC is designing an effective proposal probability $Q(\textbf{s}|\textbf{s}')$ that enables efficient sampling. Additionally, at low temperatures, the ratio of stationary probabilities $\pi(\textbf{s}') / \pi(\textbf{s})$ can become vanishingly small for transitions between states. This leads to very low acceptance rates and consequently long convergence times.
        }
Many efforts in quantum counter parts of the MCMC algorithms are targeted toward accelerating the convergence time of MCMC.
We will introduce the enhancement of MCMC using quantum computers in \cref{sec:q_enhanced_mcmc} and the quantum version of Metropolis in \cref{sec:q2ma} by utilizing the quantum phase estimation.
\subsubsection{Lindbladian dynamics}\label{sec:lindblad_classical}
The Markov chain in Eq.~\eqref{eq:mc_detail_balance} can also be expressed as~\cite{chen2023quantum}
\begin{equation}
    P_{s^{\prime} s}=\underbrace{\sum_{a \in A} p(a) \gamma\left(E(\textbf{s}')-E(\textbf{s})\right) \boldsymbol{A}_{\textbf{s}'\textbf{s}}^a}_{\text {Accept }}+\underbrace{\vphantom{\sum_{a \in A}} \boldsymbol{R}_{\textbf{s}'\textbf{s}} \delta_{\textbf{s}'\textbf{s}}}_{\text {Reject }},
\end{equation}
where $\gamma(\omega):=\min \left(1, \mathrm{e}^{-\beta \omega}\right)$ and $\omega$ is the energy difference of the transition.
$\boldsymbol{A}_{\textbf{s}'\textbf{s}}^a$ are the stochastic matrices corresponding to the transitions (\textit{e.g.,} the spin-flip in Ising model), as weighted by the Metropolis factor $\gamma(\omega)$. 
The rejection part $\boldsymbol{R}_{\textbf{s}'\textbf{s}}$ is a diagonal matrix determined by the probability preserving constraints.
A continuous-time Markov chain generator can be defined as
\begin{equation}
\begin{aligned}
    \boldsymbol{L}_{s^{\prime} s}&=\sum_{a \in A} p(a)(\underbrace{\vphantom{\sum_{a \in A}} \gamma\left(E(\textbf{s}')-E(\textbf{s})\right) \boldsymbol{A}_{\textbf{s}'\textbf{s}}^a}_{\text {Transition}}\\&-\underbrace{\delta_{\textbf{s}'\textbf{s}} \sum_{\textbf{s''}} \gamma\left(E(\textbf{s''})-E(\textbf{s})\right) \boldsymbol{A}_{\textbf{s''}\textbf{s}}^a}_{\text {Decay }})
\end{aligned}\label{eq:ft_mchain}
\end{equation}
\add{
    In Sec.~\ref{sec:ft_full_quantum}, we will introduce the quantum analog of MCMC, which employs a quantum computer to sample thermal states of an arbitrary many-body Hamiltonian $\hat{H}$. This approach follows a form similar to Eq.~\eqref{eq:ft_mchain} but constructs a Lindbladian,
}
\begin{equation}
    \mathrm{e}^{\mathcal{L}_\beta t}\left[\boldsymbol{\rho}_\beta\right]=\boldsymbol{\rho}_\beta \quad \text { where } \quad \boldsymbol{\rho}_\beta=\mathrm{e}^{-\beta \boldsymbol{H}} / \operatorname{Tr}\left(\mathrm{e}^{-\beta \boldsymbol{H}}\right)
\end{equation}
\begin{equation}
    \mathcal{L}_\beta[\boldsymbol{\rho}]=\underbrace{-\mathrm{i}[\boldsymbol{B}, \boldsymbol{\rho}]}_{\text {Coherent}}+\:(\text{"dissipative")}.
\end{equation}
Lindbladians have been combined with Monte Carlo sampling in the context of both ground and thermal state preparation, as well as real-time dynamics, as a branch of full quantum algorithms, which we will cover in \cref{sec:gs_full_quantum} and \cref{sec:ft_full_quantum}.

\subsubsection{Stochastic Series Expansion} \label{sec:sse}
Another method to evaluate thermal averages of many-body systems is the stochastic series expansion (SSE)~\cite{sandvik_1991, Sandvik1992sse}. In essence, SSE samples the trace of the $L$-th order Taylor expansion of $\exp{(-\beta \hat{H})}$. This idea goes back to the 1960s when it was first introduced by Handscomb~\cite{Handscomb_1962}.
Unlike most Monte Carlo schemes, SSE introduces no systematic errors stemming from the artificial time-discretization~\cite{Prokofev1998Aug}.  While SSE has been primarily applied to spin chains~\cite{sandvik1994disorder, sandvik1997heisenberg2d, kaul2013designer, wildeboer2020antiferro}, its framework is universal and may, in principle, be applied to generic quantum systems. However, applications to such systems are challenging due to the severity of the sign problem in the method~\cite{sandvik2019stochastic}.

Considering an $N$-body Hamiltonian $\hat{H}= -\sum_{i=1}^M \add{\hat{H}_{i}}$ with $M$ terms total,
the partition function can be computed by
\begin{align}\nonumber
    Z &= \mathrm{Tr}(\mathrm{e}^{-\beta \hat{H}}) = \sum_a \langle a | \mathrm{e}^{-\beta \hat{H}} | a\rangle \\
    &\approx \sum_a \sum_n^L \sum_{\add{\{b_i\}_{i=1}^n}} \frac{\beta^n}{n!} \langle a | \hat{H}_{b_n} ... \hat{H}_{b_1} | a \rangle, \label{eq:sse_taylor}
\end{align}
\add{where the set $\{b_i\}_{i=1}^n$ specifies a sequence of indices of length $n$ where each of $b_i$ indexes one of the terms in $\hat{H}$. Hence, $\{b_i\}_{i=1}^n$ specifies an operator string emerging from taking the $n$-th power of the $N$-body Hamiltonian.} 
\add{Following Eq.~\eqref{eq:sse_taylor},} SSE generates a random walk in a joint state- and index-space by sampling tuples \add{$(a, n, \{b_i\}_{i=1}^n)$} and evaluating corresponding matrix elements contributing to $Z$. This sampling procedure is commonly implemented via the Metropolis--Hastings algorithm, ultimately sampling relative weights adhering to the probability distribution 
\begin{align}\label{eq:sse_prob_density}
    \pi(a, n, \{b_i\}) &= \frac{\beta^n}{n!}\langle a |\hat{H}_{b_n} \cdots \hat{H}_{b_1}|a \rangle\\
    &\geq 0,
\end{align}
where the inequality is a requirement for $\pi(a, n, \{b_i\})$ to define a probability distribution. This condition is straightforwardly satisfied by shifting the system Hamiltonian by a sufficiently large constant value~\cite{lee1984sse_constant}.

For large $N$ this procedure is computationally tractable only if one imposes an additional \emph{no-branching} requirement,
\begin{equation} \label{eq:nobranching}
    \forall\: a, b \: \exists \: a': \hat{H}_b |a\rangle \propto |a'\rangle,
\end{equation}
alleviating the exponential cost of evaluating and storing all diagonal and off-diagonal operator string matrix elements. 
\add{To understand this requirement, consider the action of $\hat{H}_b$ on an arbitrary basis 
state $|a_i\rangle$ without the no-branching condition: $\hat{H}_b |a_i\rangle = \sum_l c_l |a_l\rangle$, 
where $c_l$ are expansion coefficients. 
When evaluating the operator string in Eq.~\eqref{eq:sse_prob_density}, 
each successive application of $\hat{H}_b$ would generate a superposition of states, 
leading to an exponential growth in the number of terms. The no-branching condition prevents 
this explosion by restricting each operator to map a basis state to a single basis state, 
though this limits the choice of basis states.
}

With the no-branching requirement, one can no longer shift the spectrum of the system arbitrarily, giving rise to samples with arbitrary signs. This sign problem directly manifests in,
\begin{align}\nonumber
    \langle \hat{A}\rangle_{\pi} &= \frac{\sum_{C} \pi(C) A(C)}{\sum_{C} \pi(C)}\\\nonumber
    &= \frac{\sum_{C} \mathrm{sgn}(\pi(C)) |\pi(C)| A(C)}{\sum_{C} \mathrm{sgn}(\pi(C)) |\pi(C)|} \\
    &= \frac{\langle \mathrm{sgn}(\pi(C)) A(C) \rangle_{|\pi|} }{\langle \mathrm{sgn}(\pi(C)) \rangle_{|\pi|} }, 
\end{align}
where we combined all summation indices into a configuration index, $C = (a,n,\{b_i\})$.
The average sign $\langle \mathrm{sgn}(\pi(C)) \rangle_{|\pi|}$ generally decays exponentially with inverse temperature $\beta$ and system size~\cite{henelius2000sse_sign_problem}, leading to a severe sign problem for large systems and systems at low temperatures.
Because of the sign problem, SSE simulations generally require exponential computational costs, making them unfeasible for more complicated systems. As in Sec.~\ref{sec:quantum_sse}, quantum computers can naturally represent superpositions of arbitrary basis states, allowing for dispensing with the no-branching condition.

\subsubsection{Minimally Entangled Typical Thermal States}\label{sec:classical_metts}
Another approach to compute the thermal density matrices and properties is by sampling the minimally entangled typical thermal states (METTS)~\cite{white2009minimally,stoudenmire2010minimally} obtained from the ITE of product states.
METTS is defined by
\begin{equation}
    |\phi(i)\rangle=P(i)^{-1 / 2} \mathrm{e}^{-\beta\hat{H} / 2}|i\rangle
\end{equation}
with the unnormalized probability
\begin{equation}
    P(i)=\langle i|\exp (-\beta \hat{H})| i\rangle.
\end{equation}
$|\phi(i)\rangle$ is expected to have minimal entanglement entropy if the typical states $|i\rangle$ are chosen as non-interacting classical product states (CPS).
The von Neumann entanglement entropy of CPS after bipartite decomposition is strictly zero,
\begin{equation}
|i\rangle=\left|i_1i_2i_3\ldots i_N\right\rangle
\end{equation}
where $|i_j\rangle$ is the arbitrarily chosen local basis for site $j$. 

Sampling with probability $P(i)/Z$ where $Z=\tr{e^{-\beta\hat{H}}}$, one can compute the thermal density matrix and observable expectation from an ensemble of the pure states,
\begin{equation}
    \rho_\beta=Z^{-1}\sum_i P(i)|\phi(i)\rangle\langle\phi(i)|
\end{equation}
\begin{equation}
    \langle\hat{A}\rangle=Z^{-1} \sum_i P(i)\langle\phi(i)|\hat{A}| \phi(i)\rangle
\end{equation}
The rejection-free Markov chain sampling that \add{proposes a move} from one CPS $|i\rangle$ to another $|j\rangle$  is similar to the Monte Carlo step described in \cref{sec:classical_mc}. 
The Markov chain is based on the transition probability $p\left(i \rightarrow j\right)=\left|\left\langle j| \phi(i)\right\rangle\right|^2$, leading to the detailed balance
\begin{equation}
    \frac{P(i)}{Z} p(i \rightarrow j)=\frac{P(j)}{Z} p(j \rightarrow i),
\end{equation}
recovering the correct thermal distribution. 
This approach is commonly used with matrix product state (MPS) based methods to carry out the imaginary time evolution and sample from the transition probability
~\cite{PhysRevB.92.125119,PhysRevB.95.195148}. 

As the MPS ansatz is primarily effective for low-dimensional systems thanks to the area law~\cite{PhysRevLett.100.070502,*PhysRevLett.90.227902,*RevModPhys.82.277,*orus2014practical}, making extensions to 2D and 3D quantum many-body systems challenging in terms of both storage and ITE demands.
One may hope to lift these limitations using quantum computers within METTS, which we will review in \cref{sec:metts_quantum}. 

\subsection{Real-time dynamics}
The non-equilibrium dynamics of quantum systems is studied by solving the time-dependent Schr{\"o}dinger's equation:
\begin{equation}
\mathrm{i}\partial_t |\psi(t)\rangle
=
\hat{H}(t)|\psi(t)\rangle.
\label{eq:tdse}
\end{equation}
If one is interested in pure-state dynamics at $T = 0$, this equation of motion contains all the necessary dynamical information.
For instance, the dynamics of a local observable, $\hat{O}(t)$, is given by
\begin{equation}
    \langle O(t)\rangle
    =
\langle \psi(t)
|
\hat{O}(t)
|
\psi(t)\rangle,
\end{equation}
where $|\psi(t)\rangle$ is determined by
\cref{eq:tdse}.
To approximately solve this problem, ``real-time'' wavefunction methods have been developed, such as real-time time-dependent density functional theory~\cite{ullrich2011time}, real-time coupled cluster methods~\cite{Ofstad2023Sep}, etc.

\add{
In another context, one may consider a bipartite Hamiltonian decomposed into three terms,
\begin{equation}
    \hat{H} = \hat{H}_A + \hat{H}_B + \hat{H}_{A-B}
\end{equation}
where $\hat{H}_A$ and $\hat{H}_B$ are the local Hamiltonians for subsystems $A$ and $B$ respectively, and $\hat{H}_{A-B}$ describes their interaction.
When one is only interested in the dynamics of subsystem $A$, the reduced density operator provides a complete description:
\begin{equation}
    \hat{\rho}_A(t) = \operatorname{Tr}_B\left[\hat{\rho}_{\text{total}}(t)\right]\label{eq:rhoA}
\end{equation}
The total density operator evolves according to the von Neumann equation:
\begin{equation}
    \hat{\rho}_{\text{total}}(t) = \mathrm{e}^{-i\hat{H}t}\hat{\rho}_{\text{total}}(0)\mathrm{e}^{i\hat{H}t} 
\end{equation}
where $\hat{\rho}_{\text{total}}(t)$ represents the state of the complete system at time $t$, and $\operatorname{Tr}_B$ denotes the partial trace over subsystem $B$'s degrees of freedom. 
The dynamics of $\hat{\rho}_A(t)$ depends on the full Hamiltonian $\hat{H}$, 
requiring either numerically exact evolution of the total system or effective methods like master equations to capture the reduced dynamics~\cite{zwanzig2001nonequilibrium,kubo2012statistical,fehske2007computational,tanimura2020numerically,ren2022time,Bai2024Oct,ivander2024unified}.

For thermal equilibrium initial conditions, the total density operator is given by:  
\begin{equation}
    \hat{\rho}_{\text{total}}(0) = \frac{e^{-\beta \hat{H}}}{Z}, \quad Z = \operatorname{Tr}\left(e^{-\beta \hat{H}}\right)
\end{equation}
In this case, one is also often interested in thermal correlation functions given as
\add{
\begin{equation}
\langle O(0)V(t)\rangle
=
Z^{-1}\tr_{AB}(\hat{V}(0)\hat{W}(t)\exp(-\beta\hat{H})) 
\end{equation}
where $\hat{V}$ and $\hat{W}$ act only on the system $A$. 
}
One also often considers a factorized initial state (initially non-equilibrium):
\begin{equation}
    \begin{gathered}
        \hat{\rho}_{\text{total}}(0) = \hat{\rho}_A(0) \otimes \hat{\rho}_B(0), \\
        \hat{\rho}_A(0) = \frac{e^{-\beta \hat{H}_A}}{Z_A}, \quad \hat{\rho}_B(0) = \frac{e^{-\beta \hat{H}_B}}{Z_B}
    \end{gathered}
\end{equation}
where $Z_A = \operatorname{Tr}\left(e^{-\beta \hat{H}_A}\right)$ and $Z_B = \operatorname{Tr}\left(e^{-\beta \hat{H}_B}\right)$
are the respective partition functions of subsystems $A$ and $B$.
} 

These computational tasks are usually considered part of open quantum system dynamics or impurity problems but can also be considered in mixed quantum--classical dynamics with additional approximations.

Directly calculating dynamical properties using Monte Carlo sampling is more challenging than the computation of the static properties.
The additional challenge is due to so-called the dynamical sign problem, which becomes apparent after inspecting the path integral sampling of
the real-time propagator, $\exp(\mathrm{i}\hat{H}t)$.
There is an indirect way to address this problem via analytic continuation, where one takes imaginary-time correlation functions and analytically continues to the real-time domain.
While the recent advances made are quite encouraging~\cite{PhysRevB.41.2380,PhysRevB.44.6011,Jarrell1996May,Motta2014Jan,Motta2015Oct,PhysRevE.95.061302,Bertaina2017Mar},
analytic continuation is a formally ill-conditioned inverse problem that could miss some of the key features in the target real-time observables.
This section will focus on sampling methods that work directly in real time.

\subsubsection{Classical and Mixed Quantum--Classical Methods}
For a broad range of problems found in computational chemistry, treating the entire system classically or a smaller part of it quantum mechanically while handling the rest classically can be well-suited.
Prominent examples include studying the dynamics and structure of liquid and chemical reactions in the condensed phase.
In essence, one can sample from an equilibrium distribution,
\begin{equation}
Z
=
\tr(\mathrm{e}^{-\beta \hat{H}})
=
\int \mathrm{d}\mathbf z
\exp(-\beta H(\mathbf z))
\end{equation}
where $\mathbf z$ is a classical variable that completely determines the system's energy. One could also perform the dynamics of these classical variables following Newton's equations to calculate thermal equilibrium correlation functions.
This approach is known as the molecular dynamics approach~\cite{chandler}.

One can also partition the sampling into quantum (system $A$) and classical variables (system $B$),
\begin{equation}
\begin{aligned}
Z
&=
\tr(\mathrm{e}^{-\beta \hat{H}})\\
&=
\int \mathrm{d}\mathbf z
\exp(-\beta H_B(\mathbf z))
\tr_A(
\mathrm{e}^{-\beta(\hat{H}_A+\hat{H}_{A-B}(\mathbf z))}).
\end{aligned}
\end{equation}
The Born--Oppenheimer {\it ab initio} dynamics fit into this form where $A$ is the electronic and $B$ is the nuclear degrees of freedom, respectively.
Suppose the electronic energy levels of $A$ are close, so multiple states contribute to the trace. In that case, one may consider non-adiabatic effects and use mixed quantum--classical approaches such as Ehrenfest dynamics~\cite{Li2005Aug} and surface hopping~\cite{Subotnik2016May}.

Computational bottlenecks in this area include the cost of evaluating the forces on the nuclei and the duration of the molecular dynamics simulations necessary to probe the physical and chemical questions of interest.
One may hope that quantum computers may be able to improve the accuracy or efficiency of force evaluation (i.e., the same goal as the ground state algorithms)~\cite{PhysRevResearch.4.043210,PRXQuantum.5.010343} or the efficiency of sampling the relevant regions of the phase space such that the simulation time can be shortened (similar to what enhanced sampling approaches accomplish.) 



\subsubsection{Diagrammatic Monte Carlo}\label{sec:diagmc}
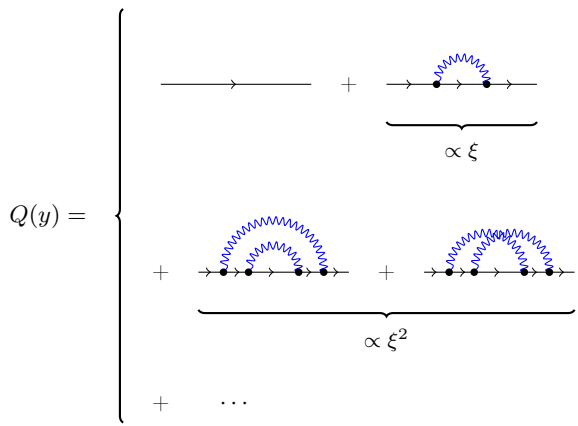
\begin{figure}
    \centering
        \begin{tikzpicture}

    \node[] at (-0.5,-1.75){$Q(y)  =$};
    \draw[ 
        decoration={markings, mark=at position 0.5 with {\arrow{>}}},
        postaction={decorate}
        ]
        (1.,0) -- (3.,0);
    \node[] at (3.5,0.){\footnotesize{+}};
    
    \draw[ 
        decoration={markings, mark=at position 0.166 with {\arrow{>}}, mark=at position 0.5 with {\arrow{>}}, mark=at position 0.833 with {\arrow{>}}},
        postaction={decorate}
        ]
        (4.,0) -- (6.,0);
    \draw[draw={blue}, style={decorate, decoration={snake, amplitude=0.5mm, segment length=2.5}}] (5.333,0) arc (0:180:0.333cm);
    \fill (4.666,0) circle (0.05); 
    \fill (5.333,0) circle (0.05); 
    \draw [
    thick,
    decoration={brace,mirror},
    decorate
    ] (4.,-0.5) -- (6.,-0.5) ;
    \node[] at (5, -0.9){\footnotesize{$\propto \xi$}};

    \draw[thick, decoration={brace}, decorate] (0.5,-4.5) node {} -- (0.5,1);

    \node[] at (1., -2.5){\footnotesize{+}};
    \draw[decoration={markings, mark=at position 0.0825 with {\arrow{>}}, mark=at position 0.2725 with {\arrow{>}}, mark=at position 0.5 with {\arrow{>}}, mark=at position 0.75 with {\arrow{>}}, mark=at position 0.93 with {\arrow{>}}},
    postaction={decorate}]
    (1.5,-2.5) -- (3.5,-2.5);
    
    \draw[draw={blue}, style={decorate, decoration={snake, amplitude=0.5mm, segment length=2.5}}] (2.833,-2.5) arc (0:180:0.333cm);
    \draw[draw={blue}, style={decorate, decoration={snake, amplitude=0.5mm, segment length=2.5}}] (3.166,-2.5) arc (0:180:0.666cm);

    \fill (2.166,-2.5) circle (0.05);
    \fill (2.833,-2.5) circle (0.05);
    \fill (3.166,-2.5) circle (0.05);
    \fill (1.833,-2.5) circle (0.05);
     \node[] at (4., -2.5){\footnotesize{+}};
     \draw[decoration={markings, mark=at position 0.0825 with {\arrow{>}}, mark=at position 0.2725 with {\arrow{>}}, mark=at position 0.5 with {\arrow{>}}, mark=at position 0.75 with {\arrow{>}}, mark=at position 0.93 with {\arrow{>}}},
    postaction={decorate}]
     (4.5,-2.5) -- (6.5,-2.5);
    
    \draw[draw={blue}, style={decorate, decoration={snake, amplitude=0.5mm, segment length=2.5}}] (5.833,-2.5) arc (0:180:0.5cm);
    \draw[draw={blue}, style={decorate, decoration={snake, amplitude=0.5mm, segment length=2.5}}] (6.166,-2.5) arc (0:180:0.5cm);

    \fill (5.166,-2.5) circle (0.05);
    \fill (5.833,-2.5) circle (0.05);
    \fill (6.166,-2.5) circle (0.05);
    \fill (4.833,-2.5) circle (0.05);
    \draw [
    thick,
    decoration={brace,mirror},
    decorate
    ] (1.5,-3.) -- (6.5,-3.) ;
    \node[] at (4., -3.4){\footnotesize{$\propto \xi^2$}};
    
    \node[] at (1, -4.25){\footnotesize{+}};
    \node[] at (2., -4.25){$\cdots$};
\end{tikzpicture}
    \caption{\add{Perturbative} expansion of $Q(y)$ in powers of the perturbation strength $\xi$. All diagrams displayed here correspond to diagonal elements of 0-boson diagrams.}
    \label{fig:perturbation}
\end{figure}

\noindent A relatively widely used QMC method for both finite temperature (imaginary-time) and real-time problems is diagrammatic Monte Carlo (DiagMC). DiagMC assumes a random variable $y$ to be distributed according to 
\begin{equation} \label{eq:diagmc_distrib}
    Q(y) = \sum_{m=0}^{\infty} \sum_{\xi_m} \int \cdots \int \mathrm{d}x_m \cdots \mathrm{d}x_1 F(\xi_m, y, x_1, \cdots, x_m).
\end{equation}
\add{A visual representation of $Q(y)$ is provided in Fig.~\ref{fig:perturbation}.}
The random variable $y$ is then sampled via a Metropolis--Hastings algorithm from a configuration space of diagram order $m$, topology $\xi_m$, and internal variables $\{x_i\}$. The algorithm is similar to sampling from multidimensional integrals but additionally accounts for the variable integration multiplicity of individual terms of the sum~\cite{prokofev1998polaron, mishchenko2000frohlich}.
\cref{eq:diagmc_distrib} can be used to sample time-dependent observables expanded in the Dyson series~\cite{Muhlbacher2008May}:
\begin{equation}
\begin{aligned}
    \langle \hat{O}(t)\rangle &= \sum_{m = 0}^\infty \sum_{n = 0}^m \int \mathrm{d} s_m \cdots \int \mathrm{d}s_1 (-1)^n \mathrm{i}^m\\ 
    &\times \left\langle \hat{H}'(s_m) \cdots \hat{H}'(s_{n+1}) \hat{O}' (t_{\text{max}})\hat{H}'(s_n) \cdots \hat{H}'(s_{1})\right\rangle
\end{aligned}
\end{equation}
where $\hat{A}'(t)$ denotes the time-dependent operator in the interaction picture $\hat{A}'(t) = \mathrm{e}^{\mathrm{i}\hat{H}_0t}\hat{A}\mathrm{e}^{-\mathrm{i}\hat{H}_0t}$, and $s_i$ is the time index ordered according to the Keldysh contour causality. Due to the $(-1)^n\mathrm{i}^m$ factor in the expansion, the average sign of the diagrams decays exponentially as the expansion order increases, which is termed as the \textit{dynamical sign problem}~\cite{Werner2009Jan}.

Recently, Cohen \textit{et al.}~\cite{cohen2015inchworm,chen2016inchworm} proposed an inchworm modification to DiagMC, effectively dealing with the dynamical sign-problem and enabling real-time propagation of the Anderson impurity model with computational
cost scales quadratically with time. To propagate for $t_i$ to $t_f$ one starts from an exact short-time propagator for a small time-segment $[t_i, t_{\scriptscriptstyle\uparrow}]$ and $t_{\scriptscriptstyle\uparrow} < t_f$. The algorithms subsequently \textit{inches} forward by $\Delta\tau$ to obtain an approximately exact propagator for the interval $[t_i, t_{\scriptscriptstyle\uparrow} + \Delta\tau]$ by sampling diagrams either completely contained in $[t_{\scriptscriptstyle\uparrow}, t_f]$, or connecting the previous approximation of the exact propagator on $[t_i, t_{\scriptscriptstyle\uparrow}]$ to the interval $[t_{\scriptscriptstyle\uparrow}, t_f]$. While the results of the Anderson impurity model are impressive, more general problems are still prohibitive to handle due to the dynamical sign problem. More details can be found in Ref.~\citenum{cohen2015inchworm}.
One may hope to reduce the dynamical sign problem by utilizing quantum computers.
This idea is explored in a quantum algorithm we will review in \cref{subsec:qcmc}.

\section{Quantum Algorithms for Ground State}\label{sec:quantum_for_gs}
\add{
This section explores quantum algorithms that leverage sampling techniques for ground state calculations. 
We primarily focus on hybrid quantum-classical algorithms, 
whose classical components are detailed in Sec.~\ref{sec:2a}. 
We also briefly discuss fully quantum algorithms designed for fault-tolerant quantum computers. 
Hybrid algorithms fall into two main categories: 
quantum-enhanced classical algorithms and 
quantum algorithms assisted with classical processings.
In quantum-enhanced classical algorithms, quantum-prepared states serve as initial guesses or trial states 
for classical algorithms, enabling faster convergence and/or better accuracy. 
In quantum algorithms with classical assistance, 
classical methods are employed to 
refine quantum results or solve intermediate steps,
enabling either larger system sizes, higher accuracy, or reduced quantum resource requirements 
like circuit depth.
We first present these algorithms grouped by their classical components,
following the structure of Sec.~\ref{sec:2a}, 
and conclude with comparison amongst them. 
}
\subsection{Quantum Algorithms for Variational Monte Carlo}
\subsubsection{Quantum-enhanced Variational Monte Carlo}
As mentioned in the introduction, VQE is a class of hybrid quantum--classical algorithms proposed as a near-term alternative to the QPE for finding the ground state of a Hamiltonian~\cite{Tilly2022Nov}. The main argument for the utility of VQE relies on its compatibility with the resource constraint on the NISQ devices \cite{cerezo2021variational}. The accuracy of the VQE is highly dependent on the choice of quantum circuit ans{\"a}tze, many of which were inspired by the classical CC theory~\cite{grimsley2019trotterized, lee2018generalized, evangelista2019exact}. 


Leveraging the VQE framework, Montanaro \textit{et al.} proposed using the output from VQE to accelerate the MCMC convergence of VMC simulation and termed this quantum-enhanced VMC (QEVMC)~\cite{montanaro2023accelerating}. Specifically, the output state from VQE is used for initial walker samples in the MCMC algorithm that generates the desired distribution $P(\bt, n)$ (\cref{eq:pn}). 
Since both VQE and VMC are performed variationally, the VMC is guaranteed to find lower energy than the original VQE input.
The hope is that if VQE output can capture the ground state to a good extent, or equivalently, the initial and final distributions have more resemblances, then the number of steps in MCMC is expected to be less. The other implicit assumption in this workflow is the efficient classical calculation of the overlap $\langle n | \Psi(\bt) \rangle$, which, however, could be hard for some $|\Psi(\bt)\rangle$~\cite{huggins2022unbiasing}, for instance, the unitary coupled-cluster quantum state~\cite{Anand2022Mar}.

Using the VQE distribution from Ref.~\citenum{stanisic2022observing}, Montanaro \textit{et al.} demonstrated that MCMC converges to a target energy more rapidly than a single determinant distribution. For Fermi-Hubbard and transverse field Ising models, Montanaro \textit{et al.} further numerically showed that the VQE distribution obtained for modest lattice sites can be used to speed up MCMC in larger site calculations. 

\add{
    Beyond the optimization challenges of many non-linear 
    parameters and vanishing gradients exhibiting barren plateaus when parameters are randomly initialized~\cite{mcclean2018barren},
    VQE also faces another more fundamental limitation: 
    it cannot leverage the zero-variance principle
    because quantum measurements inherently introdcuffe statistical noise,
    even when measuring exact eigenstates.
    Even with fault-tolerant quantum computers and highly accurate ansatz states,
    this fundamental limitation necessitates a large number of measurements
    to achieve high precision, regardless of how close the prepared quantum state
    is to the true eigenstate.
    In contrast, classical VMC benefits from the zero-variance principle~\cite{assaraf_zero-variance_1999},
    where the variance of energy estimators vanishes as the trial wavefunction approaches the exact eigenstate.
    Hence, some combination of VMC and VQE might result in a powerful alternative to VQE.
    The hybrid quantum-classical projector Monte Carlo methods we introduce in Sec.~\ref{sec:qc-pmc}
    can benefit from the zero-variance principle suppressing noises.
}

\subsubsection{Non-unitary Post-Processing via Quantum Monte Carlo}
Given the limited quantum resources in the NISQ era, classical post-processing of quantum measurements could be used to reduce the circuit depth and improve the accuracy of quantum-implemented variational wavefunction \textit{ansatz} $|\psi(\bt) \rangle = U(\bt) |\mathbf{0} \rangle$~\cite{li2021vsql}. For example, inspired by classical QMC, Mazzola \textit{et al.} proposed using the Jastrow factor as a non-unitary post-processing projector.
Specifically, they defined the state $|\psi (\bt, \boldsymbol{\phi})\rangle = \mathcal{P}{\boldsymbol{\phi}} |\psi(\bt) \rangle$, where $\mathcal{P}{\boldsymbol{\phi}}$ serves to filter out higher energy components from the suboptimal variational quantum heuristic ansatz. This approach was applied successfully to several lattice models~\cite{mazzola2019nonunitary}. 

In general, the non-unitary classical processing operator $\hat{f}$ takes in bit string $|n\rangle$ sampled from the quantum circuit and outputs \add{$f(n)$}. With this, we then consider the wavefunction \textit{ansatz} written as 
\begin{equation} \label{eq: wavefunction}
    |\psi_f (\bt, \boldsymbol{\phi}) \rangle = \hat{f}(\boldsymbol{\phi}) U(\bt) |\mathbf{0} \rangle,
\end{equation}
and in the computational basis, the classical post-processing unit $\hat{f}$ reads
\begin{equation} \label{eq: classical processing}
    \hat{f}(\boldsymbol{\phi}) = \sum_{n} f_{\phi}(n) | n \rangle \langle n |.
\end{equation}
In this framework, the optimization is carried out simultaneously for the quantum gate parameters $\bt$ and the classical post-processing parameters $\boldsymbol{\phi}$. Using efficient implementation with classical post-processing, the energy expectation value 
\begin{equation} \label{eq: expectation}
    E(\bt, \boldsymbol{\phi}) = \frac{\langle \psi_f (\bt, \boldsymbol{\phi}) | \hat{H} | \psi_f (\bt, \boldsymbol{\phi}) \rangle}{\langle \psi_f (\bt, \boldsymbol{\phi}) | \psi_f (\bt, \boldsymbol{\phi})\rangle}
\end{equation}
and its gradient (e.g., through parameter shift rule~\cite{mitarai2018quantum, schuld2019evaluating}) requires no more than the extra polynomial number of gates or measurements in addition to the original circuit $U(\bt)$. \add{Parameter shift rule relies on measuring the circuits at multiple shifted values of one parameter to yield the exact partial derivative. We note that this approach allows one to obtain analytical gradient with linear scaling measurements in the number of circuit parameters. } Since the final wavefunction is not normalized, one needs to account for the overlap explicitly. 
Following \cref{eq: wavefunction} and \cref{eq: classical processing}, the overlap can be written as 
\begin{equation} \label{eq: overlap}
    \langle \psi_f (\bt, \boldsymbol{\phi}) | \psi_f (\bt, \boldsymbol{\phi}) \rangle = \sum_{n} |f_{\boldsymbol{\phi}}(n)|^2 |\langle n |U(\bt)|\mathbf{0} \rangle|^2.
\end{equation}
$|n\rangle$ can be sampled by measuring $U(\bt)|\mathbf{0} \rangle$ in the computational basis, and the output bit string is sent to the classical unit. This requires no modification to the existing implementation of $U(\bt)$ on the quantum circuit. 

The evaluation of the energy expectation value can be done in several ways. One way is to transform $\hat{H}$ as $\hat{f}(\boldsymbol{\phi}) \hat{H} \hat{f}(\boldsymbol{\phi})$ and extra observables need to be measured from the quantum computers~\cite{benfenati2021improved}. However, this transformation introduces an exponential number of Pauli strings that must be implemented on the quantum hardware. The other method relies on an entangled copy to reconstruct the measurement probability $|\langle n |U(\bt)|\mathbf{0} \rangle|^2$ \cite{mazzola2019nonunitary}.
However, to achieve a target precision $\epsilon$, the number of samples/measurements required scales exponentially with the system size, or equivalently, the number of qubits 
~\cite{zhang2022variational}.

To overcome these exponential overheads, Zhang \textit{et al.} \cite{zhang2022variational} proposed an efficient scheme for evaluating $\langle \psi_f | H | \psi_f \rangle$ by appending a measurement circuit $M$ to the original $U(\bt)$. The following discussion assumes that $\hat{H}$ consists of only one Pauli string. In the case where $H$ has only $I$ and $Z$ operators
$M$ is identity because $H$ is diagonal in the computational basis. If the Hamiltonian has at least one spin-flip operator (e.g $X$ or $Y$), then one can isolate one qubit that is flipped and write the Hamiltonian as follows,
\begin{equation} \label{eq:H}
    \hat{H} = \sum_{\s} S(0\s) | 0\s \rangle \langle 1 \tilde{\s}| + S(1 \tilde{\s}) | 1 \tilde{\s} \rangle \langle 0\s |,
\end{equation}
where $\s$ and $\tilde{\s}$ indicate all qubits excluding the isolated qubit, and they are related by $H |0 \s \rangle = S(1\tilde{\s}) | 1\tilde{\s} \rangle$, with $S(\mathbf{s})$ be the coefficients. The coefficients have an additional constraint $S(0\mathbf{s}) S(1\tilde{\mathbf{s}}) = 1$ because $\hat{H}$ is a single Pauli string (e.g $\hat{H}^2 = \mathrm{1}$). It is straightforward to construct eigenvectors of the Hamiltonian \cref{eq:H}, denoting as $|+,\mathbf{s}\rangle$ and $|-,\mathbf{s}\rangle$ for each unique bit string $\s$ and are spanned by $\{| 0\s \rangle , | 1 \tilde{\s} \rangle \}$. The numerator in \cref{eq: expectation} is then written as
\begin{equation}
\begin{aligned}
    \langle \psi_f | H | \psi_f \rangle = \sum_{\s} | \langle +,\mathbf{s} | U(\bt) | \mathbf{0} \rangle|^2 f(0\s) f(1\tilde{\s})\\ - | \langle -,\mathbf{s} | U(\bt) | \mathbf{0} \rangle|^2 f(0\s) f(1\tilde{\s}).
\end{aligned}
\end{equation}
The measurement circuit $M$ to obtain $ |\langle +,\mathbf{s} | U(\bt) | \mathbf{0} \rangle|^2$ can then be constructed by appending a controlled rotation gate with the source being the isolated qubit and the target being all other qubits on which spin-flip operators from the Hamiltonian ($X/Y$) acted. 
The isolated qubit is then measured in the $X$/$Y$ basis, and all other qubits are measured in the $Z$ basis. The controlled gates ensure the bit string on all other qubits is flipped properly, conditioned on the outcome of the isolated qubits. The number of extra two-qubit gates depends on the number of $X$/$Y$ operators in the Hamiltonian. 

The implementation presented in \add{Ref. \cite{zhang2022variational}} is efficient if the number of these operators does not scale exponentially with the system size, or equivalently, the Hamiltonian has local interactions only. 
Furthermore, Zhang \textit{et al.} \cite{zhang2022variational} showed that the number of measurements when introducing the measurement circuit $M$ scales as $\mathcal{O}(\text{Poly}(r)/\text{Poly}(\epsilon))$, where $\epsilon$ indicates the precision (standard deviation from the measurement outcome)and $r$ is the range of the function $f_{\boldsymbol{\phi}}(n)$. 
Since the classical processing unit can be designed within a specific range, the number of measurements has polynomial scaling similar to that of the initial VQE with the same target precision. 

 Moreover, the use of quantum circuits like those presented here can be viewed as an improvement to VMC. Specifically, given that the energy expectation value in \cref{eq: expectation} can be evaluated efficiently up to an additive error with polynomial quantum resources, one can perform VMC without resorting to Monte-Carlo algorithms to sample the local energy following distribution in \cref{eq:pn}. 
With this insight, recent algorithmic developments of hybrid algorithms have introduced methods to efficiently sample the expectation values  $\langle\Psi(\bt)|\hat{H}|\Psi(\bt)\rangle$ without relying on MCMC. \add{Interested readers} are referred to Ref.~\cite{PhysRevResearch.4.033173,huang2020predicting,Hadfield2022May,wan_matchgate_2023} for more technical details.

\subsubsection{Quantum data-enhanced Variational Monte Carlo}
The accuracy of VMC heavily relies on the parametrization of the variational ans\"{a}tze. Employing more expressive ans\"{a}tze could enhance the performance of VMC calculations, an avenue for which where neural networks (NN) have been valuable. Recent years have seen the development of different NN ans\"{a}tze in VMC calculation applying to various problems, including the ground state energies of various spin models~\cite{carleo2017solving}, ground state properties of real solids~\cite{li2022ab}, ground and excited state energies of Heisenberg models for frustrated magnets~\cite{choo2019two, djuric2020efficient, roth2023high} and steady states of open quantum systems~\cite{nagy2019variational, vicentini2019variational}, \textit{etc.}

Recently, with the emergence of programmable quantum devices composed of Rydberg atom arrays~\cite{Saffman2016Oct,Endres2016Nov,Bernien2017Nov,Ebadi2021Jul}, researchers have begun to explore the ground state and phase transitions in these platforms. A Rydberg array consisting of $N$ qubits can be described using the following Rydberg Hamiltonian parametrized by Rabi frequency $\Omega$, detuning $\Delta$, and the long-range van der Waals interaction between Rydberg states $V_{ij}$:
\begin{equation}
    \hat{H} = \frac{\Omega}{2}\sum_{i = 1}^N \hat{\sigma}_i^x - \Delta \sum_{i = 1}^N\hat{n}_i  + \sum_{i < j}V_{ij}\hat{n}_i\hat{n}_j.
\end{equation}
Here $\hat{\sigma}_i^x = \ket{g}_i\bra{r}_i + \ket{r}_i \bra{g}_i$, where $\ket{g}_i$ denotes the ground state of the atom at site $i$, and $\ket{r}_i$ is the Rydberg state. The interaction between Rydberg states $V_{ij} = V_0/|\mathbf{x}_i - \mathbf{x}_j|^6$ comes from the so-called Rydberg blockade mechanism \cite{lukin2001dipole}, which can be tuned by the effective blockage range $R_{\mathrm{b}}/a$, where $R_{\mathrm{b}} \equiv (V_0/\Omega)^{1/6}$ is the blockade radius, and $a$ is the inter-atomic spacing in the array.

NN ans\"{a}tze for VMC turns out to be a suitable tool for studying this Hamiltonian, as shown in Ref. \cite{czischek2022data}. Prior to this work, generative models such as the restricted Boltzmann machine (RBM) have been proven effective for state reconstruction from the measurements using a quantum circuit \cite{torlai2019integrating}. They compared few-body observables and R'{e}nyi entropy derived from the reconstructed wavefunction and found strong agreement with both experimental outcomes and simulation results obtained using the Lindblad master equation.
Combining those efforts, Moss et. al. proposed a two-phase optimization of a neural network ansatz to target the ground state of the Rydberg Hamiltonian \cite{moss2024enhancing}, where they first trained the neural network using the raw, unprocessed experimental data obtained from the measurements of a Rydberg atom array undergoing the phase transition between disordered and checkerboard phases~\cite{Ebadi2021Jul}, and performed subsequent variational optimization with the resulting NN wavefunction. \add{Note that the states prepared by the quantum device are not the true ground states due to coherent perturbations, such as parameter miscalibration and nonadiabatic effects arising from finite preparation times, as well as incoherent errors~\cite{moss2024enhancing}. Hence, subsequent variational optimization is crucial to reach the ground state.} They employed both one-dimensional (1D) and two-dimensional (2D) recurrent neural network (RNN) to encode the ground state wavefunction, a type of universal function approximators \cite{schafer2006recurrent} that can encode arbitrarily complex functions by increasing the number of parameters. 

A notable improvement was shown using a hybrid data-driven and Hamiltonian-driven approach. Naive variational optimization of 1D RNN wavefunction struggles to reach the ground state, nevertheless, the pretraining stage significantly improves the quality of the resulting variational wavefunction, as shown by the fact that the energy calculated using the hybrid strategy aligns more closely with the true ground state energy. Although variationally optimized 2D RNN wavefunction itself can capture most of the characteristics of the ground state, the pretraining step still accelerates the convergence towards the ground state. Subsequent calculations on other observables further demonstrate the significance of the pretraining step. These results highlight the possibility of using analog quantum simulations in QMC. Namely, one can use the data from quantum computations to accelerate convergence and enhance accuracy in classical simulations.

\subsection{Hybrid Projector Quantum Monte Carlo}\label{sec:qc-pmc}
\subsubsection{Hybrid Quantum--Classical AFQMC}\label{sec:qc-afqmc}
\begin{figure*}[htbp]
    \centering
    \includegraphics[width=0.94\linewidth]{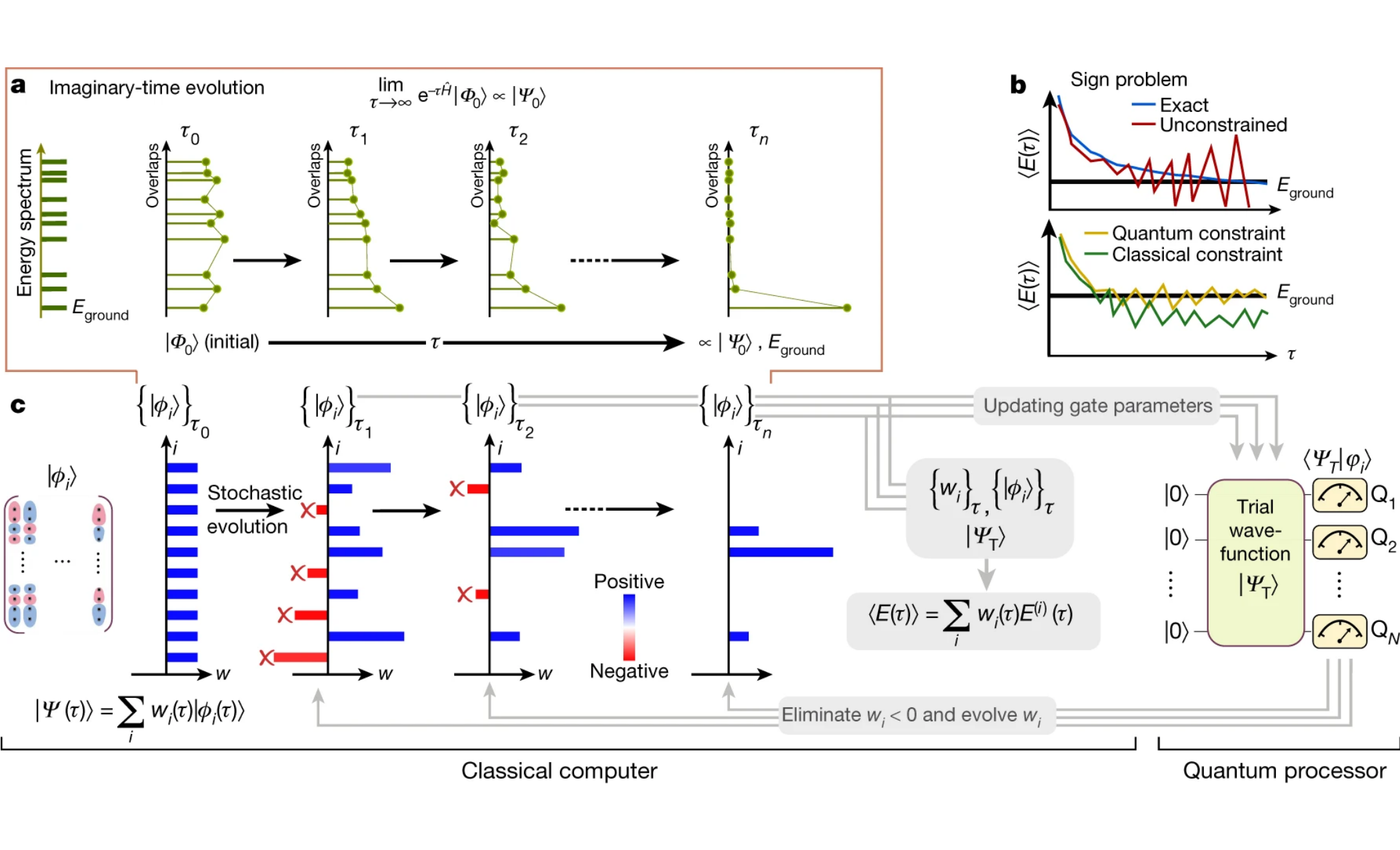}
    \caption{a. Schematic illustration of the imaginary-time evolution, which converges to the ground state in the long-time limit. b. Demonstration of the sign problem. The noise-to-signal ratio will grow exponentially if no constraint is imposed (top). Constrained QMC controls the sign problem at the expense of introducing a bias, and using quantum constraint reduces the bias. c. Summary of the QC-AFQMC algorithm. The walker wavefunctions, denoted as $\{\ket{\phi_i}\}_\tau$, are evolving in time according to the propagator $\mathrm{e}^{-\Delta \tau \hat{H}}$, which is achieved by updating the gate parameters each time step. The resulting observables are calculated on a classical computer with overlap queries to the quantum processor. Figure adapted from Ref.~\citenum{huggins2022unbiasing}.}
    \label{fig:qcafqmc}
\end{figure*}

The quantum--classical hybrid QMC (QC-QMC) method has opened new avenues for projector QMC methods, such as AFQMC, in simulating chemical systems~\cite{huggins2022unbiasing}.
Subsequent studies in refining the framework and assessing the performance~\cite{wan_matchgate_2023,amsler_quantum-enhanced_2023,kiser_classical_2023,Huang2024Apr,kiser2024contextualsubspaceauxiliaryfieldquantum} have been sharpening the path towards applying quantum--classical hybrid AFQMC (QC-AFQMC) towards more precise and experimentally relevant quantum chemical calculations.
In QC-AFQMC~\cite{huggins2022unbiasing}, 
the trial wavefunction for the phaseless approximation is prepared on a quantum device. 
At the same time, imaginary time propagation and energy estimation are performed on a 
classical computer. 
\add{Huggins \textit{et. al.}} 
prepared a CC-type trial wave function on a Sycamore processor~\cite{arute2019quantum}.  
The QC-AFQMC algorithm was shown to be robust to some of the common limitations imposed by circuit noise and hardware imperfections in the NISQ era. 
The imaginary-time evolution and sign problem in PMC methods and the workflow of QC-AFQMC are illustrated in Fig. \ref{fig:qcafqmc}.

Using this algorithm, they first calculated the atomization energy of \ce{H4}, a typical testing ground for electronic structure methods for strongly correlated systems. There are two equally important mean-field states for the ground state of this system, rendering AFQMC with a single Slater determinant trial inaccurate. However, AFQMC with a generalized valance bond perfect pairing~\cite{goddard1973generalized} (GVB-PP) trial wavefunction prepared by a quantum circuit is able to produce results with chemical accuracy (1 kcal/$\mathrm{mol}$). It is also noteworthy that although the trial prepared with the quantum circuit is far from ideal, i.e., the trial energy significantly deviates from the expected outcomes from a noiseless quantum computer. Despite this, QC-AFQMC still achieves chemical accuracy in the atomization energy. Calculating the potential energy surface of \ce{N2} and the cold curve of diamond \add{(equation of state at 0 K)} using a minimal unit cell to nearly chemical accuracy also showcased the improvement of the AFQMC energy using a more quantum circuit trial state as well as the noise-resilience of the QC-AFQMC algorithm. 

As mentioned in Section.~\ref{sec:classic_afqmc}, the key quantity in (QC-)AFQMC is the overlap between the trial wavefunction and the walkers. Shadow tomography~\cite{aaronson2018shadow, huang2020predicting} was used to efficiently measure the overlap between the quantum trial wavefunction and the walker wavefunction, enabling the prediction of properties of a quantum system with relatively few measurements.
More specifically, classical shadows were utilized to obtain the representation of the trial state on a computational basis, thus bypassing repeated overlap measurements for each time step and walker. A detailed discussion of classical shadows techniques can be found in Sec.~\ref{shadows}.

Two computational considerations are needed to scale up QC-AFQMC. The first consideration is the cost of classical post-processing. The original QC-AFQMC work used Clifford-based classical shadows, causing an exponential scaling postprocessing cost. This exponential overhead has been removed using a different channel in classical shadows~\cite{wan_matchgate_2023}.
Even with the new shadows technique, the classical post-processing per sample was found to be too high ($\mathcal O(N^8)$), but this cost was brought down to $\mathcal O(N^4)$ in a recent work~\cite{jiang2024unbiasing}. We will discuss more details in \cref{shadows}.

The second consideration is the measurement sample complexity associated with the overlap estimation.
The overlap of the walker wavefunction with a quantum trial wavefunction is expected to decay exponentially with the system size~\cite{huggins2022unbiasing}. This means for a fixed relative error one needs to refine the overlap estimation exponentially more when increasing the system size. This results into an exponential-scaling measurement complexity in QC-QMC.
On the classical computer, one can completely address this issue by directly estimating the logarithm of the overlap, which is currently not possible to achieve on the quantum computer. It was pointed out that \add{one only needs to consider }a small physical space by leveraging the virtual correlation technique~\cite{huggins2022unbiasing}, potentially postponing this orthogonality catastrophe. This way, the overlap will decay, in the worst case, exponentially with respect to the correlation length within the active space rather than the system size. Only the overlap within the active space is then computed with a quantum computer.

To further investigate the exponential challenge in overlap evaluation, Mazzola \textit{et al.}~\cite{mazzola2022exponential} conducted numerical tests on a one-dimensional transverse Ising model (TFIM)\add{, which }can be mapped to a free fermion problem where AFQMC is exact without sampling, \add{and} the authors studied this with GFMC.
They discovered that even with an exponentially increasing number of states using a Jastrow-type trial wavefunction, the statistical error in the local energy \add{is still considerably large}. Additionally, they observed that the error in the energy per site increases exponentially relative to the number of sites, even when using the exact ground state wavefunction as the trial wavefunction. However, as Lee \textit{et al.} showed in Ref.~\citenum{lee2022response}, they noted a significant reduction in error when employing more sophisticated trial wavefunctions and walker wavefunctions, specifically $\ket{\Psi_\mathrm{T}} = \mathrm{e}^{-0.05\hat{H}}\ket{\Psi_\mathrm{MC}}$, where $\ket{\Psi_\mathrm{MC}}$ is the trial wavefunction used in Ref.~\citenum{mazzola2022exponential}. 

\begin{figure}[htbp]
    \centering
    \includegraphics[width=0.48\textwidth]{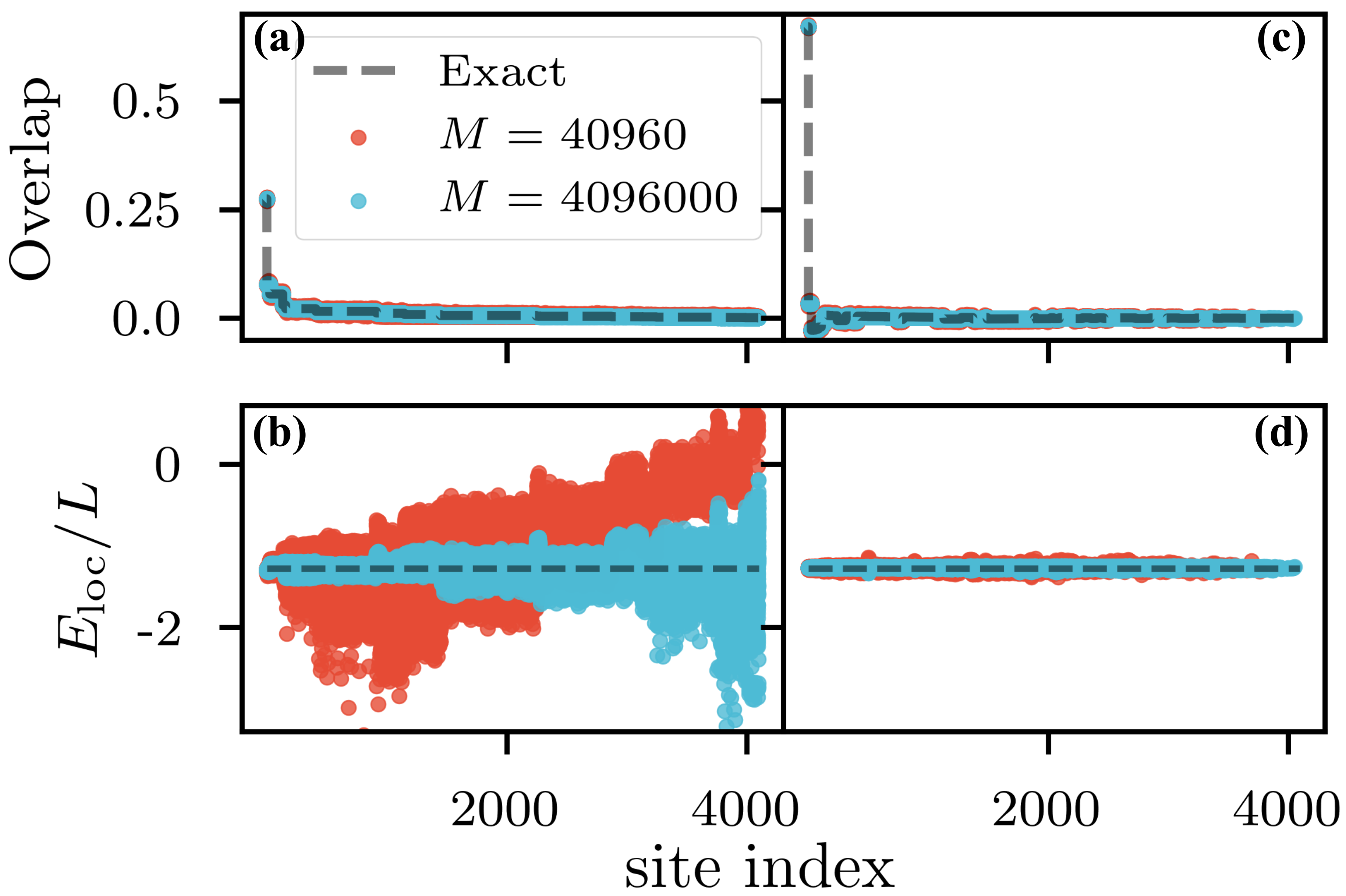}
    \caption{Distribution of overlap and local energy per site across states in 12 sites 1-dimensional TFIM for different trial and walker wavefunction as well as the number of measurements. For (a) and (b), $\ket{\Psi_{\mathrm{T}}} = \ket{\Psi_{\mathrm{MC}}}$ and single computational basis state walkers were used, while for (c) and (d), the trial wavefunction is given by $\ket{\Psi_\mathrm{T}} = \mathrm{e}^{-0.05\hat{H}}\ket{\Psi_\mathrm{MC}}$ and the walkers are generated by spin flips from $\ket{\Psi_{\mathrm{MC}}}$. 16 independent simulations were run with $M$ samples each time for the measurement of overlap and local energy. Figure replotted with the data from Ref.~\citenum{lee2022response}.}
    \label{fig:resp2}
\end{figure}

\add{As shown in Fig.~\ref{fig:resp2}, utilizing a more sophisticated walker sharpens the weight distribution and substantially reduces the variance in local energy. Like all ground-state methods exhibiting high accuracy and efficiency for certain systems but not for others, the exponential challenges associated with QC-QMC are influenced by multiple factors. These include the specific QMC method employed, the properties of the underlying system, and the form of the trial and walker wavefunctions. Future research in QC-QMC is expected to focus on identifying systems with specific combinations of these factors, paving the way toward realizing quantum computational advantages.
}

M. Amsler \textit{et al.} further tested the efficacy of the QC-AFQMC algorithm in various systems, including molecules such as \ce{H4}, ozone, and oxygen, as well as model systems such as the quasi-1D Fermi--Hubbard model derived from \ce{CuBr2}~\cite{amsler_quantum-enhanced_2023}. For \ce{H4}, they compared the AFQMC energy using VQE and RHF as the trial wavefunction, where they used a unitary coupled-cluster single and double (UCCSD) ansatz for the VQE calculation and found that both the VQE energy and AFQMC energy using VQE trial are close to FCI result. In addition, AFQMC energies using a VQE trial stay within chemical accuracy compared to FCI results except at intermediate bond lengths, and it is considerably closer to FCI results than the plain VQE approach. They also found that the relative energies of ozone and singlet oxygen compared to ground-state oxygen can be accurately predicted using AFQMC with a VQE trial wavefunction. However, this calculation yields a slightly larger error than AFQMC with a complete active space self-consistent field (CASSCF) trial wavefunction. 
They also demonstrate the capability of AFQMC to accurately compute the ground state energy of extended systems, as evidenced by calculations on the Fermi--Hubbard model.
In conclusion, quantum trial wavefunctions, such as those obtained from VQE, generally enhance the quality of AFQMC calculations. 
However, one needs to be cautious about the exponential scaling measurement overhead.

\subsubsection{Classical Shadows for QCQMC}\label{shadows}
We use $\rho$ to denote the density matrix of the $N$-qubit prepared quantum state.
Classical shadows provide a convenient and powerful framework for efficiently estimating the expectation value of general observables $O$ from measurements~\cite{huggins2022unbiasing}.
In QC-AFQMC, we want to compute expectation values of operators like $O = \vert \phi \rangle \langle 0 \vert$ where $\vert \phi \rangle$ is a Slater determinant state, and $\vert 0 \rangle$ is the vacuum state.
These observables contain many non-trivial matrix elements and are not easy to directly measure on any single basis.

The key idea of classical shadows is to combine information-dense snapshots (i.e., projective measurements of each qubit) with measurements in different bases to estimate arbitrary observables with rigorous convergence guarantees.
The shadow procedure constructs an efficient classical representation $\hat{\rho}$ of the quantum state $\rho$ from measurements, such that expectation values of generic observables can be approximately evaluated by
\begin{equation}
    \mathrm{tr}[O \hat{\rho}] \approx \mathrm{tr}[O \rho]
\end{equation}
up to an additive error $\epsilon$.
Before measurement, a unitary $U$ is applied to the state to select the measurement basis.
Then the state $U\rho U^\dagger$ is measured in the computational basis ($\left.\{|b\rangle\}_{b \in\{0,1\}^N}\right)$).
Thus the state is mapped from $\rho\rightarrow U^{\dagger}\vert b \rangle \langle b \vert U$ where $U^{\dagger}$ inverts the unitary applied pre-measurement.
In the typical shadows procedure, $U$ is sampled from a distribution $\mathcal{U}$, so we denote averaging over unitaries by $\mathbb{E}_{U \sim \mathcal{U}}$. Then, for a sampled $U$, the probability of measuring any particular $b$ is given by
\begin{equation}
p_U(b) = \langle b \vert U\rho U^{\dagger} \vert b \rangle,
\end{equation}
The map from the input state to a measured state is known as the measurement channel and can be written explicitly as
$$\mathcal{M}(\rho) = \mathbb{E}_{U \sim \mathcal{U}} \sum_b \langle b \vert U \rho U^{\dagger} \vert b \rangle U^{\dagger} \vert b \rangle \langle b \vert U$$
By inverting $\mathcal{M}(\rho)$, an estimate for $\rho$ can be constructed from a measured state $U_m^{\dagger} \vert b_m \rangle \langle b_m \vert U_m$, where $U_m$ and $b_m$ are the unitary and bitstring sampled during the $m$-th run of the experiment.

The average over measured states is known as a classical shadow and is written as
\begin{align}
\hat{\rho} &= \frac{1}{M}\sum_m \hat{\rho}_m \\
\hat{\rho}_m &= \mathcal{M}^{-1}(U_m^{\dagger} \vert b_m \rangle \langle b_m \vert U_m)
\end{align}
In general, $\mathcal{M}(\rho)$ is only invertible if the ensemble of rotations $\mathcal{U}$ is tomographically complete~\cite{huggins2022unbiasing}, \add{i.e., for any $\sigma \neq \rho$ there exist $U \in \mathcal{U}$ and $\ket{b}$ such that $\langle b |U\sigma U^\dagger| b\rangle \neq \langle b |U\rho U^\dagger |b\rangle$.}
In that case, it is easy to show that $\mathbb{E}[\hat{\rho}_m] = \rho$, the true state is recovered on average.
More generally, if $\mathcal{M}$ is not invertible, a pseudo-inverse can be used~\cite{vankirk2022hardwareefficient}, in which case $\mathbb{E}[\hat{\rho}_m]$ recovers a projection of $\rho$ onto the measurable (i.e. "visible") operators.
Then, the expectation value of (visible) observables $O$ can be estimated accurately by evaluating $\mathrm{tr}[O \hat{\rho}]$. 

Different shadow procedures are characterized by different choices of ensembles $\mathcal{U}$. There are two key characteristics that determine the performance of a shadow procedure. 
The first is the sample complexity, which depends on the variance of the estimator and quantifies how many measurements are needed to estimate observables $O$ to within an additive error $\epsilon$. 
The second is the classical post-processing complexity. For certain special ensembles, like random single-qubit rotations or random Clifford circuits, the states $U^{\dagger} \vert b \rangle \langle b \vert U$ and inverse channel $\mathcal{M}^{-1}$ take a simple classically tractable form. The complexity also depends on the type of observable being measured. For example, Pauli observables can be efficiently classically computed with respect to Clifford (i.e. stabilizer) states.

In the original QC-AFQMC proposal, the shadow tomography is performed using random unitaries sampled from the Clifford group of $N$ qubits (henceforth referred to as Clifford shadows)~\cite{huggins2022unbiasing}, but requires exponential classical post-processing cost. Wan \textit{et al.} removed the exponential overhead by replacing the Clifford group with the ensemble of random matchgate circuits~\cite{wan_matchgate_2023}. Matchgate circuits are qubit representations of fermionic Gaussian unitaries $U_Q$, which are unitaries that satisfy
\begin{equation}
    U_Q^\dagger \gamma_\mu U_Q = \sum_{\nu = 1}^{2n} Q_{\mu\nu} \gamma_\nu,
    \label{matchgate}
\end{equation}
where $\gamma_\mu$ is the $\mu$-th Majorana operator obtained by $\gamma_{2j -1} = a_j + a_j^\dagger$, $\gamma_{2j} = -i(a_j - a_j^\dagger)$. The key part of their results is the $\mathcal{O}(N^4)$ complexity for obtaining the overlap from a single snapshot.
Additionally, if there are $M$ observables, the number of classical shadow samples required to estimate every $\tr(O_i\rho), i = 1, \ldots, M$ to within additive error $\varepsilon$ with probability at least $1-\delta$ is~\cite{wan_matchgate_2023}
\begin{equation}
    N_{\text{samp}} = \mathcal{O}\left(\frac{\log(M/\delta)}{\varepsilon^2}\max_{1 \leq i \leq M} \mathrm{Var} [\hat{o}_i]\right),
\end{equation}
where $\hat{o}_i$ is defined as
\begin{equation}
    \hat{o}_i = \tr \left(O_i \mathcal{M}^{-1} (\hat{U}^
\dagger |\hat{b}\rangle\langle\hat{b}|\hat{U})\right).
\end{equation}
Here $\hat{U}$ is a random variable sampled from the distribution $\mathcal{U}$ and $\mathbb{P}[\ket{\hat{b}} = \ket{b}|\hat{U} = U] = \langle b | U\rho U^\dagger | b\rangle$. It is also proved that the variance bound of $\hat{o}_i$ is $\mathcal{O}(\sqrt{N}\log N)$. 

Combining these facts, the overall complexity for classical processing of the overlap obtained from matchgate shadows scales as $\mathcal{O}(\frac{\log(M/\delta)}{\varepsilon^2}MN^{4.5}\log N)$, and the post-processing for the energy estimates scales as $\mathcal{O}(\frac{\log(M/\delta)}{\varepsilon^2}MN^{8.5}\log N)$~\cite{Huang2024Apr}, as opposed to the exponential scaling using Clifford shadows.
Nevertheless, as Huang \textit{et al.} showed in Ref.~\citenum{Huang2024Apr}, even though the classical post-processing for matchgate shadows is asymptotically efficient, it is still practically computationally intensive. They carried out an experiment and found noise resilience in the overlap ratio obtained by matchgate shadows. However, it would require up to $10^{11}$ CPU hours to perform the classical post-processing of a (54e, 50o) calculation according to the prediction in Ref.~\citenum{Huang2024Apr}. 
Kiser \textit{et al.} also investigated the comparison of different protocols of overlap estimation~\cite{kiser_classical_2023}. 
Assuming the \add{$\mathcal{O}(N)$} number of \add{Monte Carlo} samples for maintaining the fixed additive accuracy for the energy estimator, they found that the total cost for evaluating all overlaps scales as $\mathcal{O}(N^9)$
\add{($\mathcal{O}(N^4)$ per matchgate shadow sample $\times$ $\mathcal{O}(N^4)$ overlap evaluations for local energy $\times$ $\mathcal{O}(N)$ walkers)}
which is almost intractable in real applications~\cite{Huang2024Apr}.

This computational cost in classical post-processing presents a significant challenge to the practical application of QC-AFQMC using classical shadows for overlap estimation, and addressing this issue is a crucial task for future research.  Nonetheless, there is promise in reducing the scaling to \add{$\mathcal O(N^{5})$ with an automatic differentiation} approach \add{by evaluating the derivative of the following form in the local energy~\cite{jiang2024unbiasing}:
\begin{equation} 
    E_1 = \frac{\partial\ln\langle\Psi_{\textrm{T}}|e^{\lambda\sum_{pq}h_{pq} a_p^\dagger a_q}| \psi\rangle}{\partial\lambda}\big|_{\lambda=0}
\end{equation}
and the two body energy
\begin{equation}
    E_2 = \frac{1}{2}\frac{\left\langle\Psi_{\textrm{T}}\left|\sum_{\gamma}\sum_{pqrs}L_{pq}^\gamma L_{rs}^{\gamma} a_p^\dagger a_q a_r^\dagger a_s\right| \psi\right\rangle}{\left\langle\Psi_{\textrm{T}} |\psi\right\rangle}\\
\end{equation}
where 
\begin{align}
    &\langle\Psi_{\textrm{T}}|\sum_{pqrs}L_{pq}^\gamma L_{rs}^{\gamma} a_p^\dagger a_q a_r^\dagger a_s| \psi\rangle \\\nonumber &= \left .\frac{\partial^2 \langle\Psi_{\textrm{T}}|e^{\lambda \sum_{pq}L_{pq}^\gamma a_p^\dagger a_q} | \psi\rangle}{\partial\lambda^2}\right|_{\lambda=0}.\\\nonumber
\end{align}
}

The simulation of classical shadows circuit is exponentially hard for classical computers as the random unitatires maximize the entanglement growth. One could hope to find a suitable application of classical shadows in computational chemistry to demonstrate quantum advantage.

\subsubsection{Hybrid Quantum--Classical FCIQMC}\label{subsubsec:fciqmc}
\begin{figure*}
    \centering
    \includegraphics[width=0.8\textwidth]{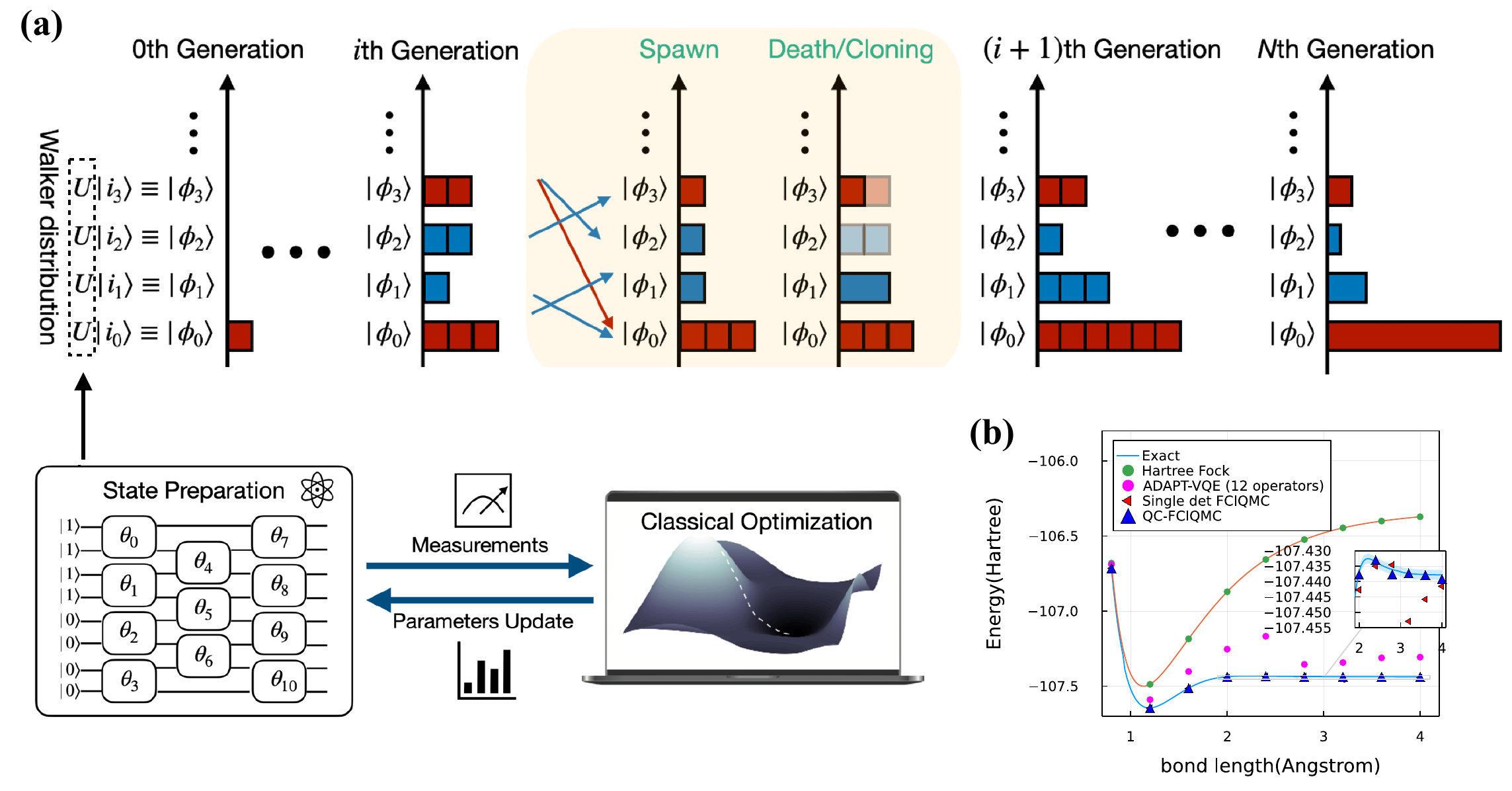}
    \caption{(a) Sketch of the spawning, death/cloning, and annihilation steps in FCIQMC with the determinants transformed with ADAPT-VQE quantum wavefunction. 
    (b) The potential energy surface for the N$_2$ with different methods
under the STO-3G basis set. Figure modified from Ref.~\citenum{zhang2022quantum}.}
    \label{fig:qc_fciqmc}
\end{figure*}
Unlike the polynomially scaling ph-AFQMC method, FCIQMC preserves the exponential scaling of FCI with exact on-average accuracy~\cite{vigor2016understanding}. The exponential scaling sample complexity of FCIQMC is due to the sign problem. 
A simple choice to reduce the sign problem in FCIQMC will be performing a suitable unitary transformation of the Hamiltonian by $\tilde{H} = U^\dagger H U$
to reduce the ``non-stoquaticity"~\cite{hangleiter2020easing}.
One can observe that an exact diagonalization leads to a fully diagonal Hamiltonian matrix free of the sign problem. This diagonalization, however, corresponds to a brute-force FCI calculation and renders any subsequent computations obsolete, as the problem is already solved.

Zhang \textit{et al.}~\cite{zhang2022quantum} prepared the quantum circuit $U$ from an ADAPT-VQE simulation to transform the basis of original Slater determinants, $U|\psi_i\rangle\rightarrow|\tilde{\psi}_i\rangle$, as shown in Fig.~\ref{fig:qc_fciqmc}(a).
This combination was found to significantly reduce the sign problem in FCIQMC, yielding results that surpass those obtained using ADAPT-VQE alone, as shown in Fig.~\ref{fig:qc_fciqmc}(b).

Kanno \textit{et al.}~\cite{kanno2023quantum} also used the solution from VQE with a hybrid tensor network (HTN) ansatz to prepare the trial wavefunction for QC-FCIQMC calculations.
Subsequently, they employed quantum wavefunction exclusively for mixed energy estimations during propagation steps, a technique referred to as quantum-assisted energy evaluation (QAEE) in Ref.\citenum{xu2023quantum}.
Such an implementation is possible because the FCIQMC approach works without applying constraints in the presence of a sign problem.

A critical step in FCIQMC is the calculation of Hamiltonian matrix elements $H_{ij} = \langle \psi_i | \hat{H} | \psi_j \rangle$ to determine the walker spawning probability.
Classical FCIQMC can easily compute the non-zero elements for a given state $|\psi_i\rangle$ using the Slater-Condon rules.
However, calculating $\tilde{H}_{ij}$ in a transformed basis as needed in QC-FCIQMC is not straightforward without prior knowledge of which indices $j$ give non-zero $\tilde{H}_{ji}$. 
As for $\hat{H}=\sum_k h_k \hat{P}_k$ where $h_k$ is the coefficient for Pauli string $\hat{P}_k$, a sampling approach was introduced to measure $\tilde{H}_{ji}$ to determine the spawning probability in QC-FCIQMC on the quantum computer~\cite{zhang2022quantum}:
\begin{equation}
    |\tilde{H}_{j i}|^2=\sum_{k k^{\prime}} h_k h_{k^{\prime}} p_{k k^{\prime}}^i(j)
\end{equation}
with
\begin{equation}
    p_{k k^{\prime}}^i(j)=\operatorname{Re}\left\langle i\left|U^{\dagger} P_k U \Pi_j U^{\dagger} P_{k^{\prime}} U\right| i\right\rangle
\end{equation}
which, for fixed $i, k, k^{\prime}$, can be viewed as the distribution over different $j$s and can be sampled with quantum circuit measurements and post-processing~\cite{zhang2022quantum}.  
The sign of $\tilde{H}_{j i}$ can be further estimated with another circuit. 
Although all the elements of $\tilde{H}_{j i}$ can be precomputed and stored prior to FCIQMC calculations, this method faces scalability challenges as system size increases.
It is important to find an efficient sampling approach for the spawning probability to scale up the capability of QC-FCIQMC beyond classical FCIQMC.

\add{
Finally, we note interesting developments in hybrid quantum-classical projector Monte Carlo methods
that have emerged since the preprint of this work was posted, including contextual AFQMC~\cite{kiser2024contextualsubspaceauxiliaryfieldquantum} 
and the aforementioned QC-FCIQMC with fixed-node approximation~\cite{Blunt2024Oct}.
Blunt et al.~\cite{Blunt2024Oct} proposed 
a QC-FCIQMC approach that employs fixed-node approximation 
and computes overlaps by sampling from Clifford shadows. While 
this avoids the exponential scaling encountered in QC-AFQMC with Clifford shadows, 
comparisons show that QC-AFQMC is more robust to noises and achieves more accurate results with 
the same quality trial wavefunctions. 
Besides, this QC-FCIQMC algorithm is currently limited to active space problems 
and requires a large number of samples even for small active spaces.
While space constraints prevent us from covering these advances comprehensively,
their emergence underscores the vibrant research activity and swift progress 
in this field. However, current evidence suggests caution regarding the prospect 
of outperforming conventional QMC approaches, as achieving chemical accuracy 
requires substantial sampling costs even for small problems.
}







\subsection{Hybrid Quantum--Classical Unitary CCMC}\label{sec:qc-ccmc}

Recently, \add{Filip proposed a Monte Carlo informed approach} to the projective quantum eigensolver (PQE) algorithm~\cite{stair2021pqe}, termed MC-PQE~\cite{filip2024fighting}. 
This method was designed to accurately estimate ground state energies and used the intrinsic noise of measurements to justify an FCIQMC-type approach. 
In this algorithm, the quantum device is used to sample Hilbert space throughout imaginary time propagation.
To encode the UCC ansatz on a quantum computer, the Jordan--Wigner decomposition of fermionic creation and annihilation operators is used, followed by the application of a Pauli gadget~\cite{cowtan2020gadget} to exponentiate the resulting Pauli strings. 
The algorithm evaluates unlinked residuals
\begin{equation} \label{eq:unlinked_residuals}
    r_{\mathbf{i}}^{u} = \langle \Phi_{\mathbf{i}} | \hat{H} - S(\tau) | \Psi(\tau)\rangle
\end{equation}
and overlaps
\begin{equation}
    s_{\mathbf{i}} = \langle \Phi_{\mathbf{i}} | \Psi(\tau)\rangle
\end{equation}
on a quantum device. As introduced in Sec.~\ref{para:fciqmc}, $\mathbf{i}$ runs over all possible excitations, and $|\Phi_{\mathbf{i}}\rangle$ are corresponding determinants. The shift $S(\tau)$ is the same as in Eq.~\eqref{eq:fciqmc_evolution} and is dynamically adjusted to stabilize the total walker number around some target number.
Instead of solving the equations defined in Eq.~\eqref{eq:unlinked_residuals} on a classical computer as proposed by Stair \textit{et al.}~\cite{stair2021pqe}, $r_\mathbf{i}^{u}$ is assumed to be a random variable. 
Each time step, all residuals are evaluated, and populations on determinants $|\Phi_\mathbf{i}\rangle$ are updated according to 
\begin{equation} \label{eq:pqe_qmc_update}
    N_{\mathbf{i}}(\tau + \Delta\tau) = N_{\mathbf{i}}(\tau) - \Delta \tau N_0 r_{\mathbf{i}}^{u},
\end{equation}
where $N_0$ is the initial number of walkers on the reference determinant. To evaluate the projected energy, one can compute $E_{\mathrm{proj}}(\tau) = r_0^{u} + S(\tau) s_0$. Repeating the evolution prescription in Eq.~\eqref{eq:pqe_qmc_update} eventually leads to an estimate of the ground state as predicted by the UCC ansatz.
While this quantum version of the UCCMC algorithm circumvents the expensive sampling of excitation and de-excitation, the encoding of UCC ansätze on quantum devices usually involves deep quantum circuits.
Furthermore, this
does not necessarily alleviate 
the sign problem inherent to CCMC.
As it is in its early exploration, it is currently unclear what advantage this would offer beyond UCC and classical CCMC.

\subsection{Entanglement Forging with Monte Carlo Sampling}
\begin{figure}
    \centering
    \includegraphics[width=0.48\textwidth]{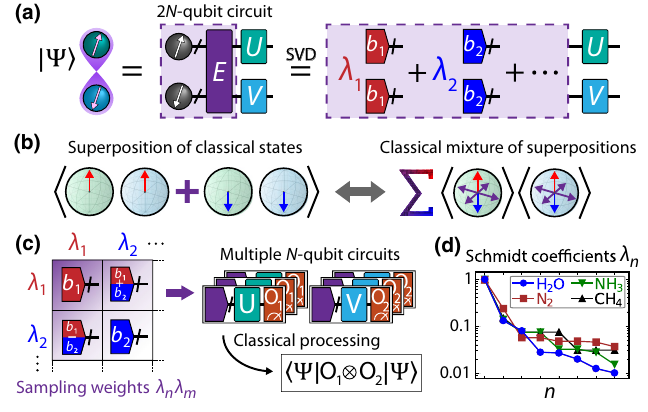}
     \caption{Entanglement forging of a bipartite system. (a) Quantum state $|\Psi\rangle$ obtained by a combination of bitstring states $\ket{b_n}$. (b) Example of a two-qubit state rewritten using superpositions of single-qubit states. (c) Reconstruction of the entangled state $\ket{\Psi}$ from $N$-qubit circuits initialized in bitstrings and pairwise superpositions. (d) The rapid or slow decay of Schmidt coefficients indicates weak or strong entanglement in the ground state of different molecules.
Figure adapted from Ref.~\citenum{PRXQuantum.3.010309}.}
    \label{fig:EF}
\end{figure}

The VQE algorithm, when combined with entanglement forging (EF)~\cite{PRXQuantum.3.010309}, represents a promising strategy that effectively simulates a $2N$-qubit system using quantum circuits with only $N$-qubit after classical postprocessing.
The Schmidt decomposition of the wavefunction is the starting point of EF for a $N+N$ qubits bipartite quantum system $\hat{H}_A\otimes\hat{H}_B$.
\begin{equation}
    |\psi\rangle=U \otimes V \sum_{n=1}^{2^N} \lambda_n|\sigma_n\rangle_A|\sigma_n\rangle_B,
\end{equation}
where $U$ and $V$ are unitary quantum circuits. 
Fig.~\ref{fig:EF}(a) graphically shows the decomposition of $2N$-qubit system.
The VQE optimization contains the parameters in the $U$ and $V$ circuits, as well as the Schmidt coefficients.
In this approach, evaluating expectation values requires the products of Schmidt coefficients, $\lambda_m\lambda_n$ (for $1 \leq m, n \leq 2^N$). We note that the number of these combinations grows exponentially with the system size.

With EF, the energy evaluation involves the expectation value of operators $O_A\otimes O_B$ with Pauli strings $O_A, O_B \in\{I, X, Y, Z\}^{\otimes N}$,
\begin{equation}
\begin{aligned}
&\langle O_A\otimes O_B\rangle= \sum_{n=1}^{2^N}(\lambda_n^2\langle \sigma_n|\tilde{O}_A| \sigma_n\rangle\langle \sigma_n|\tilde{O}_B| \sigma_n\rangle+\sum_{m=1}^{n-1} \lambda_n \lambda_m\times  \\
& \sum_{p \in \mathbb{Z}_4}(-1)^p\langle\phi_{\sigma_n \sigma_m}^p|U^{\dagger} O_A U| \phi_{\sigma_m \sigma_m}^p\rangle\langle\phi_{\sigma_n \sigma_m}^p|V^{\dagger} O_B V| \phi_{\sigma_n \sigma_m}^p\rangle)
\end{aligned}\label{eq:EF_expectation}   
\end{equation}
where $\left|\phi_{x y}^p\right\rangle=\left(|x\rangle+\mathrm{i}^p|y\rangle\right) / \sqrt{2}$ with $p\in\mathbb{Z}_4=\{0,1,2,3\}$. Fig.~\ref{fig:EF}(b) shows the $|\phi_{01}^p\rangle$ for 2 qubit case with minimal basis.
Eq.~\eqref{eq:EF_expectation} can be decomposed into measuring expectation values with independent quantum circuits of $N$ qubits.
\begin{equation}
    \left\langle\psi\left|O_A \otimes O_B\right| \psi\right\rangle=\sum_{a} \mu_a \operatorname{Tr}\left(O_A \rho_a\right) \operatorname{Tr}\left(O_B \rho_a\right)\label{eq:ER_decomp}
\end{equation}
where 
$\rho_a$ is $N-$qubit pure state quantum circuit, and $\mu_a$ is proportional to the products of two independent Schmidt coefficients. We further define
\begin{equation}
    O_{A/B}=V_{A/B}^{\dagger}\left(\sum_{x \in\{0,1\}^N} O_{A/B}(x)|x\rangle\langle x|\right) V_{A/B},
\end{equation}
where $V_{A/B}$ includes only Clifford gates that map the Pauli strings $X$ and $Y$ appearing in $O_{A/B}$ to Pauli string $Z$ and $O_{A/B}(x)$ will take the values 1 or -1 depending on the parity of the bitstring $x$.
This converts the problem of measuring the $2N-$qubit observables (Eq.~\eqref{eq:EF_expectation}) to measuring the observables on a series of $N$ qubits system. Namely, each experiment prepares a state $V_{A/B}\rho_aV_{A/B}^\dagger$ and measures it in a computational basis.

The computation of Eq.~\eqref{eq:ER_decomp} appears to be not scalable, which involves exponentially many terms
An importance sampling approach was proposed where an unbiased estimator for Eq.~\eqref{eq:ER_decomp} is obtained by $M$ samples~\cite{PRXQuantum.3.010309},
\begin{equation}
    f=\frac{\|\mu\|_1}{M} \sum_{j=1}^M \operatorname{sgn}\left(\mu_{a_j}\right) O_A\left(x^j\right) O_B\left(y^j\right)
\end{equation}
where $\|\mu\|_1 \equiv \sum_{a=1}^{\ell}\left|\mu_a\right|$.
$O_{A}(x^i)$ and $O_{B}(x^j)$ are measured by preparing the state $V_A \rho_{a_i} V_A^{\dagger}$ and $V_B \rho_{a_j} V_B^{\dagger}$ and measuring in the computational basis.
The sampling probability is given by
\begin{equation}
    \pi_a = \left|\mu_a\right| /\|\mu\|_1,
\end{equation}
which is proportional to the products of Schmidt coefficients $\lambda_m\lambda_n$.
The measurements of $2N$-qubit quantum circuit are thereby mapped to a series of $N-$qubit quantum circuit measurements by importance sampling with probability proportional to $\lambda_n\lambda_m$, as shown in Fig.~\ref{fig:EF}(c).
It was proven that $S$ measurement experiments ($S=2M$ for measuring both $V_A \rho_{a_i} V_A^{\dagger}$ and $V_B \rho_{a_i} V_B^{\dagger}$) are needed for a target precision $\epsilon$:
\begin{equation}
    S \sim\left(\frac{1}{\epsilon} \sum_{n, m}\left|\lambda_n \lambda_m\right|\right)^2=\frac{\|\vec{\lambda}\|_1^4}{\epsilon^2}
\end{equation}
Since the one-norm decreases toward 1 in the limit of weak entanglement, the overhead of entanglement forging is smaller for simulations
of states divisible into weakly entangled fragments (\textit{i.e.}, halves). As an example, Fig.~\ref{fig:EF}(d) shows the decay of leading Schmidt coefficients for molecular ground states with bipartition between
spin-up and spin-down particles.
Apart from this Monte Carlo approach, other approaches using generative neural networks~\cite{huembeli2022entanglement,de2024hybrid} were also proposed to address the sampling process.
The precise assumptions and conditions in which this method could demonstrate a practical quantum advantage in computational chemistry seem unclear as it is still in its early development stage.
\add{Finally, we also note a framework of hybrid tensor networks~\cite{PhysRevLett.127.040501} that combines 
measurable quantum states with classically contractable tensors to handle multipartite systems
that goes beyond bipartite forging.}

\subsection{Quantum Selected Configuration Interaction}\label{sec:qsci}
Recently, Kanno \textit{et al.}~\cite{Kanno_QuantumSelectedConfigurationInteraction2023} proposed a quantum version of Monte-Carlo configuration interaction~\cite{Greer_EstimatingFullConfiguration1995, Greer_MonteCarloConfiguration1998} by employing a quantum device to select determinants that are used to construct the effective Hamiltonian. 
In this quantum selected configuration interaction (QSCI) algorithm, an input state $\ket{\psi_{in}}$ and a set of $n$-qubit string basis states $\ket{i}$  are prepared on a quantum computer by VQE or other algorithms. 
One then iteratively projects $\ket{\psi_{in}}$ onto the computational basis for a total of $N_{\text{shot}}$ times to obtain
\begin{align}
    \ket{\psi} = \sum_{i=0}^{2^n-1} \alpha_i \ket{i}.
\end{align}
Then, one chooses the $R$ most frequent configurations to define the CI subspace $\mathcal S$. 
Finally, the diagonalization of the Hamiltonian in the subspace $\mathcal S$ is carried out on a classical machine, and the ground state energy is obtained. One could straightforwardly extend this approach to calculating excited states by repeating this procedure on multiple input states. 

An important technical detail in QSCI is how to prepare the input state on a quantum computer. This could be done using a quantum circuit optimized using VQE, such that QSCI can essentially be viewed as a post-processing method to refine VQE results. Related to this approach, the adaptive derivative-assembled pseudo-Trotter ansatz (ADAPT)~\cite{Grimsley_AdaptiveVariationalAlgorithm2019} was also used in a method termed ADAPT-QSCI~\cite{Nakagawa_ADAPTQSCIAdaptiveConstruction2023}. 
This algorithm provides a way of systematically preparing viable input states for QSCI with reduced gate count. 

Recently, Robledo-Moreno \textit{et al.} reported a similar hybrid quantum--classical selected CI approach with additional error mitigation techniques~\cite{Robledo-Moreno_ChemistryExactSolutions2024a}. In this algorithm, a truncated version of the local unitary cluster Jastrow (LUCJ) ansatz $\ket{\Psi}$ \cite{Motta_BridgingPhysicalIntuition2023,huggins2022unbiasing}, 
\begin{align}
    \ket{\Psi} = \prod_{\mu = 1}^{L} \mathrm{e}^{\hat{K}_\mu}\mathrm{e}^{\mathrm{i}\hat{J}_\mu}\mathrm{e}^{-\hat{K}_\mu}\ket{\Psi_\mathrm{RHF}},
\end{align}
is used to generate a set of configurations $\{\chi\}$.
Here, $\hat{K}_\mu$ is a generic one-body operator, and $\hat{J}_\mu$ is a density-density operator. This set of configurations is then subject to a classical post-processing routine where an iterative configuration recovery loop is run until convergence, and the approximated ground state energy is obtained. While they showcased that this algorithm can be used for calculating challenging chemical systems such as iron-sulfur clusters using up to 77 qubits and 6400 classical nodes, the resulting energy shows an error orders of maginitude greater than chemical accuracy (1 kcal/mol.)

QSCI could be useful for problems where sampling from a quantum state is classically hard or for reducing the subspace size needed in CI to reach a certain level of accuracy. 
It is worth noting here that since the final diagonalization step is carried out classically, the subspace size $R$ needs to be within the limits of classical computation.
Furthermore, its advantage over classical SCI methods such as CIPSI and HCI is still unclear.
\add{Lastly, this approach samples bitstrings from a quantum state and produces a variational energy using the space defined by those bitstrings. In the noiseless set-up, one may often find this approach producing energies far less accurate than the original quantum state, especially towards large systems.}

\subsection{Quantum Cooling via Lindbladians}\label{sec:gs_full_quantum}
A heuristic approach for the ground state problems is to implement Lindbladian time-evolution to simulate the cooling of an open quantum system.
For a set of jump operators $L^a$, and "coherent" Hamiltonian term $H$, the quantum state is evolved by the differential equation
\begin{align}
\frac{\mathrm{d}}{\mathrm{d}t}\rho = \mathcal{L}[\rho] = -\mathrm{i}[H,\rho] + \sum_a L^a \rho (L^a)^{\dagger} - \frac{1}{2}\{(L^a)^{\dagger}L^a, \rho\}.
\end{align}
In Ref.~\cite{ding2023singleancilla} a general prescription is given, for implementing jump operators $L^a$, such that the ground-state is the unique steady-state of $\mathcal{L}$.

\add{The jump operators $L^a$ are constructed from linear combinations of operators in the Heisenberg picture.
    Specifically, let $A^a$ denote the proposed jump operators and $A^a(s)$ represent their time evolution in the Heisenberg picture,}
\begin{align}
A^a(s) = \mathrm{e}^{\mathrm{i} H s} A^a \mathrm{e}^{-\mathrm{i} H s}.
\end{align}
Then, $L^a$ is given by,
\begin{align}{\label{eq: jump_cooling}}
L^a = \int_{-\infty}^{\infty} \mathrm{d}s \: f(s) A(s)\end{align}
for a filter-function $f(s)$, defined by
$$f(s) = \frac{1}{2\pi} \int \mathrm{d}\omega \hat{f}(\omega) \mathrm{e}^{-\mathrm{i} \omega s},$$
where $\hat{f}(\omega)$ is written in frequency space. The key intuition is that $\hat{f}(\omega)$ should be chosen to only allow cooling transitions while preventing heating transitions. 
This resembles what one obtains from ITE.
Unlike ITE, this algorithm does not need to assume the non-zero overlap between the initial state and the exact ground state.
This also offers a clear advantage over QPE for problems where preparing a state with good overlap is challenging.
As such, it should be real, non-negative, and satisfy
$$\hat{f}(\omega) = 0$$
for $\omega \geq 0$. To ensure $L^a$ can be implemented efficiently, it is crucial to choose $f(s)$ such that it is smooth and decays rapidly with the magnitude of $s$, ensuring the filter function can be approximated by a discretized and truncated version of the integral in Eq.~\eqref{eq: jump_cooling}.
Then, the linear combination of unitary techniques~\cite{Childs2012Feb}, which use ancilla qubits, can be applied to engineer $\mathcal{L}$. 

\begin{figure*}
    \centering
    \includegraphics[width=0.8\linewidth]{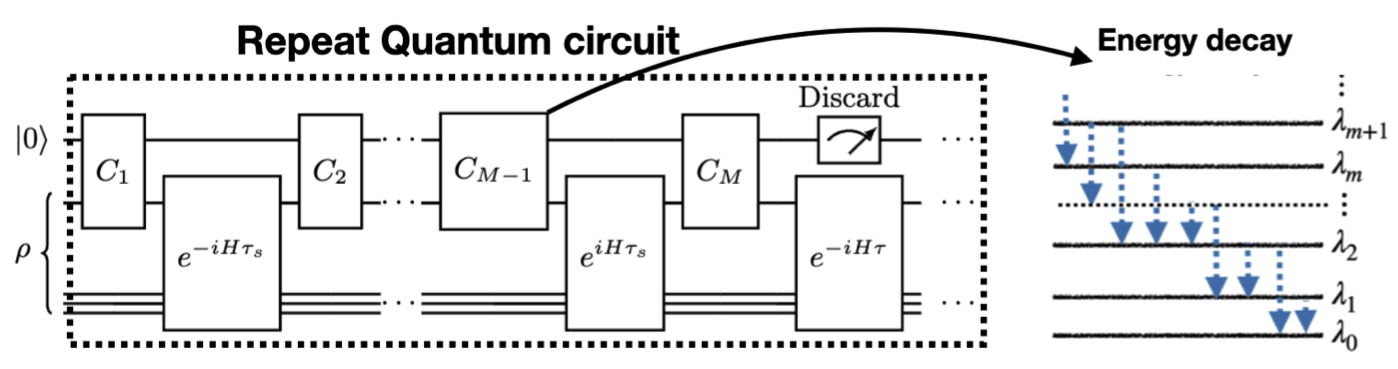}
    \caption{Single-ancilla scheme for quantum cooling algorithm from Ref.~\citenum{ding2023singleancilla}. A linear combination of Heisenberg-evolved operators is engineered by alternating controlled perturbations $C_i$ with Hamiltonian time-evolution $\mathrm{e}^{-\mathrm{i} H t}$. The ancilla is periodically discarded and reset, realizing an effective jump operator $L^a$ on the system, which generates energy-lowering transitions between system eigenstates. }
    \label{fig:qcool}
\end{figure*}

The simplest architecture requires only a single ancilla coupled to the system (see Fig.~\ref{fig:qcool}) and may be ideal for near-term devices. 
By alternating a controlled $A^a$ operation with Hamiltonian time-evolution $e^{-i H \tau_s}$, the circuit implements a Trotterized approximation to the Linbladian evolution with jump operator $L^a$.
Ultimately, the performance of a cooling scheme is mainly determined by the mixing time of the Linbladian, which depends on the choice of filter function $f(s)$ and proposed jump operators $A$. Determining fast-mixing schemes from problems of interest remains an open problem.
\add{
Applications to \textit{ab initio} electronic structure calculations have emerged in recent studies~\cite{Li2024Nov}, with their effectiveness demonstrated through classical Monte Carlo sampling techniques.
}

\add{ 
\subsection{Concluding remarks} 
To conclude this section, we present Table~\ref{tab:comparison-methods}, 
    which categorizes the quantum algorithms we introduced in this section 
    into three main types. 
    First, quantum-enhanced classical algorithms (VMC and PMCs) 
    leverage quantum computers to prepare trial wavefunctions - 
    a central object in quantum Monte Carlo.
    By preparing these states on quantum hardware, these hybrid approaches can potentially overcome 
    classical bottlenecks in wavefunction preparation, accelerate convergence, 
    and enhance accuracy of Monte Carlo sampling.
    Second, classical-enhanced quantum algorithms (primarily VQE) use 
    classical post-processing (MCMC, CCMC, selected CI) to 
    mitigate hardware limitations on noisy quantum 
    devices by reducing circuit depth, measurement requirements, and errors. 
    Third, we briefly mention the full quantum algorithm designed for fault-tolerant quantum computers. 
    For each approach, Table~\ref{tab:comparison-methods} details the quantum and classical components, 
    their synergistic benefits, and remaining technical challenges. 
    }
\begin{table*}[ht]
    \centering
    \scriptsize  
    \renewcommand{\arraystretch}{1.2} 
\add{  
    \begin{tabular}{>{\centering\arraybackslash}m{0.8cm} >{\centering\arraybackslash}m{1.6cm} 
        >{\centering\arraybackslash}m{2.6cm} >{\centering\arraybackslash}m{0.6cm} >{\centering\arraybackslash}m{4.2cm} >{\centering\arraybackslash}m{4.2cm}}  
        \toprule  
        Ref & Classical Algorithm & Quantum Role & Offline QC & Results & Challenges \\    
        \midrule
        \cite{montanaro2023accelerating} & VMC & initial samples & \ding{51} & faster convergence to target distribution. &  requires high-quality VQE samples.\\   
        \cite{huggins2022unbiasing} & AFQMC & trial wavefunction & \ding{51} & noise-resilient and unbiased QMC energy. & high scaling postprocessing; Notes$^{(a,b)}$.\\  
        \cite{mazzola2022exponential} & GFMC & trial wavefunction & \ding{51} & exponentially many samples. & noise sensitivity; Notes$^{(b)}$. \\  
        \cite{zhang2022quantum} & FCIQMC & basis transformation & \ding{55} & reduced sign problem and enhanced results on VQE. & requires classical-quantum communication at each step; Notes$^{(b)}$.\\  
        \cite{Blunt2024Oct} & FCIQMC & nodal structure & \ding{51} & polynomial scaling, but less robust than QC-AFQMC. & active space only; fixed-node approx.; requires large $\#$ of classical shadows; Notes$^{(b)}$.\\     
        \cmidrule(lr){1-6}
        \cite{filip2024fighting} & CCMC & sample variables for each QMC step & \ding{55} & fewer samples than VQE. & deep UCC circuits; sign problem; large fluctuations.\\    
        \cite{PRXQuantum.3.010309} & MCMC & entanglement forging & \ding{55} & doubled VQE simulation size. & sampling overhead; require subsystem symmetries. \\  
        \cite{Robledo-Moreno_ChemistryExactSolutions2024a} & Selected CI & CI samples & \ding{51} & improves solution accuracy & limited by ED postprocessing size; unclear advantage over classical SCI.\\
        \cite{mazzola2019nonunitary,zhang2022variational} & VMC & energy measurement & \ding{55} & Reduced circuit depth & expensive postprocessing; unclear noise resilience. \\   
        \cmidrule(lr){1-6} 
        \cite{ding2023singleancilla} & - & Lindbladian simulation & \ding{55} & avoids initial state overlap constraints & mixing time limited; fault-tolerant quantum device. \\
        \bottomrule
    \end{tabular}
    \caption{Comparison of quantum algorithms for ground state calculations, detailing their key characteristics, requirements, limitations, and demonstrated capabilities. The term "Offline QC" refers to scenarios where tasks are fully segregated between classical and quantum computers.
    a: QC-AFQMC initially required exponential processing time~\cite{huggins2022unbiasing}; this was improved to polynomial time in~\cite{wan_matchgate_2023}.
    b: The intrinsic issue of projector Monte Carlo that the overlap diminishes exponentially with the increase of system size.}
    \label{tab:comparison-methods} 
}  
\end{table*}

\section{Quantum Algorithms for Thermal State Properties}\label{sec:quantum_for_thermal}
\add{In this section, we discuss recent progress in quantum algorithms 
for finite-temperature systems. 
While quantum algorithms for ground states have been extensively studied, as discussed in the 
previous section, research on finite-temperature quantum algorithms remains relatively less explored.
The preparation of thermal states poses significant challenges, particularly at low temperatures.
Nevertheless, recent years have seen important progress in this area. 
Notable quantum algorithms~\cite{somma2008quantum,PhysRevLett.108.110502,PhysRevLett.105.170405,temme2011quantum,PhysRevLett.103.220502,Chowdhury2017Feb,
wocjan2008,Lemieux2020,yung2012quantum,Cleve2016Dec,PhysRevLett.103.220502,Rall2023Oct,Bergamaschi2024Apr,chen2023quantum,ding2024efficientthermal,Rajakumar2024Aug,PhysRevA.102.022622,
motta2020determining,PRXQuantum.2.010317}
have demonstrated polynomial quantum speedups with respect to various parameters.
In the following subsections, 
we will introduce several key approaches for preparing thermal states or
computing thermal properties on quantum computers. 
These include quantum Metropolis algorithms, 
full quantum Gibbs sampling, quantum stochastic series expansion, 
quantum imaginary time evolution, and hybrid quantum-classical Monte Carlo methods.
While we cannot cover all existing methods in detail,
we focus on representative approaches and highlight differences.
For each method, we will discuss its theoretical foundations, 
potential quantum advantages, and progress toward experimental implementations.
}
\subsection{Quantum Metropolis algorithms}
\subsubsection{Quantum Analog of Metropolis Algorithm}\label{sec:qama}
Temme \textit{et al}.~\cite{temme2011quantum} first proposed a quantum analog of the classical Metropolis sampling algorithm discussed in Section \ref{sec:classical_mc}. To generalize Metropolis sampling to the quantum setting, one needs to consider the following two aspects: performing the rejection step and proving that the fixed point of the random walk is indeed the Gibbs state. For the former, the no-cloning theorem prevents the classical solution of storing a copy of the original state and reverting back to it in this case. This issue can be handled using the Mariott--Watrous algorithm to rewind the state to the original~\cite{marriott_watrous}. For the latter, the fixed point was proven in an idealized case, where perfect phase estimation can be performed. When energy measurements are only given to a finite resolution, phase estimation does not have the same performance guarantees. A variant using a boosted shift-invariant phase estimation had certain performance guarantees, although the shift-invariant boosting has recently been proved impossible~\cite{chen2023quantum}.
More specifically, it was proved that there is no continuous family of “boosted shiftinvariant in place phase estimation” unitaries. Although this paper marks an important advancement in the field and served as an inspiration to many works that followed, such as those discussed in \cref{sec:q_enhanced_mcmc} and \cref{sec:q2ma}, the algorithm's practical feasibility remains an open question.

\subsubsection{Quantum Enhancement of Classical Markov Chain Monte Carlo}\label{sec:q_enhanced_mcmc}
In \cref{sec:classical_mc}, we briefly discussed the computational challenges associated with classical MCMC and identified aspects where quantum computing could potentially accelerate these processes. 
In this section, we review the quantum enhancements of classical MCMC.
One of the first ways of quantum enhancement was proposed by Szegedy in 2004 to accelerate the convergence of MCMC is incorporating a quantum walk instead of using classical state propagation~\cite{szegedy2004quantum}. 
Then, in 2007, Richter reported the potential for a quantum advantage using quantum MCMC with decoherent quantum walks by demonstrating potential quadratic speedups in mixing times~\cite{richter2007}. In 2008, Somma and coworkers could apply quantum walks in MCMC to solve combinatorial optimization problems with quadratic speedup~\cite{somma2008quantum}. 
The same year, Wocjan and Abeyesinghe proposed a new quantum sampling method to prepare quantum states~\cite{wocjan2008}. 
While these works all offer quadratic speedups, more recent work in 2020 by Lemieux \textit{et al.} presented an efficient implementation of Szegedy's quantum walk, which could offer better than quadratic speedups~\cite{Lemieux2020}. 

A different instance in which quantum computers can accelerate the convergence of MCMC is finding the ground state (\textit{i.e.}, zero temperature) spin configuration in the Ising model. For the Ising model, each spin configuration $\textbf{s}\in\{-1,+1\}^n$ has an associated energy (Eq.~\eqref{eq:IsingModel}).
The probability of each spin configuration can also be described using Boltzmann statistics. Sampling from this distribution is widely used, and it is often a computational bottleneck in various algorithms, especially in the low-temperature limit, because sampling from the Boltzmann distribution then approximates the challenge of finding the ground state or low energy configurations.

Recently, there have been instances of quantum algorithms for sampling and estimating partition functions demonstrating quadratic speedup in convergence times~\cite{harrow2020adaptive,Arunachalam2022}. In a work by scientists at IBM in 2023, the classical proposal step is replaced with a quantum proposal step~\cite{layden2023quantum}. The spin of each electron for any given state can be represented by a single qubit. A unitary $U$ is applied to state $\textbf{s}$ on the quantum computer under the constraint \begin{equation}
    |\langle \textbf{s}'|U|\textbf{s}\rangle| = |\langle\textbf{s}|U|\textbf{s}'\rangle|,
\end{equation}
and each qubit is measured. Then, the Metropolis--Hastings acceptance probability is used to accept or reject this proposal. The choice of unitary here guarantees that $\frac{Q(\textbf{s}|\textbf{s}')}{Q(\textbf{s}'|\textbf{s})} = 1$, and for a Boltzmann distribution, computation of $\frac{\pi(\textbf{s}')}{\pi(\textbf{s})}$ is straightforward. This Markov chain provably converges to the Boltzmann distribution but is hard to mimic classically. This is the source of potential quantum advantage. The unitary $U$ is the time-evolution operator corresponding to a Hamiltonian constructed as a sum of the classical Ising model and a term enabling quantum transitions. 

The convergence of this algorithm was probed experimentally, and they observed approximately cubic-to-quartic improvements in $\delta$ as a function of the number of qubits, where $\delta$ is defined in section \ref{sec:classical_mc}. 
While there is no precise mathematical bound on the improvement that this algorithm provides over its classical counterparts, the experimental demonstration seems promising. Because only portions of this algorithm are performed on a quantum computer with classical intermediate steps, this algorithm has lower circuit depths than purely quantum MCMC algorithms like those discussed in \cref{sec:q2ma}. This makes its implementation appealing in the near future. Finding such a quantum speed-up for computational chemistry problems using the quantum MCMC algorithm remains an open question.

\subsubsection{Quantum--Quantum Metropolis Algorithm}\label{sec:q2ma}
Yung and Aspuru-Guzik \cite{yung2012quantum} reported a Quantum--Quantum Metropolis Algorithm (Q2MA), which is a quantum generalization of the classical Metropolis sampling algorithm~\cite{metropolis1953equation}, and can be used for preparing thermal states of quantum systems with quantum speedup. The Q2MA method builds on the Markov chain quantization method by Szegedy~\cite{szegedy2004quantum}, which was formulated for classical Hamiltonians. However, a naive implementation of Szegedy's quantum walk for quantum Hamiltonians is hindered by the no-cloning theorem. \cite{wootters1982single} 

Yung and Aspuru-Guzik were able to circumvent this problem by preparing states of the form $\ket{i} \equiv \ket{\psi_i}\ket{\tilde{\psi}_i}\ket{0}$, where $\ket{\psi_i}$ is the eigenstate of the quantum Hamiltonian $H$ and $\ket{\tilde{\psi}_i}$ is its time-reversal counterpart. Hence, both states are eigenstates of $H$ with the same eigenvalue $E_i$. Leveraging this property, a generalized Szegedy operator $W$ is constructed and used to implement the projective measurement through quantum phase estimation. This routine is termed quantum simulated annealing (QSA)~\cite{somma2008quantum}. Finally, the thermal state can be prepared by applying QSA on an initial maximally entangled state to yield the coherent encoding of the thermal state:
\begin{align}
    \ket{\alpha_0} = \sum_{i=0}^{N-1} \sqrt{\mathrm{e}^{-\beta E_i} / \mathcal{Z}} \ket{i}
\end{align}
This formalism also falls within the thermofield double framework, as discussed in \cref{sec:classical_metts}.
After tracing out the ancilla qubits, this form is equivalent to the thermal state $\rho_{th} =(1/\mathcal{Z})\sum_i\mathrm{e}^{-\beta E_i}\ket{\psi_i}\bra{\psi_i}$. Here, $\mathcal{Z} = \text{Tr}(\mathrm{e}^{-\beta H})$ is the partition function.

The quadratic speedup of the Q2MA can be traced back to the $O(1/\sqrt{\delta})$ scaling when applying $W$ for the projective measurement in QSA, compared to the $O(1/\delta)$ scaling for classical Markov chain algorithms~\cite{aldous1982some}. Here, $\delta$ is the gap of the transition matrix in the Markov chain, as we mentioned in \cref{sec:classical_mc}. This quadratic speedup is the main distinction between the Q2MA method and the quantum Metropolis sampling method mentioned in \cref{sec:qama}. However, some limitations of this method include that (1) the algorithm only works when the spectrum of $H$ is non-degenerate~\cite{wocjan2023szegedy} and (2) perfect quantum phase estimation is needed~\cite{chen2023quantum}.

\subsection{Full Quantum Gibbs Sampler}\label{sec:ft_full_quantum}
A series of recent works have developed systematic techniques for preparing Gibbs states 
on quantum computers~\cite{Cleve2016Dec,PhysRevLett.103.220502,Rall2023Oct,Bergamaschi2024Apr,chen2023quantum,ding2024efficientthermal,Rajakumar2024Aug,PhysRevA.102.022622}. For a Hamiltonian $H$, which is in general composed of non-commuting terms, the Gibbs state is defined by
$$\rho_{\beta} = \mathrm{e}^{- \beta H} / \: \mathrm{Tr}[\mathrm{e}^{- \beta H}].$$ 
We introduce one approach that prepares the Gibbs state by engineering a Lindbladian, $\mathcal{L}_{\beta}$ which has $\rho_{\beta}$ as the unique stationary state,
\begin{align}{\label{eq: linbladian_fixed_point}}
\mathrm{e}^{\mathcal{L}_{\beta} t}[\rho_{\beta}] = \rho_{\beta}.
\end{align}
In nature, Lindbladians naturally occur in the dynamics of a reduced system with weak interaction between a small system and a large Markovian bath at inverse temperature $\beta$.
We briefly discussed this in the context of classical Markov chain MC in \cref{sec:lindblad_classical}
and the quantum version for ground state calculation in \cref{sec:gs_full_quantum}.
The quantum algorithm aims to use ancilla qubits to mimic the action of the bath. However, since quantum resources are limited, it is desirable to keep the ancilla overhead as small as possible.

It turns out that engineering a Lindbladian to satisfy Eq.~\eqref{eq: linbladian_fixed_point} is challenging. As in the design of classical Monte Carlo algorithms, if care is not taken, then the quantum analog of detailed balance will not be achieved. An explicit Lindbladian which satisfies quantum detailed balance for the Gibbs state can be constructed from the Davies' generators~\cite{Davies1974Jun,*Davies1976Jun,*Rivas2011Apr,*Mozgunov2020Feb}.
Given a Hamiltonian $H$, with spectrum $E_i, \vert \psi_i \rangle$, a set of jump operators $\{A^{a}\}_{a \in A}$, and Bohr frequencies $B = \{\nu | \nu = E_i - E_j\}$, the Davies generators are
\begin{equation}
\mathcal{L}_{\mathrm{Davies}}(\rho) = \sum_{a \in A} \sum_{v \in B} \gamma(v) \left(A_v^{a} \rho (A_v^a)^{\dagger} - \frac{1}{2}\{(A_v^a)^{\dagger A_v^a}, \rho \} \right),
\end{equation}
where 
\begin{equation}
A_v^{a} \propto \int_{-\infty}^{\infty} \mathrm{d}t \: \mathrm{e}^{-\mathrm{i} \nu t} \mathrm{e}^{\mathrm{i} H t} A^{a} \mathrm{e}^{-\mathrm{i} H t}.
\end{equation}
Intuitively, $A^{a}_v$ generates transitions between pairs of eigenstates, separated exactly by energy $\nu$. By considering an input state diagonal in the energy basis, it becomes clear that $\mathcal{L}_{\mathrm{Davies}}$ generates a classical Markov chain on the space of energy eigenstates. The appropriate detailed balance condition is satisfied if the function $\gamma(\omega)$ satisfies $\gamma(\omega) = \mathrm{e}^{- \omega \beta} \gamma(-\omega)$. This ensures that heating transitions are penalized relative to cooling transitions, with the ratio being determined by the target inverse temperature.

However, applying $A^{a}_{\nu}$ in a generic quantum system is challenging due to the energy-time uncertainty principle. Conceptually, the Fourier transform picks out transitions between $\omega, \omega'$ that change the energy by exactly $\nu$. As such, to approximate the Fourier transform by finite-time evolution, the evolution time must scale inversely with energy spacings. Such spacings are, in general, exponentially small in system size.
Instead, one may hope to use shorter-time approximations to the Davies' generators. However, typically, these shorter time approximations actually break the detailed balance condition, meaning that they do not have the Gibbs state as their stationary state. Nevertheless, one may still hope that these approximations prepare approximate Gibbs states.

This question was analyzed in Ref.~\cite{chen2023quantum}.
As in the ground-state cooling case, a filter function is introduced to regularize the integral.
\begin{align}
    A^a_{\nu} = \int_{-\infty}^{\infty} \mathrm{d}t f(t) \mathrm{e}^{-\mathrm{i} \nu t} \mathrm{e}^{\mathrm{i} H t} A^{a} \mathrm{e}^{-\mathrm{i} H t}
\end{align}
Then, they bound the distance between the stationary state $\rho_{\mathrm{ss}}$ of finite-time approximations to the Davies' Lindbladian and the target Gibbs state.
For a Gaussian filter function, they find that the approximation error scales as
\begin{equation}
||\rho_{\mathrm{ss}}(\mathcal{L}_{\beta}) - \rho_{\beta}||_1 = \tilde{O}\left( \frac{\beta}{\sigma_t} t_{\mathrm{mix}}(\mathcal{L}_{\beta})\right).
\end{equation}
Therefore, to achieve a target error rate $\epsilon$, the typical width of the Gaussian should scale as $\sigma_t \sim \beta t_{\mathrm{mix}}/\epsilon$. This sets the amount of Hamiltonian simulation needed for each Lindbladian time step. 
The overall scaling to prepare the Gibbs state using this approach is, therefore,
$\tilde{O}\left(\frac{\beta t_{\mathrm{mix}}}{\epsilon} \cdot t_{\mathrm{mix}}\right).$

A subsequent breakthrough was achieved in Ref.~\cite{chen2023efficient}, which showed how to construct a Lindbladian that is efficiently implementable and has the quantum Gibbs state as its steady state. The key insight was to realize that additional coherent evolution could be used to counteract some of the unwanted contributions coming from the dissipative part and restore detailed balance.
Since the detailed balance is exact, this reduces the need for long-time evolution in the implementation of a single Lindbladian time-step. As such, the resources per time-step now scale as $\tilde{O}(\beta)$.
Ref.~\cite{ding2024efficientthermal} proposed a further generalization and improvement of the algorithm, by constructing an exact Gibbs sampler using a finite number of jump operators, simplifying implementation.
Since the Linbladians used are constructed from linear combinations of Heisenberg-evolved operators, they can be implemented on a quantum processor using a linear combination of unitary techniques~\cite{Childs2012Feb}.
Finding a quantum advantage using this and other related algorithms is an active research question~\cite{Bergamaschi2024Apr}.

\subsection{Quantum Stochastic Series Expansion} \label{sec:quantum_sse}
\begin{figure*}[htbp]
    \centering
    \includegraphics[width=0.98\linewidth]{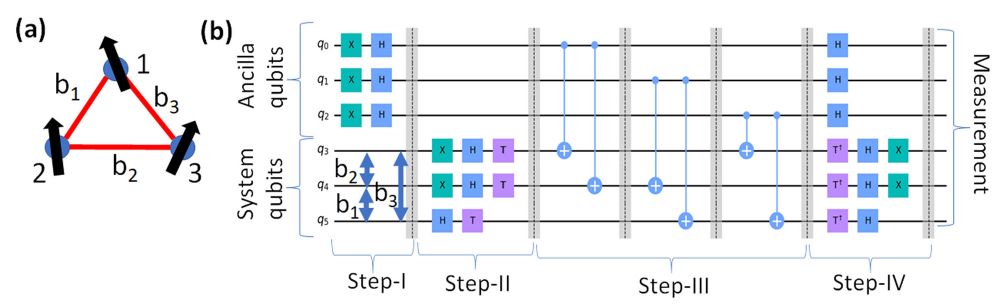}
    \caption{(a) Periodic 3-site Ising model.  (b) Circuit realizing the corresponding quantum SSE algorithm. The circuit is partitioned into $N = 3$ system qubits and $n = 3$ ancilla qubits. In step I, ancilla qubits are prepared in states $|-_{q_i}\rangle$ via the application of Pauli X and Hadamard gates. In step II the system qubits are prepared in a sampled state $|a\rangle_A$. Step III consists of the sequential application of controlled unitaries defined in Eq.~\eqref{eq:sse_unitary}. Finally, in step IV, qubits are rotated back and measured. Adapted from Ref.~\citenum{tan2022sign}.}
    \label{fig:sse_circuit}
\end{figure*}
Similar to classical SSE (Sec.~\ref{sec:sse}), quantum SSE~\cite{tan2022sign} evaluates a thermal expectation value of observable $\hat{O}$,
\begin{equation}\label{eq:thermal_avg_sse}
    \langle \hat{O}\rangle_{\beta} = \frac{1}{Z} \sum_{n,a,\{b_i\}} \frac{\beta^n}{n!} \langle a | \hat{H}_{b_n}\cdots \hat{H}_{b_1}|a\rangle \langle a | \hat{O} | a\rangle,
\end{equation}
where 
the tuple of the perturbation order, state, and index $\{(n, a, \{b_i\})\}$, as detailed in \cref{sec:sse}, is sampled.
On quantum computers, superpositions of states are represented naturally, potentially enabling efficient evaluation of the pertinent matrix elements in arbitrary bases. In particular, Ref.~\citenum{tan2022sign} introduces a quantum circuit (Fig.~\ref{fig:sse_circuit}) evaluating these matrix elements while sampling the state-index space classically. Therefore, on the quantum device, the no-branching condition (see \cref{sec:sse}) is not needed, and one may proceed to add a sufficiently large constant to the Hamiltonian, evading negative weights.

Let us consider the quantum SSE procedure for a simple Ising model,
\begin{equation}
    \hat{H} = -\sum_{\langle ij\rangle} J_{jk} \sigma_j^x \sigma_k^x.
\end{equation}
where the spectrum of the Hamiltonian is shifted by a sufficiently large number $K = \sum_{\langle jk \rangle} |J_{jk}|$.
One can define an initial state $|\Psi_{\mathrm{in}}\rangle$ on a ($N + n$)-qubit circuit,
\begin{equation}
    |\Psi_{\mathrm{in}}\rangle = |a_A\rangle |+_{B_1}\rangle \cdots |+_{B_n}\rangle,
\end{equation}
with $n$ being the order of the SSE expansion, $N$ being the number of particles and$|+_{B_i}\rangle = (|0\rangle + |1\rangle) / \sqrt{2}$.
Accommodating the additional shift of the Hamiltonian, one defines a controlled unitary 
\begin{align} \label{eq:sse_unitary}
    &U_{A,B_i} |a_A\rangle |X_{B_i}\rangle = \nonumber \\
    &\begin{cases}
        \qquad \qquad \qquad \quad   1_A \:\: |a_A\rangle |X_{B_i}\rangle \quad &\text{if} \quad |X_{B_i}\rangle = |0_{B_i}\rangle\\
        \left[ \mathrm{sgn}(h_b) \otimes_{j=1}^N \sigma_{b^j}^{A_j} \right] |a_A\rangle |X_{B_i}\rangle \quad &\text{if} \quad |X_{B_i}\rangle = |1_{B_i}\rangle,
    \end{cases}
\end{align}
which extracts the relative weight of a state-index configuration by evaluating the estimator
\begin{equation}
    |\langle \Psi_{\mathrm{in}} | U_{A, B_i} | \Psi_{\mathrm{in}} \rangle|^2 = \left|\frac{\langle a_A | \hat{H}_{b_n} \cdots \hat{H}_{b_1} | a_A \rangle}{2^n h_{b_n} \cdots h_{b_1}}\right|^2.\label{eq:qab}
\end{equation}
To evaluate the matrix elements in Eq.~\eqref{eq:qab} up to a target additive precision $\epsilon$, the circuit needs to be measured a total number of $\mathcal{O}(\frac{1}{\epsilon^2})$ times.
The full circuit to compute the expectation values is provided in Fig.~\ref{fig:sse_circuit}.
Although quantum SSE is able to alleviate the sign problem for arbitrary Hamiltonians successfully, its circuit depth depends on the inverse temperature $\beta$. Low temperatures lead to non-negligible contributions from increasingly long operator strings, limiting the applicability of quantum SSE.
It is yet to be seen whether quantum SSE offers a substantial advantage over classical SSE, as it does not control the sign problem either.

\subsection{Quantum Minimally Entangled Typical Thermal States}\label{sec:metts_quantum}
QMETTS, the quantum analog of METTS (see \cref{sec:classical_metts}) was realized with quantum imaginary time evolution (QITE)~\cite{motta2020determining}.
The METTS algorithm was originally integrated into the finite temperature density matrix renormalization group (DMRG), which is the \textit{de facto} method for low-dimensional systems. One aims to extend the applicability of METTS beyond low-dimensional many-body systems using QMETTS.
The basic procedure of QMETTS is summarised in \cref{alg:QMETTS}. A related stochastic approach was also developed for finite-temperature dynamical observables~\cite{PRXQuantum.2.010317}. 
\begin{algorithm}[H]
\caption{QMETTS Algorithm}
\begin{algorithmic}[1]
\State Start from a product state and perform imaginary time evolution (QITE) up to time \(\beta\).
\State Measure the expectation value of an operator \(\hat{O}\) to obtain its thermal average.
\State Measure a product operator, such as \(\hat{Z}^1 \hat{Z}^2 \cdots \hat{Z}^N\), to collapse back to a random product state.
\State Repeat from step 1.
\end{algorithmic}
\label{alg:QMETTS}
\end{algorithm}
The main bottleneck in QMETTS is the QITE process, which requires quantum resources with the quantum circuit depth exponentially increasing with system correlation length and linearly growing with $\beta$.
QITE was improved by several algorithmic improvements: exploiting symmetries to minimize quantum resources, optimizing circuits to decrease the depth, and implementing error-mitigation techniques to enhance the quality of raw hardware data~\cite{PRXQuantum.2.010317}.
Recently, an adaptive variational version of QITE, using the McLachlan variational principle~\cite{mcardle2019variational,yuan2019theory,PhysRevA.106.062416,gomes2021adaptive} 
for time propagation, has been incorporated into QMETTS. This approach dynamically generates 
compact and problem-specific quantum circuits, making them suitable for noisy intermediate-scale 
quantum (NISQ) hardware~\cite{getelina2023adaptive}.
To demonstrate a quantum advantage of QMETTS, one needs to find a chemical example with a limited correlation length that a classical tensor network method cannot efficiently simulate. 
This question appears to be open at the time of writing. 


\add{
\subsection{Quantum time series assisted Monte Carlo}
Quantum imaginary time evolution typically requires deep quantum circuits,
which limits its practical implementation on current quantum devices~\cite{motta2020determining,PRXQuantum.2.010317}.
To address this limitation, heuristic time-series algorithms~\cite{PRXQuantum.2.020321,PhysRevB.107.L140410} have been developed that combine quantum evolution with classical Monte Carlo post-processing to compute equilibrium properties in both microcanonical and canonical ensembles~\cite{PRXQuantum.2.020321}.
The effectiveness of this approach was recently demonstrated experimentally on a $2\times 8$ Fermi-Hubbard model~\cite{PRXQuantum.5.030323}.

The time-series algorithm introduced in Ref.~\cite{PRXQuantum.2.020321} 
employs a filtering operator analogous to quantum phase estimation (QPE) 
to exclude energies outside a target energy window~\cite{ge2019faster,PhysRevB.101.144305}.  
Rather than directly applying this filter to prepare physical states, 
the algorithm extracts physical observables through interferometric measurements. 
This approach avoids the need for explicit filtered state preparation while still 
yielding expectation values that converge to the microcanonical ensemble 
average with minimal variance.
The filtered state has local density of state:
\begin{equation}
    D_{\delta, \psi}(E)=\langle\psi| P_\delta(E)|\psi\rangle,
\end{equation}
with the filtering operator approximated as~\cite{ge2019faster,PRXQuantum.2.020321}
\begin{equation}
    P_\delta(E) \simeq e^{-(\hat{H}-E)^2 / 2 \delta^2} 
\end{equation}
where $E$ is the energy of the target state $|\psi\rangle$ and $\delta$ 
is the width of the filter.
The filtering operator can be decomposed into a sum of time-evolution operators,
\begin{equation}
        \hat{P}_\delta(E) \approx \sum_{-R}^R c_m e^{-i(H-E) t_m}
\end{equation}
where $c_m=\left(1 / 2^M\right)\binom{M}{M / 2-m}$, $R=\lfloor x / \delta\rfloor$
in which $x$ is a scale controlling the truncation, and the time series
$t = 2m/N$ where $N$ is the system size. 
It was also shown that reduced number of measurements is possible by choosing
$R\propto 1/\sqrt{N}$ if the initial states are product states.
One can compute the microcanonical observables of $\hat{O}$,
\begin{equation}\label{eq:microcano}
    \langle\hat{O}_\delta(E)\rangle=\frac{\sum_{\left|\psi_p\right\rangle} D_{\psi_p, \delta}(E) O_p}{\sum_{\left|\psi_p\right\rangle} D_{\psi_p, \delta}(E)}
\end{equation}
where $O_p=\left\langle\psi_p\right| \hat{O}\left|\psi_p\right\rangle$ 
within the product state basis $\left\{\left|\psi_p\right\rangle, p=1, \ldots, 2^N\right\}$. 
It turns out that the local density of states $D_{\psi_p, \delta}$ can be obtained by 
computing for the Loschmidt amplitudes $\langle\psi| e^{-i H t}|\psi\rangle$ 
on a quantum device (either with quantum computers or analog quantum simulators).
Eq.~\eqref{eq:microcano} can be sampled with Metropolis-Hastings Monte Carlo with the accept probability 
\begin{equation} 
\min \left(1, \frac{D_{\left|\psi^{\prime}\right\rangle,\delta}(E)}{D_{|\psi\rangle,\delta}(E)} \frac{P_{\psi \rightarrow \psi^{\prime}}}{P_{\psi^{\prime} \rightarrow \psi}}\right) 
\end{equation} 
Finally, the finite temperature equilibrium properties can also be obtained from
importance sampling according to $\textrm{e}^{-\beta E} D_{|\psi\rangle,\delta}(E)$.

Although in practice the algorithm does not require evolution for long times,
it still requires a significant number of measurements to obtain reliable results.
A detailed discussion of the practical quantum advantage and resource requirements
of this algorithm can be found in Ref.~\cite{PRXQuantum.5.030323}.
}

\section{Quantum Algorithms for Quantum dynamics}\label{sec:quantum_for_dynamics}
\add{
While quantum computers are naturally suited for simulating real-time dynamics, 
significant challenges remain in practice. Preparing physically relevant initial states 
can be computationally hard even for ideal quantum computers~\cite{Kempe2006Jul}. 
Additionally, current noisy quantum devices face substantial hardware limitations, 
including decoherence and gate errors, that restrict their ability to implement 
long-time quantum evolution.
Several hybrid quantum-classical algorithms~\cite{Yang_AcceleratedQuantumMonte2021b,eklund2024hybrid,Hamilton2017Oct,Wang2019Dec,Neville2017Dec,Zhong2020Dec,Zhong2021Oct,Madsen2022Jun,
dutta2024simulating,PhysRevLett.129.120505} have been developed that combine the unique capabilities
of quantum computers with classical methods to enhance the simulation of quantum dynamics.
In this section, we dive into and analyze the ideas behind 
several representative approaches, while examining their underlying remaining challenges.
}

\subsection{Quantum Circuit Monte Carlo}\label{subsec:qcmc}
Much like most of the methods mentioned earlier, QMC suffers from the sign problem when applied to real-time dynamics. Yang \textit{et al.} proposed a new way of mitigating the sign problem by using a hybrid quantum--classical algorithm termed quantum-circuit Monte Carlo (QCMC)~\cite{Yang_AcceleratedQuantumMonte2021b}. Quantum computers are used to tackle the sign-problem-inducing parts of the time evolution, while the rest of the algorithm is still carried out on a classical device. On the quantum computer, the time evolution operator is implemented as a summation of unitary operators $U(s)$ over auxiliary fields $s$:
\begin{align}
    \mathrm{e}^{-\mathrm{i}H\Delta t} = \sum_s c(s) U(s)
\end{align}
Specifically, in this work, the authors proposed summation formulae based on extensions of the Lie--Trotter--Suzuki product~\cite{Suzuki_1990319, Yoshida_1990262}, including expansions with Pauli operators and replacing the leading order with rotation operator, with the goal of minimizing the number of gates. These formalisms can be viewed as perturbative expansions of the Lie-Trotter-Suzuki product, which parallels the idea of DiagMC (Sec.~\ref{sec:diagmc}). 
\add{We note the recent work on perturbative quantum simulations (PQS)~\cite{PhysRevLett.129.120505}, 
which aims to simulate the dynamics of large quantum systems by evolving their smaller subsystems.
Like DiagMC, PQS samples trajectories consisting of local operations that approximate 
the time evolution over infinitesimal time steps. However, PQS and QCMC employ different 
approaches for sampling and expanding the real-time propagator.}
On a fault-tolerant quantum computer, the circuit depth of the time evolution simulation scales polynomially using these algorithms.

More specifically, they partition the real-time propagator into a product of its first-order Trotter-Suzuki $S_1(\Delta t)$ approximation and a correction term $V_1(\Delta t)$ according to the first-order product formula,
\begin{equation} \label{eq:trotter_prop}
    \mathrm{e}^{-\mathrm{i} H \Delta t} = V_1(\Delta t) S_1(\Delta t),
\end{equation}
with
\begin{align}
    S_1(\Delta t) = \mathrm{e}^{-\mathrm{i} H_M \Delta t} \cdots \mathrm{e}^{-\mathrm{i} H_1 \Delta t}
\end{align}
and $H = \sum_{i=1}^{M} H_i$, where $H_i$ are Hermitian. From the expression for $S_1(\Delta t)$ we can readily obtain 
\begin{equation} \label{eq:realtime_correction}
    V_1(\Delta t) = \mathrm{e}^{-\mathrm{i} H \Delta t} \mathrm{e}^{\mathrm{i} H_M \Delta t} \cdots \mathrm{e}^{\mathrm{i} H_1 \Delta t}.
\end{equation}
Taylor expanding each exponential in Eq.~\eqref{eq:realtime_correction} and regrouping the resulting terms leads to
\begin{equation}
    V_1(\Delta t) = \sum_{k=0}^{\infty} F^{(k)}_1(\Delta t).
\end{equation}
These terms $F^{(k)}_1(\Delta t)$ admit to the summation
\begin{align}
    &F^{(k')}_1(\Delta t) = \sum_{k, k_1, \cdots, k_M}^{\infty} \delta_{k',k + \sum k_i } \left[\prod_{j=1}^M \frac{(\mathrm{i} h_j \Delta t)^{k_j}}{k_j !} \right] \nonumber \\
    &\times \sum_{j_1, \cdots, j_k}^{M} \frac{\prod^k_{a=1} (-\mathrm{i} h_{j_a} \Delta t)}{k!} \sigma_{j_k} \cdots \sigma_{j_1} \sigma_{1}^{k_1} \cdots \sigma_{M}^{k_M}.
\end{align} 
To evaluate this expression, indices $k$ and $k'_j, k_j$ are sampled from appropriate Poisson distributions for all $j$.  
It is this evaluation that resembles previously discussed ideas from DiagMC most closely.
Ultimately, this sampling prescription can be used to evaluate the correlation function $\langle \psi_f | \mathrm{e}^{\mathrm{i} H t} O \mathrm{e}^{-\mathrm{i} H t} | \psi_i \rangle$ on a quantum device. Wavefunctions $|\psi_i\rangle$ and $|\psi_f\rangle$ correspond to initial and final state, $O$ is an arbitrary operator.
As this approach still suffers from the sign problem (unless the entire dynamics is performed on the quantum computer), the applicability and scope of the method are currently limited.
Nonetheless, recently, a similar QCMC algorithm was used to perform imaginary-time evolution by sampling random quantum circuits with finite depth~\cite{Huo2023Feb}. In this work, the authors demonstrated the promise of QCMC even without a fully fault-tolerant quantum computer by applying it to ground state calculations.

\subsection{Hybrid Path Integral Monte Carlo for Time Correlation Functions}
Recently, Eklund and Ananth reported a hybrid Path Integral Monte Carlo (hPIMC) algorithm to calculate real-time thermal correlation functions~\cite{eklund2024hybrid}. The classical Path Integral Monte Carlo (PIMC) algorithm was first proposed by Barker in 1979~\cite{barker1979pimc}. While the method can be used in high-dimensional computations, there are still system size limitations in the computation of the time-evolution operator matrix elements.

In the work from Eklund and Ananth, these time-evolution operator matrix elements are computed on a quantum computer, and they demonstrate that this can be performed accurately using the probabilistic imaginary-time evolution (PITE) algorithm~\cite{2022pite}. The symmetrized real-time time-correlation function between two operators $A$ and $B$ is given by 
\begin{equation}
    C_{AB}(t) = \frac{1}{Z}\text{Tr}\left[U^\dag(t_c)AU(t_c)B\right]
\end{equation}
where $U(t_c) = \mathrm{e}^{-\mathrm{i}Ht_c}$, and $t_c = t-\mathrm{i}\beta/2$. In PIMC, $U(t_c)$ is discretized into the Trotter product of $N$ time evolution operators such that \begin{equation}
    \tilde{C}_{AB}(t) = \frac{1}{Z}\text{Tr}\left[\left(\prod_{k=1}^N \tilde{U}^\dag(\Delta t_c)\right)A\left(\prod_{k=1}^N\tilde{U}(\Delta t_c)\right)B\right]
\end{equation}
where $\tilde{U}$ is the approximation for the unitary $U$ and $\Delta t_c = t_c/N$. If the Hamiltonian takes the form $H = H_1 + H_2$, where $[H_1, H_2] \neq 0$, $U(t_c)$ is approximated with a 2$k$-th order Trotter--Suzuki product. 

If we expand out the operator in a basis $\{\phi_j\}$, each Monte Carlo iteration $k$ involves computing matrix elements $\{\langle \phi_{j_{k+1}}|\tilde{U}(\Delta t_c)|\phi_{j_k} \rangle\}$ and $\{\langle \phi_{j_{k+1}}|\tilde{U}^\dag(\Delta t_c)|\phi_{j_k} \rangle\}$. In classical PIMC, these matrix elements are computed by diagonalizing the Hamiltonian in the basis, so for basis size $M$, we require $O(M^{3d})$ cost for a $d$-dimensional system. The overall runtime for classical PIMC is $O(N\tau + M^{3d})$ where $\tau$ is the number of Monte-Carlo iterations determined by the desired error, and the space requirement is $O(M^{2d})$. In hPIMC, each matrix element is computed on the quantum computer, eliminating the need for diagonalization. Instead, $2N\tau$ calls are made to the oracle, which computes the matrix elements. Therefore, the total runtime of hPIMC is $O(N\tau Q(\tilde{U}))$, where $Q$ is the cost of executing the oracle. The space requirement is $O(dn) + \text{ Anc.}(\tilde{U})$ where $n = \log_2(M)$ is the number of qubits required, and $\text{Anc.}(\tilde{U})$ is the number of ancilla qubits required to implement the time-evolution unitary. The efficiency of the algorithm, therefore, depends on the efficiency of the oracle, and there is some potential for significant speedup in runtime compared to a standard PIMC implementation.
Nevertheless, this algorithm also suffers from the dynamical sign problem, so its practical applicability and scope to large-scale problems still need to be examined.
\subsection{Boson Sampling for Molecular Vibronic Dynamics}
Boson sampling~\cite{Aaronson2011Jun} and its variant, Gaussian boson sampling (GBS)~\cite{Hamilton2017Oct,Wang2019Dec}, have demonstrated the quantum advantage in experiments for artificial problems~\cite{Neville2017Dec,Zhong2020Dec,Zhong2021Oct,Madsen2022Jun}.
However, it remains an open question whether they can be applied to real-world problems of practical interest and demonstrated to have a quantum advantage over classical methods~\cite{dutta2024simulating}.

The first theoretical attempt to apply GBS to study chemical problems was made with the computation of molecular vibronic spectra based on the GBS device made of squeezed states of light coupled to a boson sampling optical network~\cite{Huh2015Sep}.
With Fermi's golden rule and the Condon approximation, the optical absorption and emission spectra at zero temperature can be written as
\begin{equation}\label{eq:abs_emi}
S(\omega)=\left|\mu\right|^2 \sum_{n}\left|\langle\phi^i_0(\mathbf{q}_{\mathrm{i}}) | \phi^f_{n}(\mathbf{q}_{\mathrm{f}})\rangle\right|^2
\delta\left(\omega+\omega_0-\omega_n\right),
\end{equation}
where the frequency-dependent prefactor is omitted, $\mu$ is the transition dipole moment, and  $|\phi^i_0\rangle$ is the initial state of the transition, which corresponds to the ground state or the lowest excited state with energy $\omega_0$ for the absorption or emission spectra, respectively.
$|\phi_n^f\rangle$ corresponds to the final states of the transition.
The normal modes of the initial state and final states' potential of energy surfaces (PES) are shifted by $\Delta \mathbf{q}$ after applying the Duschinsky rotation $\mathbf{S}$ for mode mixing,
\begin{equation}
    \mathbf{q}_{\mathrm{f}}=\mathbf{S} \mathbf{q}_{\mathrm{i}}+\Delta \mathbf{q}.
\end{equation}
The key quantity to solve \cref{eq:abs_emi} is the Franck--Condon factor (FCF)
\begin{equation}
\textrm{FCF} = \left|\langle\phi_0^i(\mathbf{q}_{\mathrm{i}}) | \phi_{n}^f(\mathbf{q}_{\mathrm{f}})\rangle\right|^2,    
\end{equation}
which corresponds to a vibronic transition between states of two harmonic oscillators.
Solving in \cref{eq:abs_emi} involves $d^N$ summations if the maximal boson occupation for each mode is truncated at $d$ for $N$ vibrational modes in total, which is naively exponentially hard for classical computers.
Since FCFs with larger amplitudes will contribute more to the spectra intensity, one only needs to sample that $ n$ having large FCFs.
However, it is hard for classical algorithms to determine which $n$ contributes the most importance unless computing their FCF directly~\cite{Santoro2007Feb}.

Huh \textit{et al.}~\cite{Huh2015Sep} proposed to directly measure the FCF on a photonic quantum device, 
and naturally, the more important $n$ has a larger probability of being sampled.
The quantum measurement of FCF is performed with the Doktorov rotation circuit and the photon Fock states,
\begin{equation}
    \textrm{FCF}=|\langle\boldsymbol{m}|\hat{U}_{\text {Dok }}| \boldsymbol{n}\rangle|^2,
\end{equation}
where $|\boldsymbol{n}\rangle=\left|n_1, n_2, \ldots, n_N\right\rangle$ describes the ground state with $n_i$ excitations for the $i-$th mode (for zero temperature, $|\boldsymbol{n}\rangle$ is the ground state (mathematically a vacuum state), and $|\boldsymbol{m}\rangle$ denotes the excited state). 
After this original theoretical work, a series of experimental works using GBS with a special focus on molecular vibronic spectra~\cite{Shen2018Jan,Clements2018Nov,Wang2020Jun} were implemented. Additional theoretical studies addressed more complex models, such as the non-Condon effect~\cite{Jnane2021Jul}, anharmonic PESs~\cite{Wang2022Jul}, and vibrational dynamics~\cite{Jahangiri2020Nov}.

Demonstrating that the GBS algorithm offers a quantum computational advantage for calculating molecular vibronic spectra requires solving the challenges in computing spectra (\cref{eq:abs_emi}), a task that is exponentially challenging for classical computers using a brute-force sum-of-states approach. 
However, an exact classical method exists for efficiently computing the time correlation function in the time domain, applicable when there is no anharmonicity. This method, known as the thermal vibrational correlation function (TVCF)~\cite{Peng2007Aug,*Peng2007Mar,Shuai2020Nov}, challenges the potential advantages of boson sampling.
\add{A recently developed classical sampling algorithm 
demonstrates better performance in simulating the 
ideal distribution compared to experimental implementations, 
further challenging the suggested quantum advantage of GBS~\cite{Oh2024Sep}.}



\section{Summary}\label{sec:conclusion}

In this review, we have explored an interface between quantum computing and computational chemistry, focusing particularly on the complex sampling tasks essential for studying chemical systems' ground state, equilibrium, and non-equilibrium properties. 
Due to their common occurrences in computational chemistry, tackling these tasks is a good target for quantum algorithms. 
Current quantum hardware still grapples with noise, coherence times, and qubit scalability limitations. Nonetheless, these challenges are driving innovative developments in both fully quantum and quantum--classical hybrid algorithms as we summarized in this Review.

We have reviewed various sampling approaches, from hybrid quantum--classical algorithms to fully quantum algorithms, with a special emphasis on the Monte Carlo methods and other relevant sampling techniques.
We have summarized the theoretical frameworks of quantum and classical algorithms and discussed the practical challenges of these algorithms. 
The rapid advancement of quantum simulation presents a moving target, and this Review captures only a transient moment in the ongoing progress. 

We hope our Review offers new perspectives and useful background to quantum information scientists and computational chemists developing quantum and classical algorithms for complex sampling problems in chemistry, materials science, and physics.
While we have not yet identified a good use case for these quantum algorithms with a definitive quantum advantage,
this continued effort in developing and assessing new algorithms
will bring us one step closer to a practical quantum advantage.




\begin{acknowledgments}
This work was supported by Harvard University’s startup funds. Additionally, the work of T.J., H.D., and J.L. was supported by the DOE Office of Fusion Energy Sciences ``Foundations for Quantum Simulation of Warm dense Matter'' project and Google's research gift. N.M. acknowledges support by the Department of Energy Computational Science Graduate Fellowship under award number DE-SC0021110.
\end{acknowledgments}

\bibliography{references}
\end{document}